\documentclass[twocolumn]{emulateapj}

\lefthead{Law et al. 2011}
\righthead{WFC3 morphologies}

\slugcomment{DRAFT: \today}

\begin{document}

\newcommand{\msun}{\ensuremath{\rm M_\odot}}
\newcommand{\msunyr}{\ensuremath{\rm M_{\odot}\;{\rm yr}^{-1}}}
\newcommand{\Ha}{\ensuremath{\rm H\alpha}}
\newcommand{\Hb}{\ensuremath{\rm H\beta}}
\newcommand{\lya}{\ensuremath{\rm Ly\alpha}}
\newcommand{\Ntwo}{[\ion{N}{2}]}
\newcommand{\kms}{\textrm{km~s}\ensuremath{^{-1}\,}}
\newcommand{\ztwo}{\ensuremath{z\sim2}}
\newcommand{\zthree}{\ensuremath{z\sim3}}
\newcommand{\feh}{\textrm{[Fe/H]}}
\newcommand{\afeh}{\textrm{[$\alpha$/Fe]}}
\newcommand{\nifeh}{\textrm{[Ni/Fe]}}
\newcommand{\othree}{\textrm{[O\,{\sc iii}]}}
\newcommand{\otwo}{\textrm{[O\,{\sc ii}]}}
\newcommand{\ntwo}{\textrm{[N\,{\sc ii}]}}

\newcommand{\sitwo}{\textrm{Si\,{\sc ii}}}
\newcommand{\oone}{\textrm{O\,{\sc i}}}
\newcommand{\ctwo}{\textrm{C\,{\sc ii}}}
\newcommand{\sifour}{\textrm{Si\,{\sc iv}}}
\newcommand{\cfour}{\textrm{C\,{\sc iv}}}
\newcommand{\fetwo}{\textrm{Fe\,{\sc ii}}}
\newcommand{\altwo}{\textrm{Al\,{\sc ii}}}
\newcommand{\hetwo}{\textrm{He\,{\sc ii}}}

\newcommand{\hst}{{\it HST}-WFC3}

\shortauthors{Law et al.}
\shorttitle{Morphological Properties of $z\sim1.5-3.6$ Star Forming Galaxies}

\title{An HST/WFC3-IR Morphological Survey of Galaxies at $z = 1.5-3.6$: I.  Survey Description and Morphological Properties of Star Forming Galaxies}\thanks{Based 
in part on data obtained at the W. M. Keck Observatory, which is operated as a scientific partnership among the California Institute of Technology, the University of 
California, and NASA, and was made possible by the generous financial support of the W. M. Keck Foundation.}

\author{David R.~Law\altaffilmark{1,2}, Charles C. Steidel\altaffilmark{3}, Alice E. Shapley\altaffilmark{2}, Sarah R. Nagy\altaffilmark{2}, Naveen A. Reddy\altaffilmark{1,4}, \& Dawn K. Erb\altaffilmark{5}}

\altaffiltext{1}{Hubble Fellow.}
\altaffiltext{2}{Department of Physics and Astronomy, University of California, Los Angeles, CA 90095;
drlaw, aes@astro.ucla.edu, snagy@ucla.edu}
\altaffiltext{3}{California Institute of Technology, MS 249-17, Pasadena, CA 91125; ccs@astro.caltech.edu}
\altaffiltext{4}{National Optical Astronomy Observatories, 950 N. Cherry Ave., Tucson, AZ 85719}
\altaffiltext{5}{Department of Physics, University of Wisconsin-Milwaukee, P.O. Box 413, Milwaukee, WI 53201}

\begin{abstract}

We present the results of a 42-orbit {\it Hubble Space Telescope} Wide-Field Camera 3 ({\it HST}/WFC3) survey of the rest-frame optical morphologies
of  star forming galaxies with spectroscopic redshifts in the range $z=1.5-3.6$.  The survey consists of 42 orbits of F160W imaging 
covering $\sim 65$ arcmin$^2$ distributed widely across the sky and
reaching a depth of  27.9 AB for a $5\sigma$ detection within a 0.2 arcsec radius aperture.
Focusing on an optically selected sample of 306 star forming galaxies with stellar masses in the range $M_{\ast} = 10^9 - 10^{11} M_{\odot}$, we find that 
typical circularized effective half-light radii range from $\sim 0.7 - 3.0$ kpc and 
describe a stellar mass - radius relation as early as $z\sim3$.
While these galaxies are best described by an exponential surface brightness profile (Sersic index $n \sim 1$), 
their distribution of axis ratios  is strongly inconsistent with a population of inclined exponential disks 
and is better reproduced by triaxial stellar systems with minor/major and intermediate/major axis ratios $\sim 0.3$ and 0.7 respectively.
While rest-UV and rest-optical morphologies are generally similar for a subset of galaxies with {\it HST}/ACS imaging data, differences are more pronounced at higher masses
$M_{\ast} > 3 \times 10^{10} M_{\odot}$.  
Finally, we discuss galaxy morphology in the context of efforts to constrain the merger fraction, finding that 
morphologically-identified mergers/non-mergers generally have insignificant differences
in terms of physical observables such as stellar mass and star formation rate, 
although merger-like galaxies selected according to some criteria have statistically smaller effective radii and correspondingly larger $\Sigma_{\rm SFR}$.

\end{abstract}


\section{INTRODUCTION}

In recent years our understanding of the broad global characteristics of galaxies in the young universe has grown considerably.  Using rest-frame UV and optical spectroscopy and 
multi-wavelength broadband photometry it has been possible to estimate their stellar and dynamical masses, average metallicities, ages, and star formation rates across cosmic time 
from $z>6$ to the present day (e.g., Shapley et al. 2005; Cowie \& Barger 2008; Maiolino et al. 2008; Stark et al. 2009).   Such studies indicate that the majority
of structures observed in the local universe were
already in place at $z\sim1$ (Papovich et al. 2005) and point to the era spanned by the redshift range $1.5<z<3$ as the peak epoch of both the cosmic star formation rate density
(Dickinson et al. 2003; Reddy et al. 2008) and AGN activity in the universe (e.g., Miyaji et al. 2000).

In contrast to our knowledge of the global characteristics of such galaxies from ever-expanding samples however, our understanding of their 
internal structure and evolution has been limited by their small angular size.  
It has therefore been challenging to constrain the major mode of mass assembly in these galaxies (i.e., from major/minor mergers, hot mode or cold filamentary gas accretion, etc.).
With typical half-light radii $\sim 0.2-0.3$ arcsec at $z\sim2$ (Bouwens et al. 2004; Nagy et al. 2011), 
such galaxies are barely resolved in the $\sim1$ arcsec FWHM ground-based imaging and spectroscopy that form the backbone of the observational data.

Significant efforts have therefore been invested in imaging studies capitalizing on the 
high angular resolution afforded by the {\it Hubble Space Telescope} ({\it HST}).
Early efforts to characterize the morphologies of galaxies at $z\sim1.5-3$ (e.g., Abraham et al. 1996; Giavalisco et al. 1996; Lowenthal et al. 1997;
Bouwens et al. 2004; Conselice et al. 2004; Lotz et al. 2006; Papovich et al. 2005; Law et al. 2007b; and references therein)
used the visible-wavelength surveying efficiency of the ACS (Advanced Camera for Surveys)
to demonstrate that star forming galaxies typically have irregular, clumpy morphologies unlike the well-known Hubble sequence
that has been established since $z\sim1$ (e.g., Conselice et al. 2005, Oesch et al. 2010).
Indeed, 
rest-UV luminosity 
and morphology for such galaxies appears to be only
poorly correlated with other physical observables such as stellar mass, outflow characteristics, and characteristic rotation velocity (e.g., Shapley et al. 2005; Law et al. 2007b).

Recent technological developments have permitted additional insights to be gleaned from ground-based observations using 
adaptive-optics (AO) fed imagers or integral field unit (IFU) spectrographs on 10m-class telescopes
(e.g., Law et al. 2007a, 2009; Melbourne et al., 2008ab, 2011; Stark et al., 2008; F{\"o}rster-Schreiber et al. 2009; Wright et al. 2009; Jones et al., 2010).
Such IFU spectroscopy mapping rest-frame optical nebular line emission  (redshifted into the near-IR at $z>1$) from star forming galaxies 
has suggested that high redshift star forming galaxies often have dispersion-dominated kinematics at odds with the classical picture of galaxy formation via rotationally supported thin gas disks (e.g., F{\"o}rster-Schreiber et al. 2009;
Law et al. 2009).
Instead, the dynamical evolution of these systems may be driven by gravitational instabilities within massive gas-rich clumps or low angular-momentum cosmological gas flows (e.g., Kere{\v s} et al. 2005; Bournaud et al. 2007; Genzel et al. 2008 ).  
What is immediately clear is that we do not yet understand the dynamical state of galaxies during the period when they are forming the majority of their stars.

Both early rest-UV imaging and AO IFU observations of high-redshift galaxies tend to trace regions of active star formation however, and in order to 
understand these galaxies we also wish to map the regions in which the bulk of the underlying stellar population live.
While young and old populations may have a generally similar distribution for lower-mass galaxies (e.g., Conselice et al. 2011), more significant differences exist for galaxies with larger stellar mass
(e.g., Dickinson 2000; Papovich et al. 2005; and \S \ref{uvopt.sec}).
Efforts to characterize rest-optical galaxy morphologies using ground-based instruments and/or
the {\it HST}/NICMOS camera have been made by (e.g.) Papovich et al. (2005), Franx et al. (2008), Toft et al. (2009), van Dokkum et al. (2010),
and Mosleh et al. (2011), generally finding that galaxies at $z\sim2$ 
were significantly smaller at fixed stellar mass than in the local universe.
Additionally,  {\it HST}/NICMOS work by Kriek et al. (2009) has demonstrated that
star-forming and quiescent galaxies differ substantially from each other in relative compactness of their
rest-optical morphologies,
and both differ from their kin in the local universe.

Given the narrow field of view of both ground-based AO-fed imagers (e.g., Carrasco et al. 2010) and the
{\it HST}/NICMOS camera (e.g., Conselice et al. 2011a) however, it is only recently with the 
advent of the new WFC3 camera onboard {\it HST} that it has become practical to perform wide-field morphological surveys in the near-IR that
trace rest-frame optical emission from galaxies at $z\gtrsim1$.   
The results of the first such studies
in the UDF have been reported recently in the literature (e.g., Cameron et al. 2010; Cassata et al. 2010;
Conselice et al. 2011b).
Our recent survey has greatly extended these early results by obtaining {\it HST}/WFC3-IR morphological data for 306
 $z = 1.5-3.6$ galaxies in 10 fields widely distributed across the sky for which we have obtained dense spectroscopic sampling.
 
 Preliminary results for the evolution of the stellar mass - radius  relation were presented in Nagy et al. (2011).
In this first contribution of a series of papers using the full sample, we introduce our survey and describe a selection of results concerning  evolution of the characteristic size, shape, and major merger fraction
for actively star forming galaxies.
Future contributions (Law et al. 2012, in preparation) will discuss the relation between morphology and low-ionization gas-phase kinematics, treat quiescent galaxies and AGN, and discuss 
the morphology of uniquely interesting galaxies (e.g., Q2343-BX442; Law \& Shapley 2012, in preparation) in greater detail.
This paper is organized as follows:  In \S \ref{obs.sec} we describe the {\it HST}/WFC3 observing program and review the properties of the star forming galaxy sample.
In \S \ref{stats.sec} we present postage-stamp morphologies of the galaxy sample and discuss our morphological analysis techniques.
An extended discussion of the robustness of the morphological statistics and the systematic variations
between measurement systems commonly adopted in the literature is
presented in the Appendix.
\S \ref{basiccharac.sec} summarizes the basic morphological characteristics (luminosity profile, relation to rest-UV imaging, and intrinsic 3D shape) 
of the galaxy sample, and the implications of our data
for the evolution of the stellar mass - effective radius relation
are discussed in \S \ref{massrad.sec}.
Finally, we use a variety of morphological statistics to constrain the major merger fraction and its evolution with redshift in \S \ref{mergers.sec}.
We summarize our results in \S \ref{discussion.sec}.

Throughout our analysis, we adopt a standard $\Lambda$CDM cosmology based on the seven-year WMAP results (Komatsu et al. 2011) in which
$H_0=70.4$ km s$^{-1}$ Mpc$^{-1}$, $\Omega_{\rm M}=0.272$, and $\Omega_{\Lambda}=0.728$.













\section{OBSERVATIONAL DATA}
\label{obs.sec}

\subsection{Observations and Data Reduction}
\label{obsdr.sec}

Data were obtained using the WFC3/IR camera onboard the {\it Hubble Space Telescope} (\hst) as part of the Cycle 17 program GO-11694 (PI: Law).
This program was comprised of 42 orbits using the F160W filter ($\lambda_{\rm eff} = 15369$\, \AA, which traces rest-frame $5123/3824$\, \AA\
at $z = 2/3$ respectively), divided amongst fourteen pointings in ten different survey fields (see Table \ref{fields.tab})
for a combined sky coverage of $\sim 65$ arcmin$^2$
centered on lines of sight to bright ($V \sim 17$) background QSOs.\footnote{Two fields (Q1623+26 and Q2343+12) had additional pointings
in order to include sightlines to additional bright background QSOs and to include the uniquely interesting systems Q2343-BX415 (Rix et al. 2007) and Q2343-BX418 (Erb et al. 2010).}
Each pointing had a total integration time of 8100 seconds composed of nine 900 second exposures
dithered using a custom nine-point sub-pixel offset pattern designed to uniformly sample the PSF.

The data were reduced using the MultiDrizzle (Koekemoer et al. 2002) software package to clean, sky subtract, distortion correct, and combine
the individual frames.  
The raw WFC3 frames are undersampled with a pixel scale of 0.128 arcsec; these frames
were drizzled to a pixel scale of 0.08 arcsec pixel$^{-1}$ using a pixel droplet fraction (pixfrac) of 0.7.
This combination of parameters was found to give the cleanest, narrowest point-spread function (PSF) while ensuring that the RMS variation of the final weight map
was less than $\sim 7$\% across the $136 \times 123$ arcsec field of view.  Using nine isolated and unsaturated stars in the Q1623+26 field
we estimate that the FWHM of the PSF is $0.18 \pm 0.01$ arcsec (i.e., Nyquist sampled by the 0.08 arcsec drizzled pixels), varying by less than 4\% across the detector and from field-to-field.

\subsection{The Galaxy Sample}

Our fourteen individual pointings are located within ten survey fields centered on lines of sight to bright background QSOs ($z_{\rm QSO} \sim 2.7$).
In the present contribution, we focus on actively star forming galaxies
drawn from rest-UV color-selected catalogs of $z \sim 1.5 - 3.5$ star forming galaxy candidates
constructed according to the methods described by Steidel et al. (2003, 2004) and Adelberger et al. (2004).  These catalogs are based on deep ground-based imaging
and therefore select galaxies with ${\cal R} \lesssim 27$ independent of morphology or surface brightness (since even the largest galaxies are nearly unresolved
in these seeing-limited images).  
Extensive ancillary information is available in these survey fields.  In addition to deep ground-based $U_n G {\cal R}$ optical imaging and rest-UV spectroscopy,
many of the fields also have deep ground-based $J/K_s$ imaging, {\it Spitzer} IRAC/MIPS photometry, and for Q1549+19/Q1700+64 respectively
spatially resolved {\it HST}/WFC3-UVIS and {\it HST}/ACS rest-UV imaging.  All galaxy candidates in these catalogs are detected with WFC3 at $>10\sigma$ down to $\sim 27.5$ AB.

Rather than relying on photometric redshifts, which typically have large uncertainties ($\Delta z/(1+z) \gtrsim 0.06$ at $z>1.5$; van Dokkum et al. 2009), we restrict
our attention to the subsample of galaxies with ${\cal R} \leq 25.5$ that
have been spectroscopically confirmed using Keck/LRIS rest-UV spectra
to lie in the 
redshift range $1.5 < z < 3.6$; i.e., the ``BM'' ($\langle z \rangle = 1.70 \pm 0.34$), ``BX'' ($\langle z \rangle = 2.20 \pm 0.32$), and ``LBG'' 
or $U_n$-dropout ($2.7 < z < 3.6$)
samples defined by Steidel et al. (2003, 2004).
Systemic redshifts for the majority of our galaxies were derived from rest-UV absorption/emission line centroids using the prescriptions of Steidel et al. (2010);
for 51 galaxies that have been successfully observed to date with either long-slit (Erb et al. 2006b) and/or IFU spectroscopy
(13 galaxies,  F{\"o}rster Schreiber  et al. 2009; Law et al. 2009; Wright et al. 2009) systemic redshifts were derived
from rest-optical nebular emission lines (e.g., H$\alpha$, \othree).\footnote{Nebular emission line redshifts are better indicators of the systemic redshift than
UV interstellar features at the $< 100$ \kms level; see discussion by Steidel et al. 2010).}

\begin{figure*}
\plotone{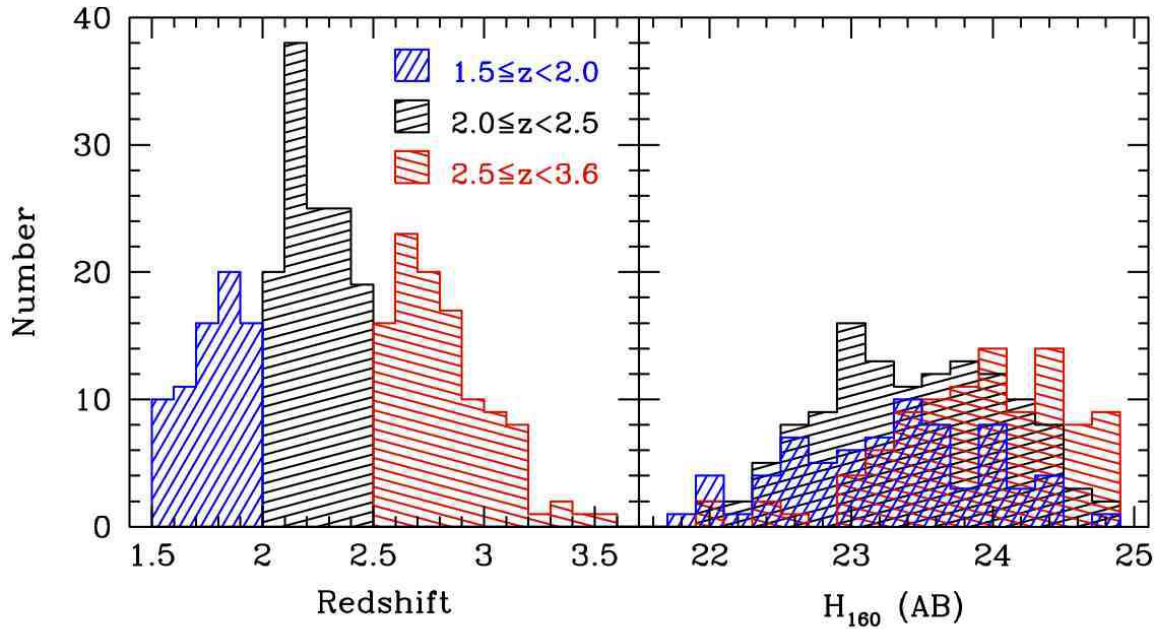}
\caption{Histograms of spectroscopic redshift and observed $H_{160}$ magnitude for the 306 star forming galaxies
in our sample.}
\label{pophist.fig}
\end{figure*}

Additionally, we omit from our sample any galaxies that lie within $\sim$ 1.5 arcsec of the edge of the WFC3-IR detector (where our dither coverage is incomplete),
or which are known to contain AGN on the basis of rest-UV spectroscopy
(24 systems; 12 bright QSOs with $H_{160} < 19$ AB, and 12 faint AGN with $H_{160} > 19$ AB).
We discuss the morphological properties of these AGN in detail in a forthcoming contribution
(Law et al. 2012, in preparation).  
The redshift and F160W magnitude distribution of the final sample of 306 galaxies are shown in Figure \ref{pophist.fig}.
As detailed in Table \ref{fields.tab} the galaxies are roughly evenly distributed amongst the 10 fields (with additional pointings in Q1623+26 and Q2343+12).
Motivated by the redshift ranges of the photometric selection criteria we loosely divide our galaxies into the three redshift ranges $z=1.5-2.0$, $z=2.0-2.5$, and $z=2.5-3.6$,
containing 72/127/107 galaxies respectively.  Although we include galaxies up to $z=3.6$ in our analysis we note 
that the galaxy sample is very sparse for $z>3.2$, as shown in Figure \ref{pophist.fig}.

\begin{deluxetable*}{lrrrcccc}
\tablecolumns{8}
\tablewidth{0pc}
\tabletypesize{\scriptsize}
\tablecaption{WFC3 Imaging Fields}
\tablehead{
\colhead{} & \colhead{R.A.} & \colhead{Decl.} & \colhead{} & \colhead{} & \colhead{} & \colhead{}\\
\colhead{Field} & \colhead{(J2000)} & \colhead{(J2000)} & \colhead{Date Observed} & \colhead{$N_{z1}$\tablenotemark{b}} & \colhead{$N_{\rm z2}$\tablenotemark{c}} & \colhead{$N_{\rm z3}$\tablenotemark{d}} & \colhead{$N_{\rm AGN}$}\tablenotemark{e}}
\startdata
Q0100+13	& 	01:03:11	& +13:16:30 & Oct 23, 2010 & 2 & 6 & 8 &   3\\
Q0142-09		& 	01:45:17	& -09:45:04  & Nov 2, 2010 & 4 & 13 & 13 & 1\\
Q0449-16		&	04:52:14	& -16:40:17  & Nov 19, 2010 & 6 & 10 & 10 & 0\\
Q1009+29	&	10:11:55	& +29:41:44 & Jan 11, 2010 & 10 & 7 & 10 & 1\\
Q1217+49	&	12:19:30	& +49:40:59 & Oct 18, 2009 & 9 & 7 & 6 & 0\\
Q1549+19	&	15:51:53	& +19:11:02 & Aug 8, 2010 & 4 & 9 & 12 & 4\\
Q1623+26\tablenotemark{a}	&	16:25:48	& +26:47:04 & Aug 6, 2010 & 7 & 36 & 24 & 1\\
...	&	16:25:58	& +26:44:49 & Aug 28, 2010 &  ... & ... & ... & ...\\
...	&	16:25:48	& +26:44:38 & Oct 8, 2010 & ... & ... & ... & ...\\
...	&	16:25:55	& +26:49:39 & Jul 8, 2010 &  ... & ... & ... & ...\\
Q1700+64	&	17:00:59	& +64:12:09 & Jan 21, 2010 & 5 & 7 & 9 & 0\\
Q2206-19		& 	22:08:53	& -19:43:56  & Oct 2, 2010 & 13 & 7 & 7 & 0\\
Q2343+12\tablenotemark{a}	&	23:46:29	& +12:48:42 & Jun 13, 2010 & 10 & 25 & 10 & 2\\
...	&	23:46:22	& +12:48:13 & Jun 14, 2010 & ... & ... & ... & ...\\
\hline
TOTAL	          &      ...                &          ...        &     ...                 & 72 & 127 & 107 & 12\\
\enddata
\label{fields.tab}
\tablenotetext{a}{Multiple overlapping pointings in the Q1623+26 and Q2343+12 fields.}
\tablenotetext{b}{Number of star forming galaxies in the range $1.5 \leq z < 2.0$.}
\tablenotetext{c}{Number of star forming galaxies in the range $2.0 \leq z < 2.5$.}
\tablenotetext{d}{Number of star forming galaxies in the range $2.5 \leq z < 3.6$.}
\tablenotetext{e}{Number of faint ($H_{160} >$ 19) broad and narrow-lined AGN in the redshift range $1.5 \leq z < 3.6$.}
\end{deluxetable*}


\subsection{Initial Segmentation Map}
\label{initialmap.sec}

Reduced {\it HST}/WFC3 images were registered to the same world coordinate system (WCS) as our deep ground-based optical/near-IR data
using $\sim 10-15$ stars per pointing.
Source Extractor (Bertin \& Arnouts 1996) was then used to perform automated object detection
(with no smoothing kernel)
and produce an initial segmentation map in which each source is assigned a unique identifier.
We set the source detection threshold to  $1.5\sigma$ with a required minimum of 10 pixels
above threshold for analysis, 32 deblending thresholds, and a minimum deblending contrast of 1\%.\footnote{Adopting reasonable
alternative values for the smoothing kernel and deblending thresholds makes an imperceptible difference to our derived morphological
statistics.}
We adopt an RMS map proportional to the inverse square root of the weight map produced by
MultiDrizzle, scaling by a
correction factor $F_A = 0.3933$ (see discussion by Casertano et al. 2000) to account for the fact that the MultiDrizzle process introduces 
correlation in the pixel-to-pixel noise.

The initial segmentation map was manually inspected for each galaxy in our sample
to ensure both that no spurious pixels were assigned to the galaxies and that each galaxy was not artificially broken into multiple objects.
Since galaxies in the redshift range $z \sim 2-3$ are well-known to be clumpy (e.g., Conselice et al. 2005; Law et al. 2007b; and references therein), 
this latter goal is non-trivial and Source Extractor frequently classifies multi-component galaxies as separate sources
(see, e.g., Colley et al. 1996).
While some neighboring clumps are likely to be physically associated with each other (if, for instance, they are
embedded in a common envelope of low surface brightness emission), it is not always obvious which clumps are part of the target
source and which are unassociated low- or high-redshift interlopers along the line of sight.
Generally, we assume that all clumps that lie within a 1.5 arcsec ($\sim 12$ kpc at $z\sim2-3$) radius about the ${\cal R}$-band centroid (i.e., the original detection image)
are physically associated with a given galaxy unless there is evidence to the contrary
(e.g., different spectroscopic redshifts, or dramatically different $U_n G {\cal R} J H K$ colors), and combine them under a single identifier.
We discuss the validity of this method with respect to the incidence of genuine vs apparent pairs in \S \ref{pairs.sec}.

In Figure \ref{postage.fig}, we present postage-stamp images of the galaxy sample
(all pixels
identified with sources other than the target $1.5 < z < 3.6$ galaxy sample have been cosmetically masked out by Gaussian random noise matched to the noise
characteristics of the background sky).  As expected on the basis of previous rest-UV morphological studies there is considerable diversity among the morphologies, which range from
compact isolated sources to multi-component systems with extended regions of diffuse emission.
While this initial segmentation map is adequate for estimating total source magnitudes and constructing postage-stamp images, it is inadequate for calculating quantitative morphologies;
we discuss construction of second-pass segmentation maps in Section \ref{segmaps.sec}.

\begin{figure*}
\plotone{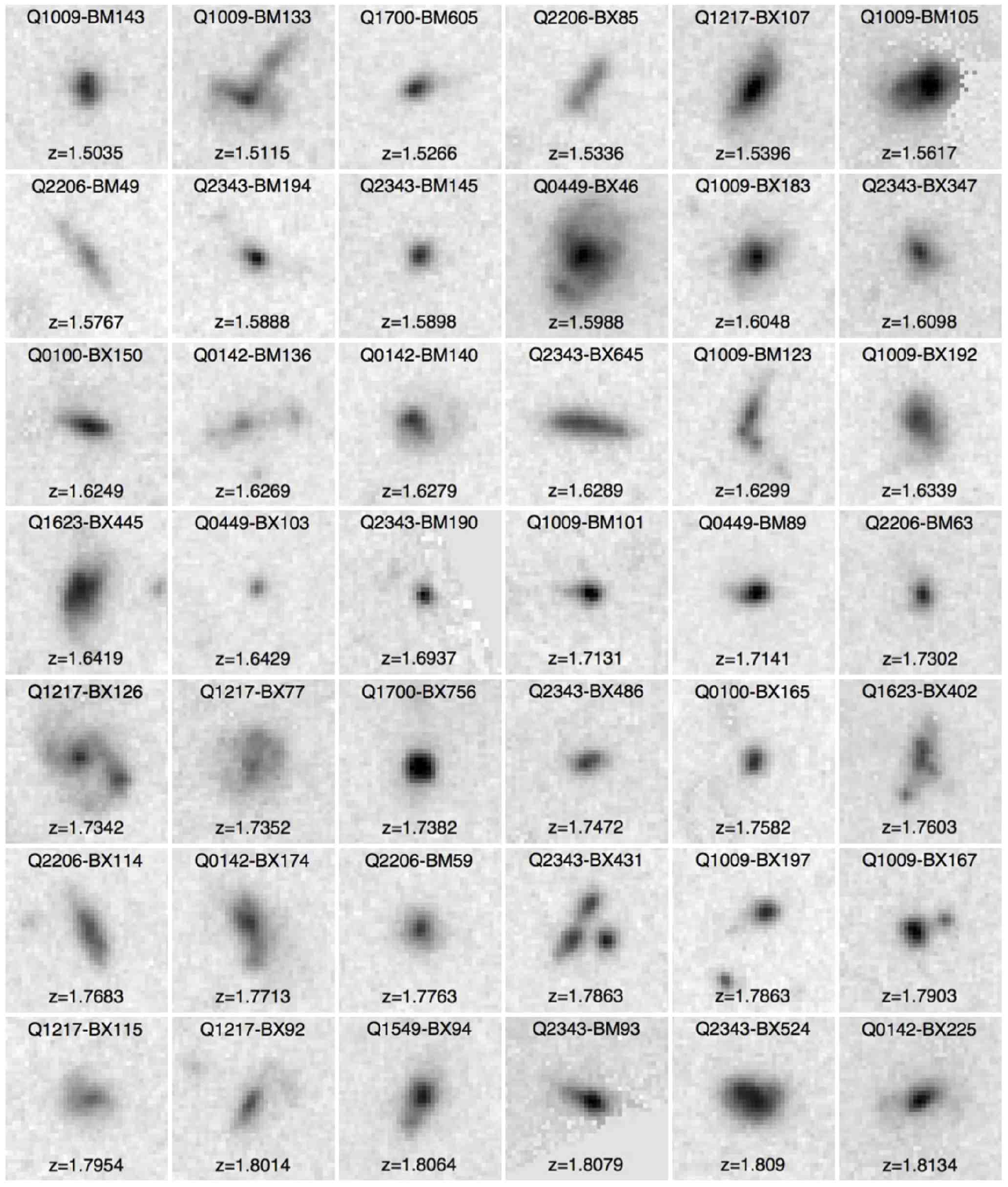}
\caption{{\it HST} WFC3 F160W rest-optical morphologies of the 306 systems in the $z=1.5-3.6$ star forming galaxy sample, sorted in order of increasing redshift.
Images are 3\arcsec\ to a side, oriented with north up and east to the left and centered on the F160W flux centroid.  
The color map has been inverted, and uses an arcsinh stretch with the black point set to 27.3 AB pixel$^{-1}$
(21.8 AB arcsec$^{-2}$).}
\label{postage.fig}
\end{figure*}

\addtocounter{figure}{-1}
\begin{figure*}
\plotone{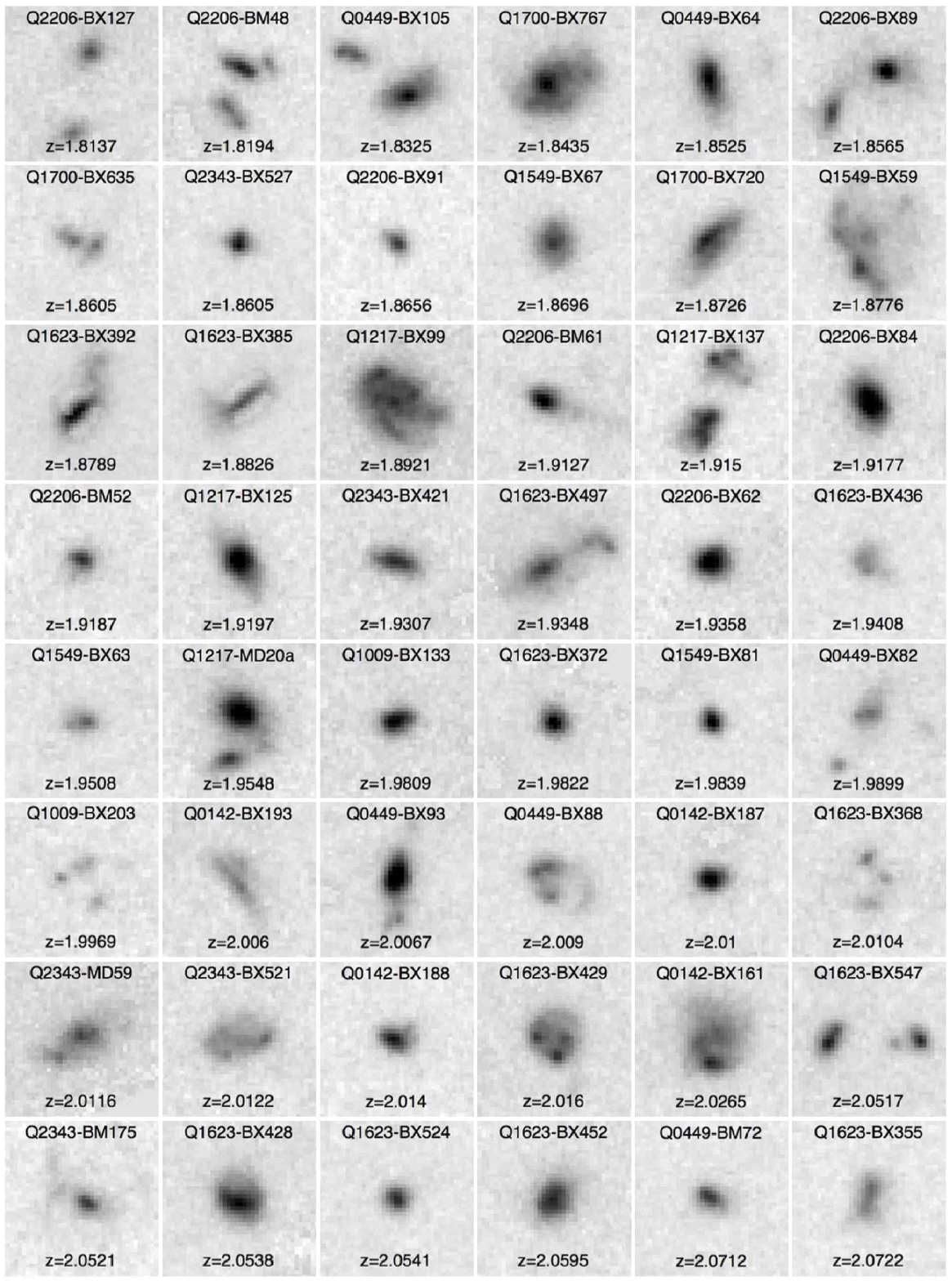}
\caption{{\it Continued}}
\end{figure*}

\addtocounter{figure}{-1}
\begin{figure*}
\plotone{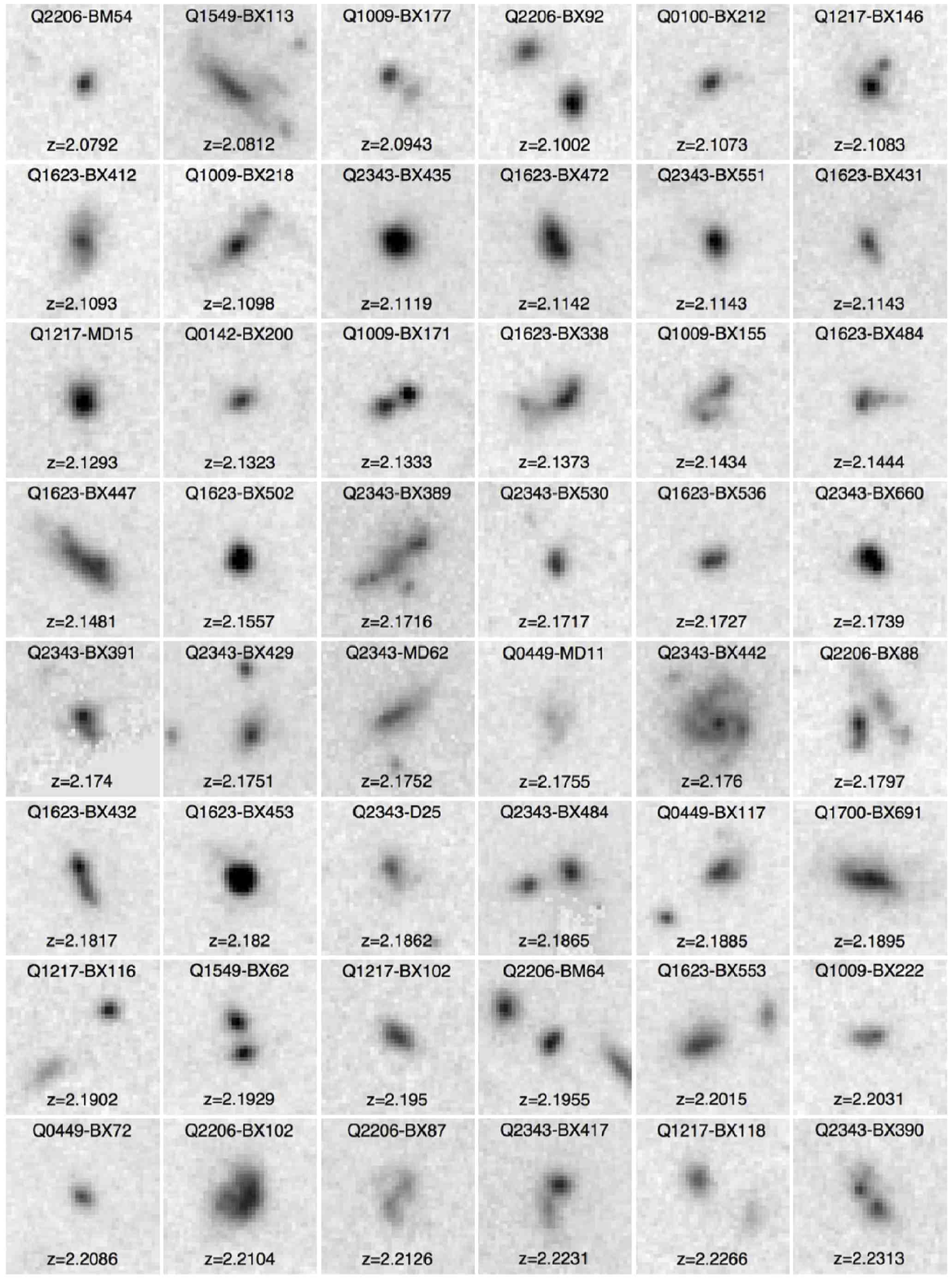}
\caption{{\it Continued}}
\end{figure*}

\addtocounter{figure}{-1}
\begin{figure*}
\plotone{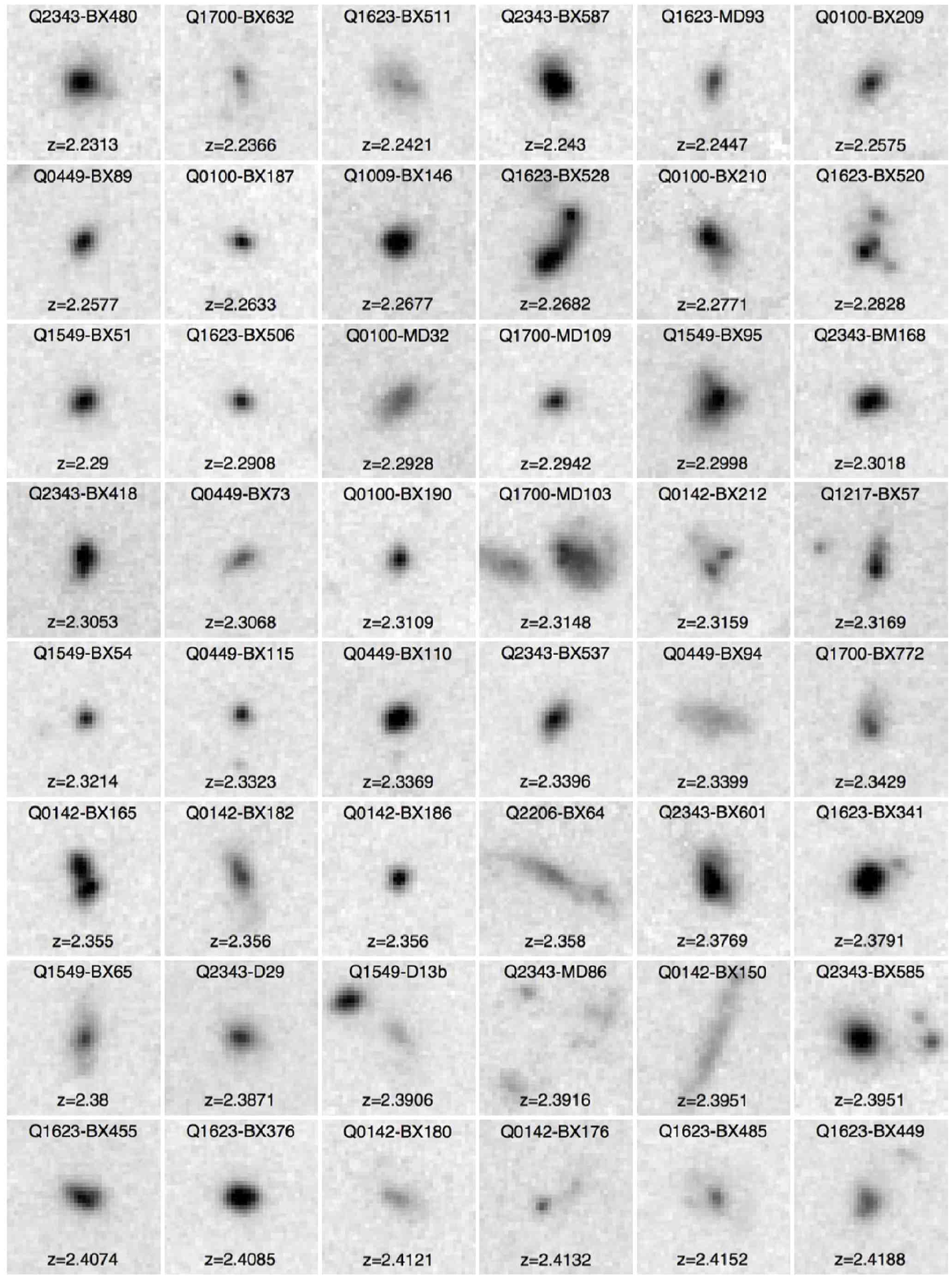}
\caption{{\it Continued}}
\end{figure*}

\addtocounter{figure}{-1}
\begin{figure*}
\plotone{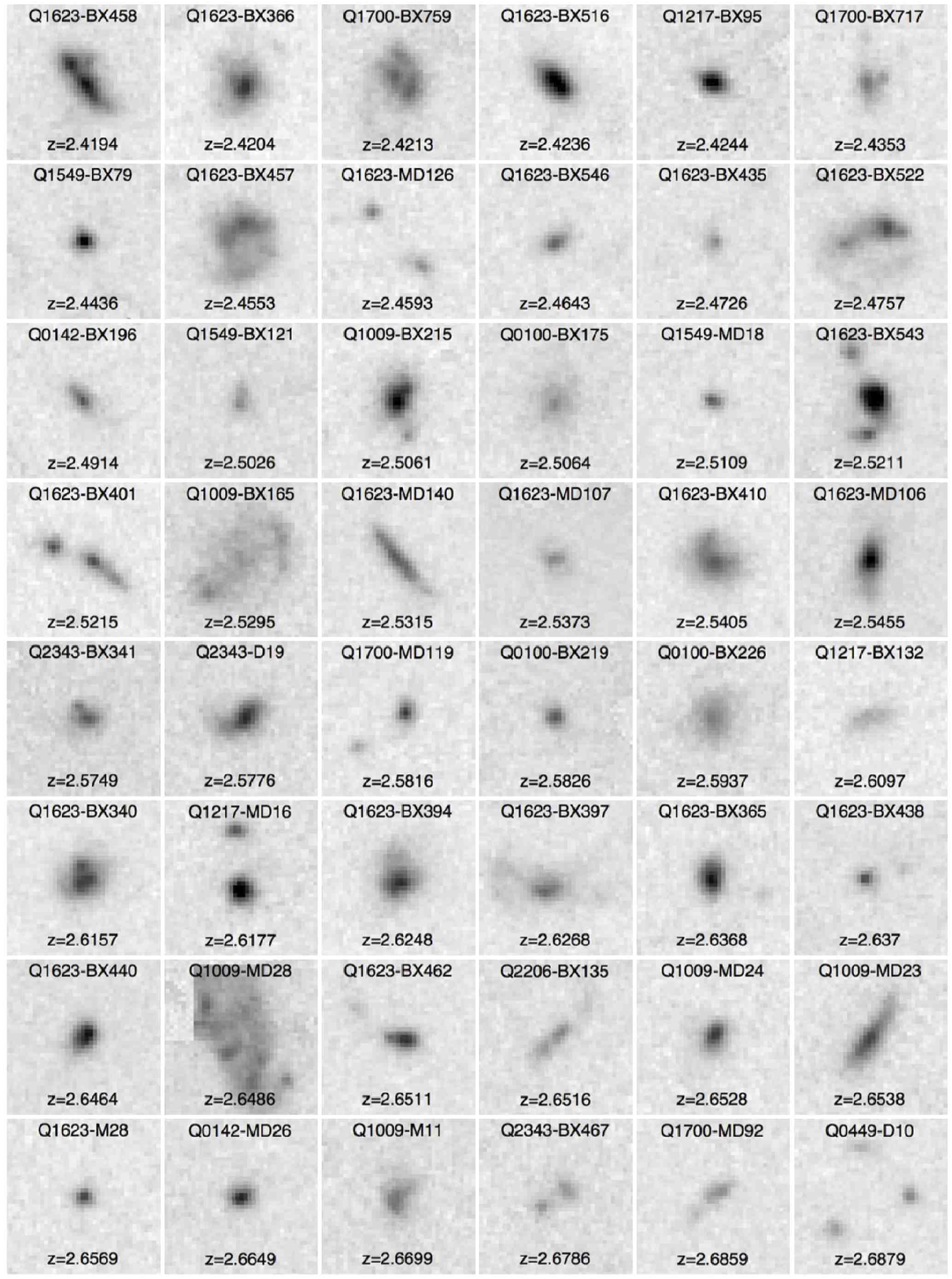}
\caption{{\it Continued}}
\end{figure*}

\addtocounter{figure}{-1}
\begin{figure*}
\plotone{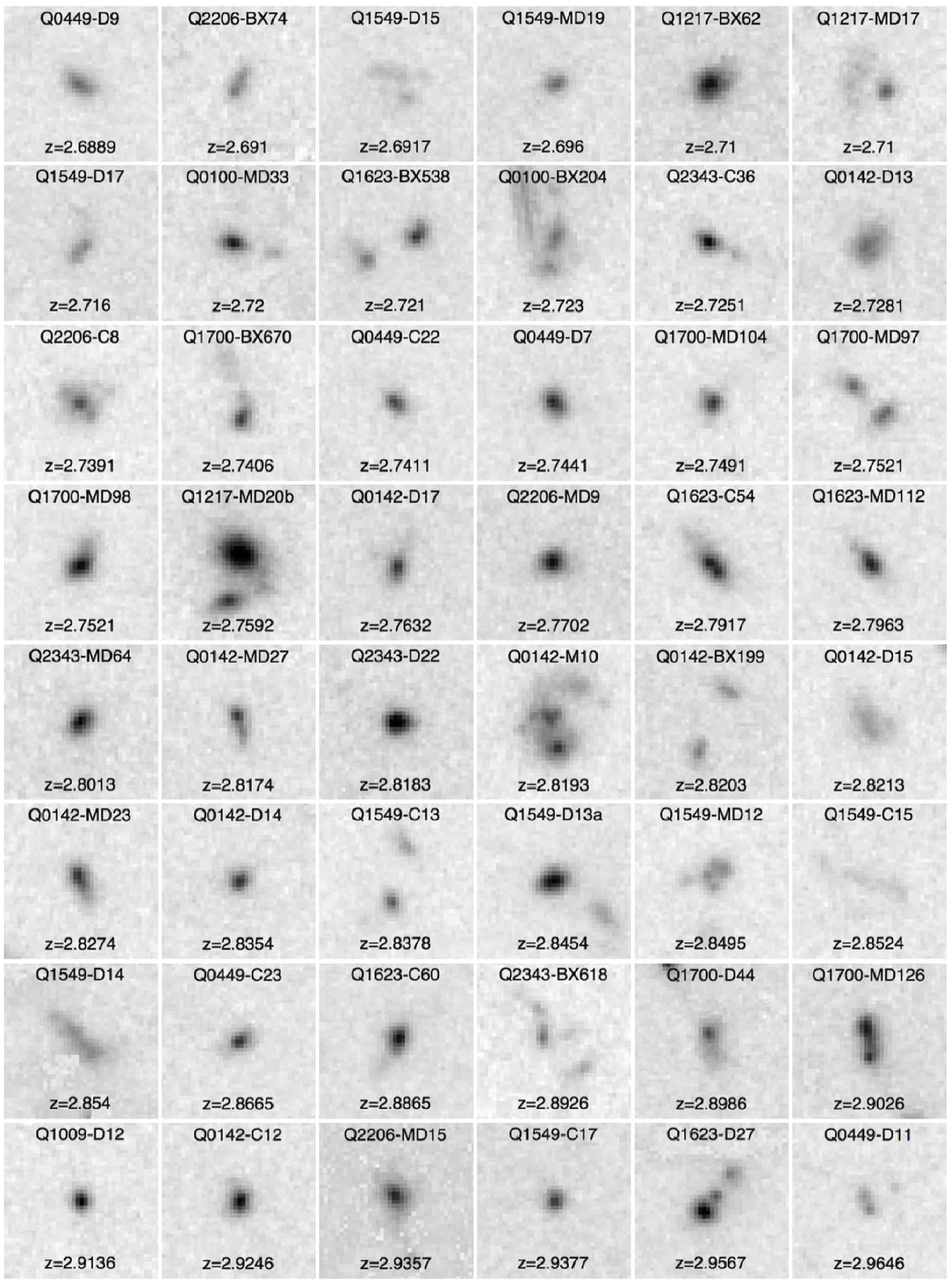}
\caption{{\it Continued}}
\end{figure*}

\addtocounter{figure}{-1}
\begin{figure*}
\plotone{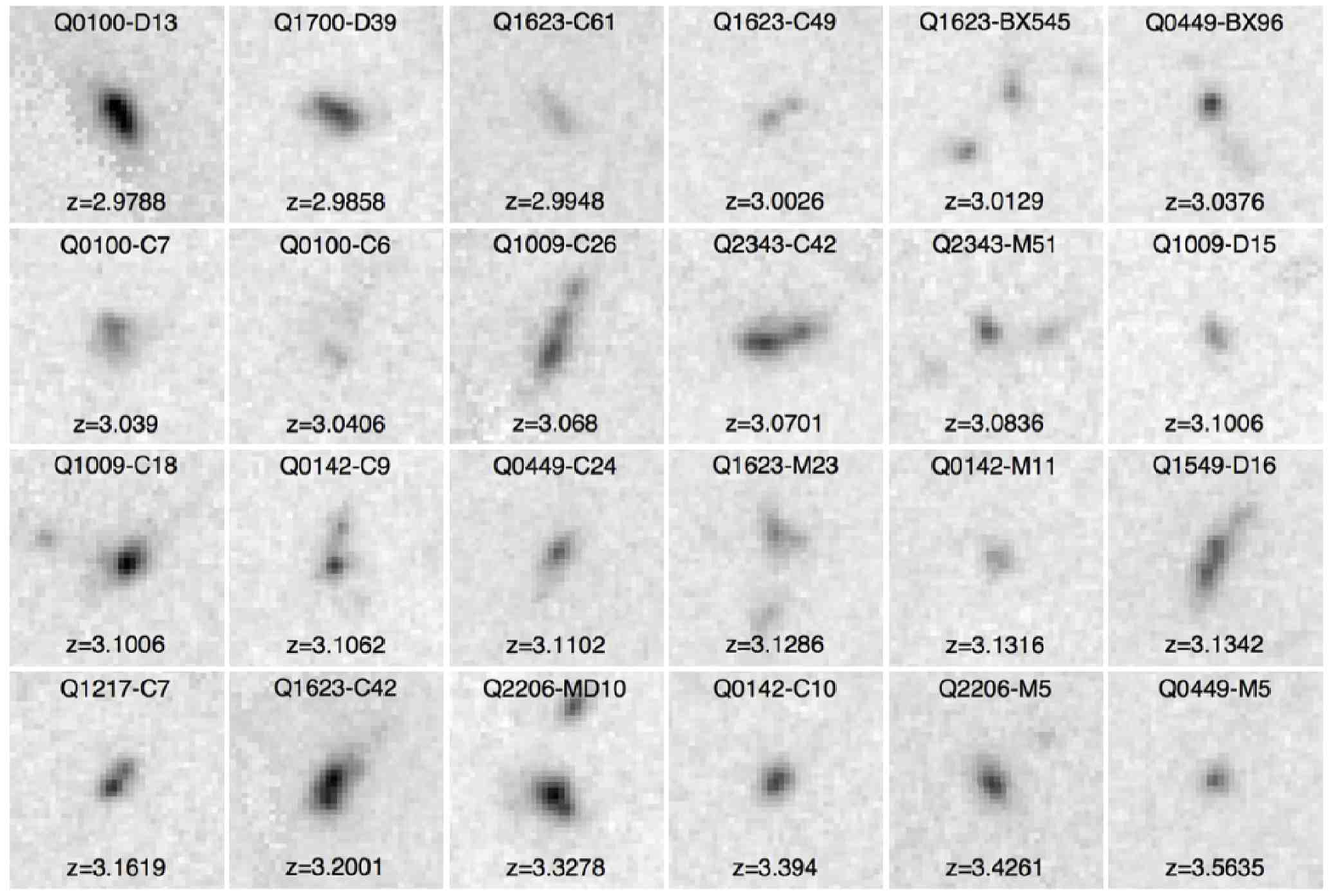}
\caption{{\it Continued}}
\end{figure*}

\subsection{Photometry}
\label{photometry.sec}

We photometrically calibrated our data using the zeropoint magnitude of 25.96\,AB
given for the F160W filter in the {\it HST}/WFC3 data handbook.
Masking all pixels identified with luminous sources using Source Extractor (Bertin \& Arnouts 1996), we use a $3\sigma$ clipped mean to estimate
that our drizzled images typically reach a limiting depth of 27.9 AB for a $5\sigma$ detection within a 0.2 arcsec radius aperture,
or $3\sigma$ surface brightness sensitivity of 25.1 AB arcsec$^{-2}$.\footnote{For comparison,
the HUDF09 program (GO 11563; Bouwens et al. 2010) covered an area of 4.7 arcmin$^2$ to a $5\sigma$ depth of 28.8 AB,
the GOODS-NICMOS survey (GNS; Conselice et al. 2011a) covered 45 arcmin$^2$ to a depth of 26.8 AB,
and the ERS/GOODS-S program (GO 11359; Windhorst et al. 2011) covered an area of $\sim 40$ arcmin$^2$ to a depth of 27.2 AB.}



\begin{deluxetable}{lcc}
\tablecolumns{3}
\tablewidth{0pc}
\tabletypesize{\scriptsize}
\tablecaption{Results of Monte Carlo Photometry Tests}
\tablehead{
\colhead{AB Magnitude} & \colhead{$\Delta H_{\rm ZP}$\tablenotemark{a}} & \colhead{$\sigma_{H}$\tablenotemark{b}}}
\startdata
$21.75 < H_{160} \leq 22.25$ & 0.06 & 0.03 \\
$22.25 < H_{160} \leq 22.75$ & 0.05 & 0.04 \\
$22.75 < H_{160} \leq 23.25$ & 0.08 & 0.05 \\
$23.25 < H_{160} \leq 23.75$ & 0.07 & 0.08 \\
$23.75 < H_{160} \leq 24.25$ & 0.07 & 0.11 \\
$24.25 < H_{160} \leq 24.75$ & 0.06 & 0.16 \\
\enddata
\label{mctests.tab}
\tablenotetext{a}{Bias between the measured and simulated photometry.}
\tablenotetext{b}{Statistical uncertainty in the recovered magnitudes.}
\end{deluxetable}

Initial estimates of the F160W magnitudes of the galaxies are obtained from the Source Extractor corrected isophotal magnitudes (MAG\_ISOCOR), which are consistent
to within 0.04 mag
with estimates obtained from matched-aperture photometry from images smoothed to the angular resolution of the ground-based 
${\cal R}$-band survey images for well-defined, isolated sources.
We perform Monte Carlo tests of the statistical uncertainty and photometric biases in these magnitudes by inserting 1000 artificial galaxy models
with known total magnitudes into randomly selected blank-field regions of the images
and calculating the accuracy with which their magnitudes are recovered using Source Extractor.  
The galaxy models are constructed using GALFIT
(see \S \ref{galfit.sec}) to model the light profiles of five
real galaxies in the Q1700+64 field that span a wide range of effective radii and Sersic index.  These tests are performed for 0.5 magnitude bins spanning the range 
$H_{160}= 22 - 25$ AB of the galaxy sample, and suggest (Table \ref{mctests.tab}) that 
MAG\_ISOCOR systematically underestimates the brightness of objects by $\Delta H_{\rm ZP} = 0.05 - 0.08$ mag.  After correcting
for this systematic offset, we find that the magnitudes of 
10 isolated, bright ($H_{160} \sim 15-16$ AB), unsaturated stars in our target fields all agree with values published in the 2MASS point source catalog
(Skrutskie et al. 2006) to within the photometric uncertainty of the catalog.

\subsection{SED Fitting}
\label{sed.sec}

Stellar masses, ages, and star formation rates were calculated by fitting the broadband spectral energy distribution (SED)
of the galaxies with stellar population synthesis models using a customized IDL code (Reddy et al. 2012, in preparation).
In addition to the {\it HST}/F160W and ground-based $U_n G {\cal R}$ photometry
many galaxy models also incorporate $J/K_s$-band data, and in some cases {\it Spitzer} IRAC photometry.
The SED fitting process is described in detail by Shapley et al. (2001, 2005), Erb et al. (2006c),
and Reddy et al. (2006, 2010); in brief, we use Charlot \& Bruzual (2011, in preparation) population synthesis
models, a Chabrier (2003) initial mass function (IMF), and a constant ($\tau = \infty$) star formation history.

Although the statistical uncertainty of the  $H_{160}$ magnitudes is small (see Table \ref{mctests.tab}), 
the true uncertainty in the {\it continuum} magnitudes $H_{\rm cont}$ is significantly larger due to the uncertain contribution from nebular line emission
that falls within the F160W bandpass ($\lambda\lambda 14028 -16711$ \AA).
In order to ensure that the  $H_{160}$ magnitudes do not unduly influence the SED fit with their small formal uncertainties
(see also discussion by McLure et al. 2011) 
we attempt to quantify
the additional uncertainty due to nebular emission in a physically motivated manner
by bootstrapping approximate line fluxes from broadband scaling laws and typical nebular line ratios.
We use the ground-based $U_n G {\cal R}$ magnitudes to estimate the rest-frame monochromatic luminosity $L_{\nu}$ at 1500 \AA, and convert this to a UV star formation rate
using the Kennicutt (1998) relation.  This UV SFR is corrected for extinction by estimating the UV slope $\beta$ from the $U_n G {\cal R}$ photometry, and converting to an
estimated extinction $E(B-V)$ using the Meurer et al. (1999) relation in combination with a Calzetti et al. (2000) attenuation law (motivated by comparison
with direct indicators of the dust emission at 24$\micron$).  We then assume that the extinction-corrected
UV SFR is equal to the H$\alpha$ SFR (see discussion by Erb et al. 2006b), and use the Kennicutt (1998) relation to estimate the corresponding H$\alpha$ nebular emission
line flux.  Based on standard atomic physics and 
the observations of Maiolino et al. (2008) and Erb et al. (2006a), we assume that the other strong rest-optical nebular emission lines have typical flux ratios
given by: 
$H\alpha/H\beta = 2.9$,
$\othree \lambda5007/H\beta = 4.6$,
$\othree \lambda5007/\othree \lambda4959 = 3.0$,
$\otwo \lambda3727/H\beta = 1.5$,
$\ntwo \lambda6585/H\alpha = 0.16$.
All of these estimated emission line fluxes are converted to observed values using the extinction coefficients described above.

The combined flux  $\Delta H_{\rm line}$ of emission lines that fall within the F160W bandpass 
at the redshift of each galaxy is added to the photometric bias corrected $H_{160}$ magnitude
to obtain an estimate of the continuum magnitude $H_{\rm cont} = H_{160} + \Delta H_{\rm line}$.
There are significant uncertainties associated with almost every step of our estimate of the nebular line-emission correction described above, not least of which
is the strong variation in line flux ratios (e.g., $\othree \lambda5007/H\beta$) with metallicity.  We therefore conservatively estimate the uncertainty in the continuum magnitudes
as $\sigma_{\rm cont} = \sqrt{\sigma_{H}^2 + \Delta H_{\rm line}^2}$.  Typical values of this uncertainty  are generally in the range $0.1 < \sigma_{\rm cont} < 0.3$ mag, but values
as high as 0.5 mag can occur in 10\% of cases (and 1.0 mag in 1\% of cases).  Due to the downweighting of the WFC3 data point when $\sigma_{\rm cont} \approx 0.5$, derived
stellar masses in such cases differ by only 1\% on average from stellar masses derived by omitting the WFC3 data point from the SED fit entirely.






\section{DEFINING THE MORPHOLOGICAL STATISTICS}
\label{stats.sec}

Many efforts have been made to quantify the morphologies of predominantly-irregular high redshift galaxies by using a combination of qualitative visual
analyses, parametric Sersic model fits, and non-parametric numerical statistics (e.g., `$CAS$'; Conselice 2003).
Here we explore all of these methods and discuss the physical inferences that can be gleaned from each.

We describe below our methods for visual classification, Sersic profile fitting, and calculating the 
Gini coefficient $G$, the second order moment of the light distribution $M_{20}$, the concentration $C$, 
asymmetry $A$, and multiplicity $\Psi$ statistics.\footnote{We do not calculate the smoothness parameter (i.e, the `$S$' in `$CAS$') because it is not 
robustly defined for galaxies as small and poorly resolved as those at $z\sim 2-3$ (see discussion by Lotz et al. 2004).}
In our discussion of the non-parametric numerical statistics we define
$f_i$ as the fluxes of the $N$ individual pixels in the segmentation map (see \S \ref{segmaps.sec})
with physical location $x_i, y_i$, where $i$ ranges from 1 to $N$.

\subsection{Visual Classification}
\label{vclass.sec}

\begin{figure}
\plotone{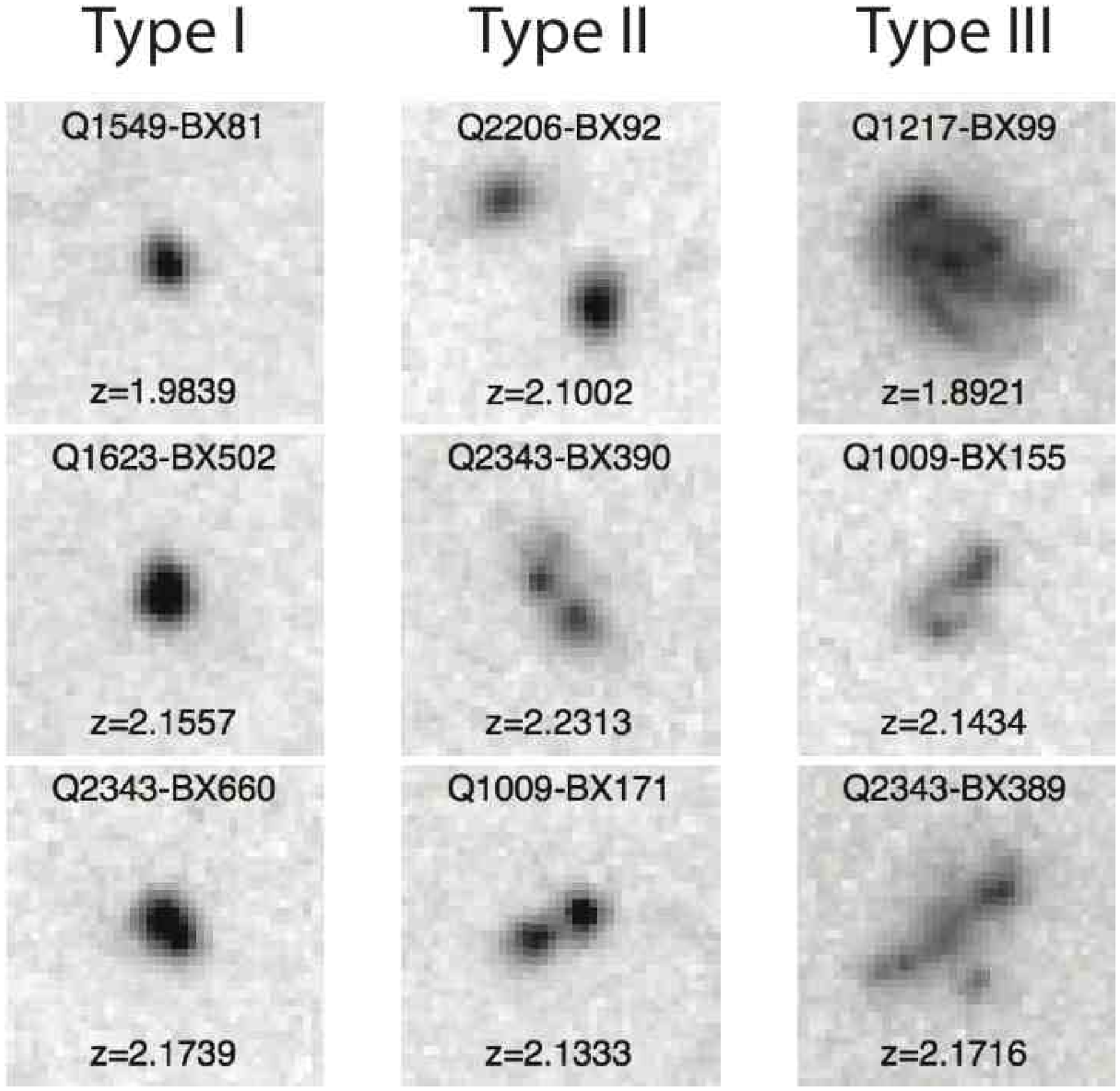}
\caption{Visual classification scheme illustrated by three sample galaxies for each of the three types:
Type I (single nucleated source), Type II (multiple well-defined nucleated sources), Type III (diffuse and extended emission, possibly hosting multiple clumps).}
\label{vclass.fig}
\end{figure}

Our first morphological classification  groups galaxies visually based on the apparent nucleation of their light profiles and the number of distinct components.
As illustrated in Figure \ref{vclass.fig}, we group galaxies from Figure \ref{postage.fig} into three general classes:

\begin{description}
\item[Type I] Single, nucleated source with no evidence for multiple luminous components or extended low surface brightness features.  127 galaxies in our sample.
\item[Type II] Two or more distinct nucleated sources of comparable magnitude, with little to no evidence for extended low surface brightness features.  56 galaxies in our sample.
\item[Type III] Highly irregular objects with evidence of non-axisymmetric, extended, low surface-brightness features.  123 galaxies in our sample.
\end{description}

Type I galaxies appear consistent
with being regular and isolated systems, while Type II galaxies may represent either early-stage mergers between two such formerly isolated systems
or intrinsically clumpy systems with little continuum emission between the clumps.
Type III galaxies in
contrast may represent later-stage mergers with bright tidally induces disturbances, or clumpy concentrations within a single extended system (e.g., Bournaud et al. 2007).
Of course, there is significant overlap between the three classes, and degeneracy in the classes to which a given galaxy may be assigned.  Galaxies with identical luminosity
profile but different surface brightness may, for instance, be assigned to either Type I or Type III depending on whether the low surface brightness features are above or below
the limiting surface brightness of the data, and the division between Types II and III is similarly unclear.
The goal of these visual classifications is not to provide decisive quantitative divisions however, but simply as a reference point to describe the general qualitative
appearance of galaxies throughout the following discussion.

\subsection{Sersic Profiles}
\label{galfit.sec}

In the local universe the surface brightness profiles of galaxies can often by well-fit by Sersic (1963) models over a large dynamic range in luminosities
(e.g., Kormendy et al. 2009).
While regular ellipsoidal models are clearly an incomplete description of the irregular galaxy morphologies illustrated in Figure \ref{postage.fig}, such models
nonetheless provide a useful description of the characteristic sizes and surface brightness profiles of the major individual clumps.
We therefore use GALFIT 3.0 (Peng et al. 2002, 2010) to fit the galaxy sample with two-dimensional Sersic profiles described by the functional form
\begin{equation}
\Sigma(r) = \Sigma_e {\textrm{exp}} \left[ -\kappa \left( \left (\frac{r}{r_{1/2}}\right ) ^{1/n} -1 \right ) \right ]
\end{equation}
convolved with the observational PSF.  These models are characterized by the effective half-light radius $r_{1/2}$ and the radial index $n$ of the profile.
GALFIT actually calculates the effective half-light radius along the semi-major axis ($r$); following a common practice in the literature (e.g., Shen et al. 2003; 
Trujillo et al. 2007; Toft et al. 2009) we convert this to a circularized
effective radius $r_{\rm e} = r \sqrt{b/a}$, where $b/a$ is the minor/major axis ratio.
As described in Peng et al. (2002, see their Figure 1), two of the most commonly observed
values of the radial index in the nearby universe are $n=1$, which corresponds to the exponential disk profile, and $n=4$, which corresponds to a classical de Vaucouleurs profile
with steep central core and relatively flat outer wings typical of elliptical galaxies and galactic bulges.

We use a median-combined stack of isolated, bright ($H_{160} < 20$ AB), unsaturated stars from across our WFC3 imaging fields to define the PSF model.  While the structure
of the PSF varies slightly across a given field, and from field-to-field with the \hst\ roll angle,
we find the details of our PSF model have little effect on the derived physical properties  of our faint and extended galaxies (see also discussion by Szomoru et al. 2010).
Since GALFIT convolves physical models with the observational PSF it is able to determine effective radii down to extremely small spatial scales.
Following the method described by Toft et al. (2007), we use  a variety of stellar point sources as PSF models to fit Sersic models to 11 stars in our WFC3 imaging fields, finding that
the mean estimated size of known point sources is $0.16 \pm 0.25$ pixels.  We therefore adopt a $3\sigma$ limit for unresolved point sources of $0.16 + 3 \times 0.25 = 0.91$ pixels, or 0.073 arcsec,
corresponding to 0.62 kpc at redshift $z=2.0$.

Our procedure for fitting Sersic models to individual galaxies is as follows.  
We used the Source Extractor segmentation map to mask out all objects not associated with the target galaxies, replacing these pixels with Gaussian
random noise matched to the noise characteristics of the image.  We then cut out a $5\times5$ arcsec region surrounding each galaxy and subtracted from it a `local sky' estimated from the
median of pixels excluded from the segmentation map.  GALFIT is then used to fit the minimum number of axisymmetric (we do not introduce bending or Fourier modes)
components required to satisfactorily reproduce the observed light distribution.
For the majority of galaxies shown in Figure \ref{postage.fig} we use a single component, unless there are clearly multiple spatially distinct clumps or significant asymmetry in the light
distribution.  All GALFIT models were inspected by two of us (DRL \& SRN) in order to verify that a consistent approach was taken throughout the galaxy sample.

Unlike the non-parametric morphological statistics (which represent an integrated quantity over the entire light distribution of
a galaxy), $r_{\rm e}$ and $n$ can formally be multi-valued for galaxies fit by multiple Sersic components
(i.e., Type II and some Type III galaxies).
We adopt the convention of describing such multi-component galaxies by the $r_{\rm e}$ and $n$ of the brightest individual component; as we discuss in \S \ref{mrcaveats.sec}, this assumption
does not significantly bias our conclusions.
For the few cases for which a reasonable model cannot be obtained with $\lesssim 5$ Sersic components (e.g., Q1009-MD28, which is in close physical proximity to the bright Q1009 QSO and turns out to be a Ly$\alpha$ blob based on recent narrowband imaging)
we consider $r_{\rm e}$ and $n$ to be undefined.

\subsection{Gini coefficient $G$}
\label{gini.sec}

The Gini coefficient ($G$; Gini 1912) was introduced into the astronomical literature by Abraham et al. (2003) and further developed by Lotz et al. (2004).
$G$ measures the cumulative flux distribution
of a ``population'' of pixels and is insensitive to the actual spatial distribution of the individual pixels.

Formally $G$ is defined (Glasser 1962) in the range $G= 0 - 1$ as
\begin{equation}
G = \frac{1}{\bar{f} N (N-1)}\sum_{i=1}^N(2i -N -1)f_i
\end{equation}
where $\bar{f}$ is the average flux and the $f_i$ pixel fluxes are sorted in increasing order before the summation over all $N$ pixels in the segmentation map.
High values of $G$ represent the majority of the total flux being concentrated in a small number of pixels, while low values represent a more uniform distribution of flux.

\subsection{Second order moment $M_{20}$}

The spatial distribution of the light may be quantified via the second order moment of the light distribution, $M_{20}$, introduced in this context by Lotz et al. (2004).
$M_{20}$ is defined as the second order moment of the brightest pixels that constitute 20\% of the total flux in the segmentation map, normalized by the
second order moment of all of the pixels in the segmentation map.  Mathematically,
\begin{equation}
M_{20} = \textrm{log} \left(\frac{\sum_i M_i}{M_{\rm tot}}\right), \textrm{while} \sum_i f_i < 0.2 f_{\rm tot}
\end{equation}
where
\begin{equation}
M_{\rm tot} = \sum_i^N M_i = \sum_i^N f_i [(x_i - x_c)^2 + (y_i - y_c)^2]
\end{equation}
Following Lotz et al. (2004, 2006) we adopt the position that minimizes $M_{\rm tot}$
as the center ($x_c, y_c$) of the light distribution.

Typical values of $M_{20}$ range from $\sim -1$ (most irregular, often with multiple clumps) to $\sim -2$ (most regular).

\subsection{Concentration $C$}

The concentration index $C$ (Kent 1985; Abraham et al. 1994; Bershady et al. 2000; Conselice 2003) measures the concentration of flux about a central point in the galaxy.  
While slightly different versions have been introduced by various
authors, we adopt the `$C_{28}$' standard:
\begin{equation}
C = 5 \textrm{log} \left(\frac{r_{80}}{r_{20}}\right)
\end{equation}
where $r_{20}/r_{80}$ are the circular radii containing 20\%/80\% respectively of the total galaxy flux within the segmentation map.
Following Conselice et al. (2008), we adopt the flux-weighted centroid of the segmentation map as the center for the concentration calculation.
While in many cases this corresponds naturally to a peak in the flux distribution, it is not necessarily the case
for extremely irregular galaxies without well-defined central flux concentrations.

Typical concentration values range from $\sim 1$ (least compact) to $\sim 5$ (most compact).  We note, however, that galaxies with two or more clumps 
(e.g., Type II galaxies) that are each individually
compact are not generally compact in a global sense.


\subsection{Asymmetry $A$}

The asymmetry $A$ (Schade et al. 1995; Conselice et al. 2000) quantifies the $180^{\circ}$ rotational asymmetry of a galaxy.
Mathematically, $A$ is calculated by differencing the original galaxy image with a rotated copy:\footnote{We note that Schade et al. (1995) and Conselice et al. (2000) included a
factor of two in the denominator of Eqn. \ref{Aeqn.eqn}, while more recent work by Lotz et al. (2004) and Conselice et al. (2008) do not.  We follow the convention of the more recent
literature by neglecting this factor.}
\begin{equation}
A = \textrm{min} \left(\frac{\sum | f_{0,i} - f_{180,i}|}{\sum | f_{0,i} |}\right) -  \textrm{min} \left(\frac{\sum | B_{0,i} - B_{180,i}|}{\sum | f_{0,i} |}\right)
\label{Aeqn.eqn}
\end{equation}
where $f_{0,i}$ represents flux in the original image pixels and $f_{180,i}$ flux in the rotated image pixels.  
Following Conselice et al. (2000, 2008) we determine the rotation center iteratively by allowing it to walk about an adaptively spaced grid with 0.1 pixel resolution
until converging on the point that minimizes $\Sigma \left | f_{0,i} - f_{180,i} \right |$.
The $B_{0,i}$ and $B_{180,i}$ terms represent fluxes in nearby background pixels
to which we have applied an identical segmentation map, and are included to subtract the contribution of noise to the total galaxy asymmetry.  
As discussed by Conselice et al. (2008), the background sum is minimized similarly to the original image.

Typical values of $A$ range from 0 for the most symmetric galaxies to 1 for galaxies with the strongest $180^{\circ}$ rotational asymmetry.

\subsection{Multiplicity $\Psi$}
\label{mult.sec}

The multiplicity coefficient, introduced by Law et al. (2007b), calculates the effective ``potential energy'' of the light distribution
\begin{equation}
\psi_{\rm actual} = \sum_{i=1}^{N}\sum_{j=i+1}^{N}\frac{f_i f_j}{r_{ij}}
\end{equation}
where $f_i$ and $f_j$ are the fluxes in pixels $i/j$ respectively, $r_{ij}$ is the separation between pixels $i$ and $j$, and
where the sum runs over all of the $N(N-1)/2$ $i$ --- $j$ pixel pairs.\footnote{Note that $\psi_{\rm actual}$ and $\psi_{\rm compact}$ were defined incorrectly in Law et al. (2007b)
with the sum double-counting each pixel pair.  These factors of 2 would, however, cancel out upon constructing the final statistic $\Psi$ from the 
ratio of $\psi_{\rm actual}$ to $\psi_{\rm compact}$.}
This is compared to the most compact possible rearrangement of pixel fluxes, that by analogy with a gravitational system would require the most ``work'' to pull apart.  This
compact map is constructed by rearranging the positions of all $N$ galaxy pixels so that the brightest pixel is located in the center of the distribution, and the surrounding pixel fluxes
decrease monotonically with increasing radius.  Calling $r'_{ij}$ the distance between pixels $i$ and $j$ in this compact map,
\begin{equation}
\psi_{\rm compact} = \sum_{i=1}^{N}\sum_{j=i+1}^{N}\frac{f_i f_j}{r'_{ij}}
\end{equation}

The multiplicity coefficient $\Psi$ measures the degree to which the actual distribution of pixel fluxes differs from the most compact possible arrangement, i.e.
\begin{equation}
\Psi = 100 \times \textrm{log}_{10} \left(\frac{\psi_{\rm compact}}{\psi_{\rm actual}}\right)
\end{equation}

As discussed by Law et al. (2007b), 
values for $\Psi$ can range from 0 (i.e., for which the galaxy pixels are already in the most compact possible arrangement) to $\gtrsim 10$ for extremely irregular sources.
Generally, we find that isolated, regular galaxies in our sample may be described by $\Psi \leq 1$, galaxies with some morphological irregularities by $1 < \Psi < 2$, and galaxies
with strong morphological irregularities or multiple components by $\Psi \geq 2$.

\subsection{Detailed Segmentation Maps}
\label{segmaps.sec}


The preliminary segmentation maps constructed in \S \ref{initialmap.sec}  above assign pixels to a given galaxy 
based on a constant surface brightness threshold tied to the noise characteristics of the WFC3 data.  While such a segmentation map is sufficient for 
estimating total source magnitudes, it is inadequate for calculating quantitative morphologies using the nonparametric statistics defined in \S \ref{gini.sec} - \ref{mult.sec}
since surface-brightness based pixel selection  produces results that vary 
with total source luminosity, redshift, and limiting survey magnitude.  
Multiple methods have been adopted in the literature for defining robust segmentation maps;
in the Appendix we discuss four such methods (Conselice et al. 2000; Lotz et al. 2004; Abraham et al. 2007; Law et al. 2007b)
and calculate values for $G$, $M_{20}$, $C$, $A$, and $\Psi$ in each.

In part, this Appendix is provided so that our results can be directly translated to the readers preferred choice of segmentation map, but it is also instructive to consider
how the calculated values of the morphological parameters depend upon this choice.
While we find that the values of $G$, $M_{20}$, $C$, $A$, and $\Psi$
are well-correlated between different segmentation maps, there can be significant systematic offsets in dynamic range (particularly for $G$; see also Lisker 2008) between the systems.
We discuss the implications of such offsets in Section \ref{mergers.sec} below.

Throughout the following analysis, we choose to calculate our baseline morphologies
using the Abraham et al. (2007) quasi-Petrosian method with isophotal threshold  $\eta=0.3$
as this method is arguably most well suited to the irregular morphologies of our target galaxies.
Using the transformation relations presented in the Appendix however, we convert our values to the 
Lotz et al. (2004, 2006) systematic reference frame in order to compare to both recent observational results and numerical simulations (e.g., Lotz et al. 2010ab).

\subsection{Robustness}

The robustness of our morphological indices has been discussed in the literature many times before (e.g., Bershady et al. 2000; Lotz et al. 2006; Lisker 2008; Gray et al. 2009).
Generally speaking, such work suggests that morphological statistics are relatively robust for large, bright galaxies but that they can become unreliable
at faint magnitudes and for galaxies that are small with respect to the observational PSF.
Most of these previous studies are tailored to the analysis of deep {\it HST}/ACS imaging in public survey fields however, and in order to understand the effects of systematic biases on our
WFC3 imaging data (and on our specific galaxies) it is necessary to perform many robustness tests anew. 

The details of our analysis exploring the robustness of each of the five quantitative morphological statistics, and the Sersic parameters $r_{\rm e}$ and $n$, 
to total source magnitude $H_{160}$, the size of the observational
PSF, and our choice of pixel scale are presented in the Appendix.  
In brief, we find that:
\begin{enumerate}
\item The derived values of six of the seven indices are fairly robust for galaxies with magnitudes $H_{160} \leq 24.0$ (roughly corresponding to total S/N $> 100$), but become less reliable at fainter magnitudes.
The exception is the concentration parameter $C$, for which the small sizes of many of our galaxies cause the inner 20\% flux isophote to be unresolved at all magnitudes and therefore $C$ to be unreliable
(see also Bershady et al. 2000).
We therefore omit $C$ from detailed discussion, and restrict our analyses in the following sections (except where indicated) to the subsample of galaxies with $H_{160} \leq 24.0$, resulting in a sample of 
206 galaxies, 59/95/52 in the $z=1.5-2.0$, $z=2.0-2.5$, and $z=2.5-3.6$ redshift bins respectively.
The physical implications of this self-imposed apparent magnitude limit, and of systematic variations with the observational PSF, are discussed in the relevant sections below.
\item Six of the seven indices (except $C$) are robust to our choice of a 0.08 arcsec pixel scale; our conclusions would be unchanged if we had
drizzled our data to 0.06 arcsec or 0.1 arcsec pixels instead.
\item Given the small size of many of our galaxies, the nonparametric statistics $G$, $M_{20}$, $C$, $A$, and $\Psi$ can vary systematically with the observational PSF as morphological features become
more or less well-resolved (see also discussion by Lotz et al. 2004, 2008b).  
In particular, these five statistics will have less dynamic range to their values than in high-resolution imaging as it becomes progressively more difficult to distinguish them from point sources.
This complicates quantitative comparisons to data obtained at different wavelengths or local comparison samples,
but is less significant for comparisons within the $z\sim 1.5-3.6$ population.
In contrast, the Sersic parameters $r_{\rm e}$ and $n$ are relatively robust to the PSF because the modeling process convolves theoretical models with the observational
PSF.
\item The uncertainty in each of the seven indices is calculated via Monte Carlo simulations placing GALFIT model galaxies atop different blank-field regions of the WFC3 footprint in order to compare different
realizations of the noise statistics.  This uncertainty varies as a function of both source magnitude and morphological type; averaged over these considerations, typical uncertainties are
3\% in $G$, 4\% in $M_{20}$, 11\% in $C$, 22\% in $A$, 21\% in $\Psi$, 2\% in $r_{\rm e}$, and 15\% in $n$.
\end{enumerate}


\section{BASIC MORPHOLOGICAL CHARACTERISTICS}
\label{basiccharac.sec}

\begin{figure}
\plotone{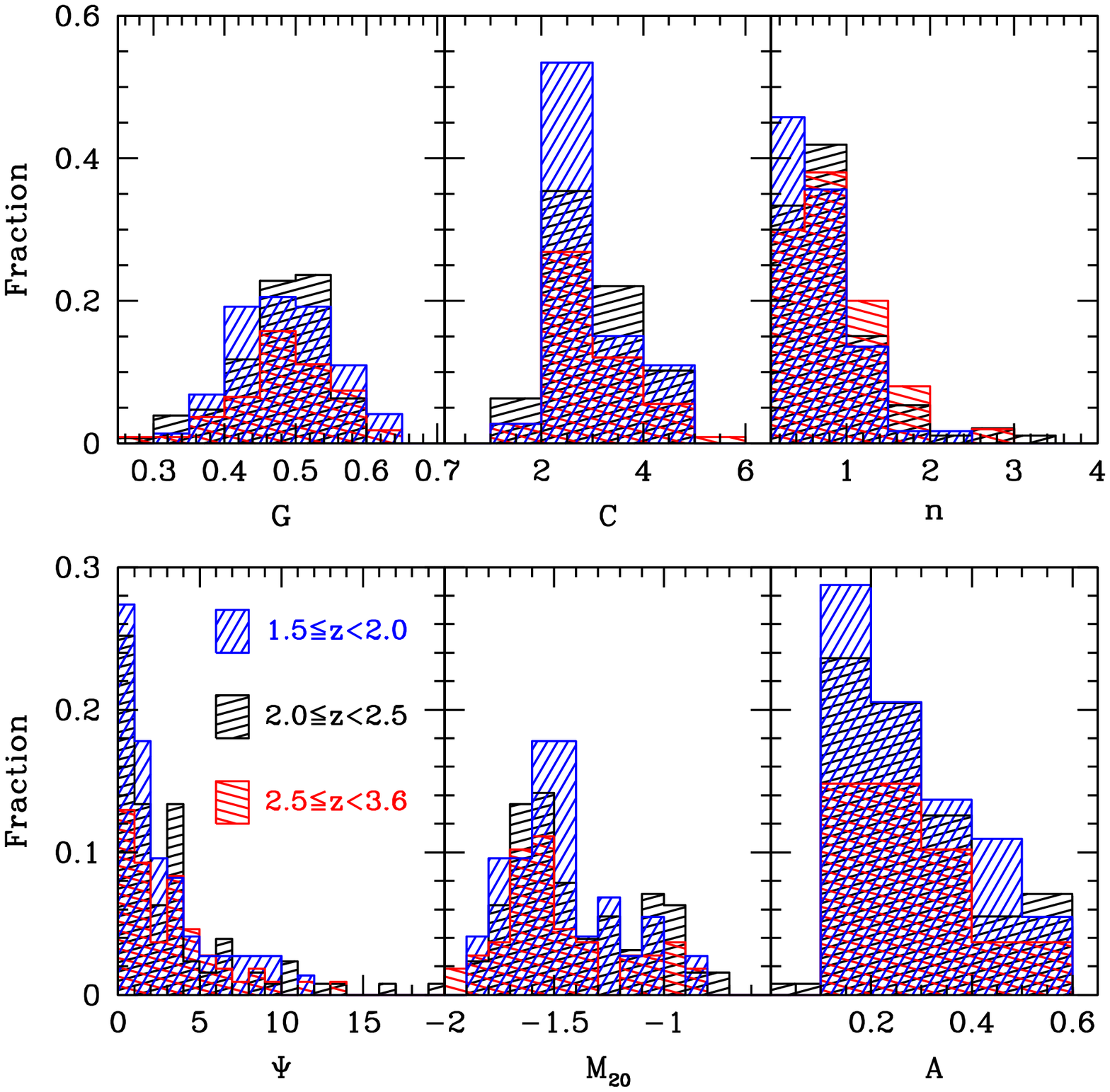}
\caption{Histograms of morphological parameters divided according to spectroscopic redshift and normalized by the total number of galaxies in each sample.  Galaxies in all three redshift bins are
statistically consistent with $G$, $C$, $n$, $\Psi$, $M_{20}$, and $A$
being drawn from identical distributions (minimum confidence in the null hypothesis 7\%).}
\label{morphhist.fig}
\end{figure}

In Figure \ref{morphhist.fig} we plot histograms of the five nonparametric morphological statistics ($G$, $C$, $\Psi$, $M_{20}$, and $\Psi$) and the Sersic index $n$ divided according to redshift
(we discuss the evolution of the characteristic effective radius in detail in \S \ref{massrad.sec}).
The typical star forming galaxy is best represented by a Sersic profile of index $n\sim1$, $G\sim0.5$, $C\sim 3$, $\Psi \sim 2$, $M_{20} \sim -1.5$, and $A\sim 0.25$.   
Despite the range of rest wavelengths probed by the F160W filter across the redshift range of our sample ($\sim 3800- 6100$ \AA), there is no evidence to suggest
systematic variation with redshift across our sample (whether due to evolution or to a variable morphological k-correction).
Applying a Kolmogorov-Smirnoff (KS) test  suggests that all six indices are consistent at $>7$\% confidence with the null hypothesis that they are drawn from
the same distribution at all redshifts.

\begin{figure}
\plotone{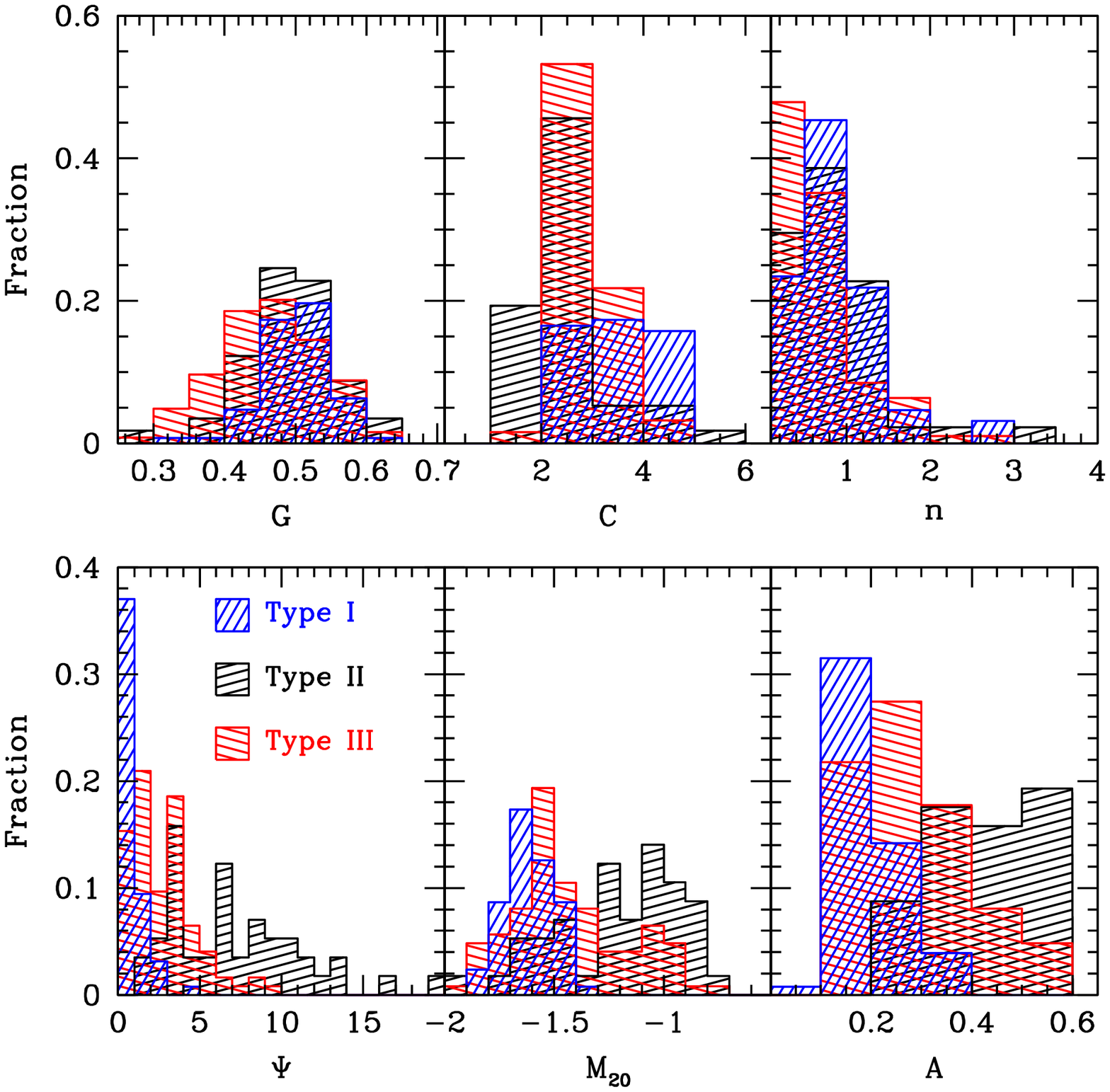}
\caption{Histograms of morphological parameters divided according to visual morphological classification and normalized by the total number of galaxies in each sample.  
Galaxies in the three visual classes have drastically different
automated morphological statistics, with only two pairings ($G$ and $n$ for Type I and Type II galaxies) consistent with the null hypothesis at greater than 2\% confidence.}
\label{morphhistVM.fig}
\end{figure}

In contrast, there is significant difference between many of the morphological indices when divided according to their apparent visual morphology
(Figure \ref{morphhistVM.fig}), indicating the underlying correlation between visual and numerical classification techniques.
In particular, we note the strong correlation between visual type and the `irregularity' statistics $\Psi$, $M_{20}$, and $A$.
Broadly speaking, Type I galaxies have $\Psi < 2$, Type III galaxies have $2 < \Psi < 5$, and Type II galaxies have $\Psi > 5$.
While $\Psi$ is the statistic most strongly correlated with visual estimates of irregularity, qualitatively similar results are apparent for both $A$ and $M_{20}$.
As expected from our definition of Type III galaxies, this galaxy sample also has significantly lower mean values of $G$, and slightly shallower Sersic indices.

\subsection{Composite Luminosity Profile}

As illustrated by Figures \ref{morphhist.fig} and \ref{morphhistVM.fig}, typical galaxies are best described by an $n\sim1$ Sersic profile.
Indeed, only six galaxies have $n>2.5$, of which four have estimated radii 
less than our $3\sigma$ resolution estimate, suggesting that these galaxies are simply too small to robustly determine their structure.
Focusing our attention on the $200$ galaxies with $n<2.5$ we find a mean $\langle n \rangle = 0.63$ with standard deviation of 0.39,
corresponding to flat inner regions intermediate between a Gaussian ($n=0.5$) and an exponential profile ($n=1.0$), and a steeply declining profile at larger radii
(similar to previous results by, e.g., Ravindranath et al. 2006; Conselice et al. 2011b; F{\"o}rster Schreiber et al. 2011).

\begin{figure}
\plotone{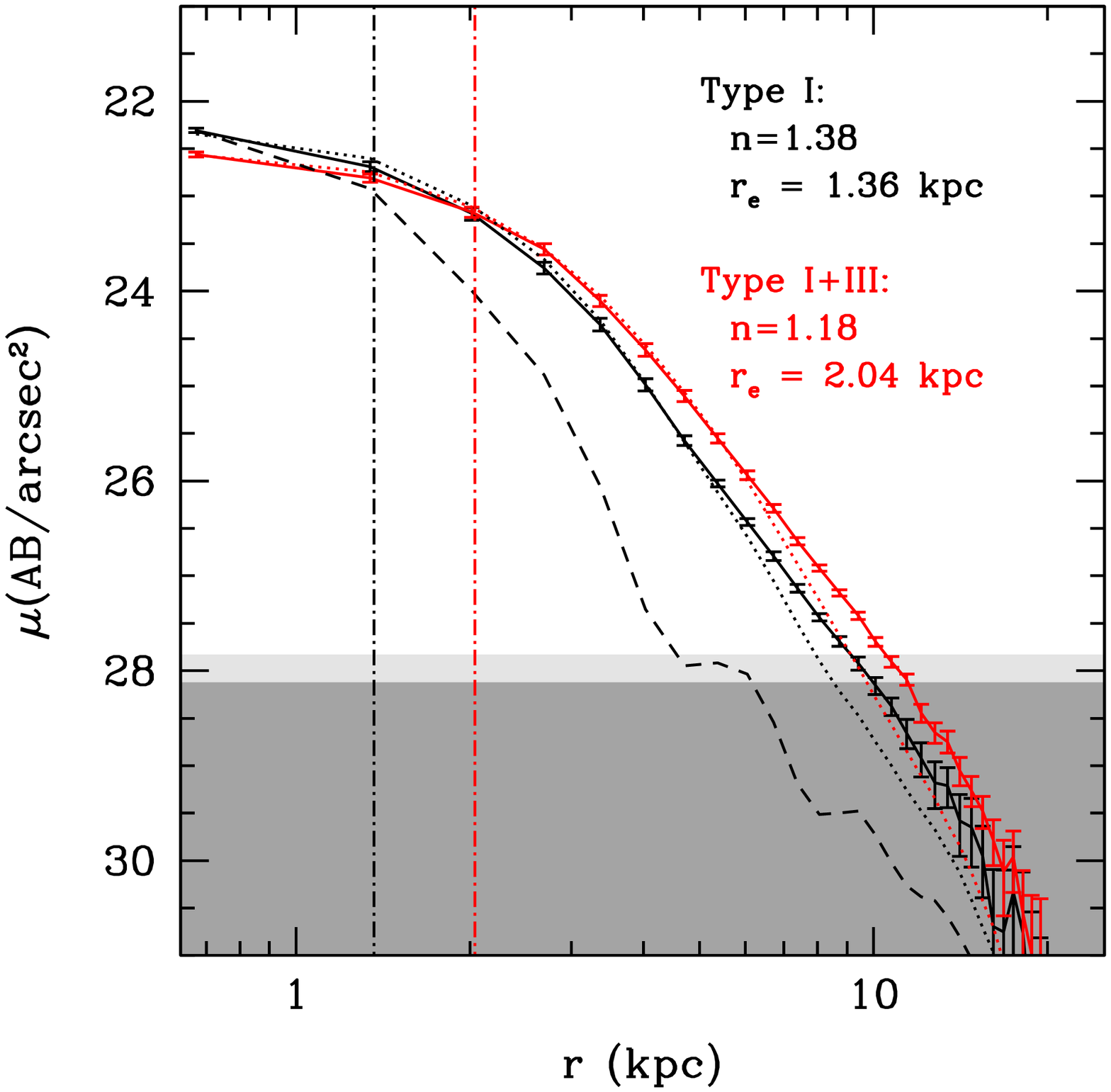}
\caption{Radial profiles of stacked galaxy samples.  The solid black line represents the radial profile of a stack of 127 type I galaxies, the solid red line
represents the radial profile of a stack of 250 type I+III galaxies.  Uncertainties at each point represent uncertainties in the mean.
The dotted red/black lines represent the radial profile of the best-fit Sersic model convolved with the observational PSF (dashed line), the parameters
of each Sersic model are given.  
The vertical dot-dashed lines indicate the effective radius of the stacked images.  The light/medium grey shaded areas represents the $3\sigma$ sky background for the stack of 127/250
galaxies respectively.}
\label{stackprofile.fig}
\end{figure}

In order to investigate the faint extended structure of our star forming galaxy sample we create a composite stack of galaxies (irrespective of redshift and total $H_{160}$ magnitude).
We cut out $5\times5$ arcsec regions around each galaxy, align all of the flux-weighted image centroids using sub-pixel bilinear interpolation, and stack the individual images
together using a $3\sigma$-clipped mean algorithm.  The resulting stack for our 127 galaxies of morphological type I (i.e., those galaxies whose morphologies are most regular and well-defined)
reaches a $3\sigma$ limiting surface brightness of 27.8 AB arcsec$^{-2}$ (31.2 AB for a $5\sigma$ detection in a 0.2 arcsec radius aperture).

As illustrated in Figure \ref{stackprofile.fig} the stacked radial profile is well 
described by an $n=1.38$ Sersic model with effective radius $r_e = 1.36$ kpc.
The profile is a good match to the Sersic model out to at least 6 kpc ($> 4 r_e$) and deviates only moderately from the model out
to the detection limit at $r \sim 15$ kpc.
As expected, including the Type III galaxies (which, by definition, are more extended) in the stack results in a slightly larger characteristic effective radius $r_e = 2.04$ kpc
but a similarly good match to an $n \sim 1$ Sersic model.

We caution, however, that while Figure \ref{stackprofile.fig} confirms that $n\sim1$ models are a fairly good representation of $z =1.5-3.6$ star forming galaxies, the stacked profile
does not account for variability in the size or orientation of its component galaxies.  By effectively discarding information about the projected ellipticity the stack
overestimates the mean effective radius of the sample by a factor $\sim \sqrt{\langle b/a \rangle}$.
We discuss the characteristic sizes of the star forming galaxies in detail in \S \ref{massrad.sec}.

%
%
%

\subsection{Distribution of Axial Ratios}
\label{axial.sec}

\begin{figure}
\plotone{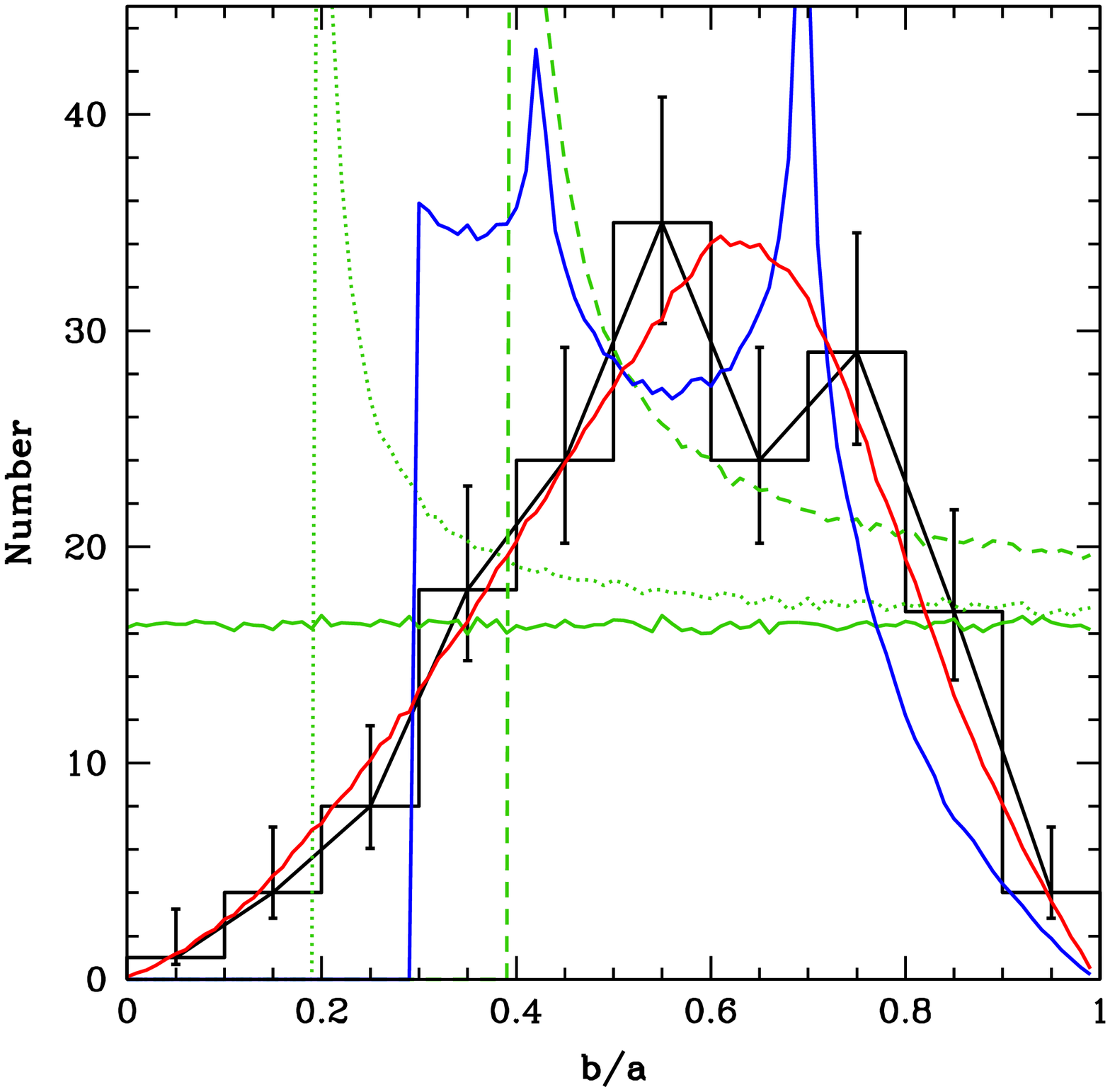}
\caption{Histogram of axis ratios $b/a$ for the galaxy sample (solid black line).  Error bars represent Bayesian confidence intervals (e.g., Cameron 2011) on the number of galaxies in each bin.
Green lines indicate the distribution of $b/a$ expected for inclined disk models based on Monte Carlo simulations; solid, dotted, and dashed green lines
represent intrinsic disk thicknesses $r_0 = 0.0$, 0.2, and 0.4 respectively ($\chi^2 = 43.0, 21.5, 16.9$).  The solid blue line indicates the expected distribution for a triaxial ellipsoid population with
intermediate/major and minor/major axis ratios of 0.7 and 0.3 respectively ($\chi^2 = 6.2$), while the solid red line indicates the  expected distribution assuming a gaussian distribution
of intrinsic intermediate/major and minor/major axis ratios with mean 0.7 and 0.3, and $1\sigma$ width 0.1 and 0.2 respectively ($\chi^2 = 1.2$).
The high frequency noise in the model distributions represents statistical scatter in our Monte Carlo results.}
\label{boahist.fig}
\end{figure}

In Figure \ref{boahist.fig} we plot a histogram of $b/a$ for galaxies with $H_{160} < 24.0$, $n<2.5$, and both major and minor axis lengths well resolved (a total sample of
164 galaxies).\footnote{Since the K-S test indicates a greater than 50\% likelihood of the null hypothesis that the $1.5 \leq z < 2.0$, $2.0 \leq z < 2.5$, and $2.5 \leq z < 3.6$ samples are drawn from the same distribution
we simply combine these three subsamples.  Statistically indistinguishable results are obtained if we exclude Type II galaxies from our analysis,
or include galaxies with resolved major but unresolved minor axes.}  
The distribution\footnote{We assess the reliability of our $b/a$ measurements using Monte Carlo simulations.  Artificial galaxies
with magnitude, radius, Sersic index, and position angle drawn at random from the observed distributions
and $b/a$ uniformly distributed in the range $0 - 1$ are created using GALFIT and placed within our WFC3 fields.
We find that the mean error $\langle | (b/a)_{\rm model} - (b/a)_{\rm measured} | \rangle  = 0.02 - 0.03$ for values of $(b/a)_{\rm model} > 0.3$
and $\langle | (b/a)_{\rm model} - (b/a)_{\rm measured} | \rangle  = 0.07 - 0.1$ for values of $(b/a)_{\rm model} \leq 0.3$.}  
is strongly peaked about $(b/a)_{\rm peak} \approx 0.6$ with tails extending to both extremes $b/a = 0$ and $b/a = 1$.
As we demonstrate below, such a distribution is strongly inconsistent with a population of thick exponential disks  as is commonly assumed in the literature
(e.g., Genzel et al. 2008)
and much more consistent with a population of triaxial ellipsoids.



As discussed by Padilla \& Strauss (2008, and references therein) for a large sample of local galaxies drawn from the Sloan Digital Sky Survey (SDSS),
a population of spiral galaxies with random orientations defines a distribution in $b/a$ that is relatively flat above some minimum value corresponding to the edge-on thickness of the disks.
Taking $i$ to be the inclination of such a disk to the line of sight (where $i = 0^{\circ}$ represents a disk viewed face-on), the observed axial
ratio ($b/a$) of a flattened axisymmetric system is given by (see, e.g., Hubble 1926; Tully \& Fisher 1977):
\begin{equation}
{\rm cos}^2 i = \frac{(b/a)^2-r_0^2}{1-r_0^2}
\label{hubble.eqn}
\end{equation}
where $r_0$ is the intrinsic minor/major axis ratio for a perfectly edge-on system.  In the thin-disk approximation $r_0 = 0$ and Equation \ref{hubble.eqn} reduces to the familiar $b/a = {\rm cos} i$.
In the local universe, typical values for $r_0$ range from 
$r_0 \sim 0.20$ for Sa-type to $\sim 0.08$ for Sd-type galaxies (Guthrie 1992; Ryden 2006), although
variations can also occur with wavelength (e.g., Dalcanton \& Bernstein 2002).
At redshifts $z>1.5$ however, star forming galaxies are known to have significant 
vertical velocity dispersion (e.g.,  Law et al. 2009;  F{\"o}rster Schreiber et al. 2009), and analysis of five of the most disk-like objects (based on velocity maps derived from
integral field spectroscopy) indicates that the median $r_0 \approx 0.34$ (Genzel et al. 2008).
For more typical dispersion-dominated galaxies ($v/\sigma \leq 1$) $r_0$ might be expected to be even larger.

We perform Monte Carlo tests in which we artificially observe a sample of $10^6$ flattened axisymmetric disks from a random distribution of inclinations.\footnote{Strictly, 
we observe a single model galaxy from random viewing angles in the spherical polar coordinate system
$(\theta,\phi)$, where the random viewing positions are distributed uniformly in the azimuthal coordinate $0^{\circ} \leq \theta < 360^{\circ}$ and the cosine of the polar coordinate
$-90^{\circ} \leq \phi \leq 90^{\circ}$, thereby uniformly covering the sky as seen from the perspective of the model galaxy.}
Formally, we quantify the difference between the observational data and the model by the statistic
\begin{equation}
\chi^2 = \frac{1}{\nu} \sum_{i=1}^{B} \frac{(N_{\rm model}-N_{\rm obs})^2}{N_{\rm obs}}
\end{equation}
where $N_{\rm obs}$ is the number of galaxies observed in each of our $B=10$ bins in $b/a$, $\nu=8$, and $N_{\rm model}$ is the number of galaxies expected
in each bin according to the assumed model.
We overplot the distribution of $b/a$ obtained using such flattened axisymmetric disk models on the observational data in Figure \ref{boahist.fig}.
Regardless of the value of $r_0$ adopted, it is not possible  to satisfactorily explain the observed distribution of $b/a$;
$r_0 = 0.0$, 0.2, and 0.4 models have  $\chi^2 = 43.0, 21.5,$ and 16.9 respectively.

In contrast, the peaked distribution of $b/a$ is exactly the form expected for a population of randomly
oriented triaxial ellipsoids such as that found by van den Bergh (1988) for a sample of local irregular galaxies.
We therefore repeat our Monte Carlo analysis assuming the the galaxies can be characterized as triaxial ellipsoids with axis lengths $r_x, r_y, r_z$.
Calculating the projected minor/major axis ratio $b/a$ of a triaxial ellipsoidal surface viewed in an arbitrary orientation is an interesting problem in its own right, and we discuss the
details of this calculation in Appendix \ref{batriax.sec}.  Since we are only interested in axial ratios rather than the absolute lengths we set $r_z = 1$ and consider a grid of
values in the range $r_x, r_y = 0.1, 0.2,..., 1.0$.

\begin{figure}
\plotone{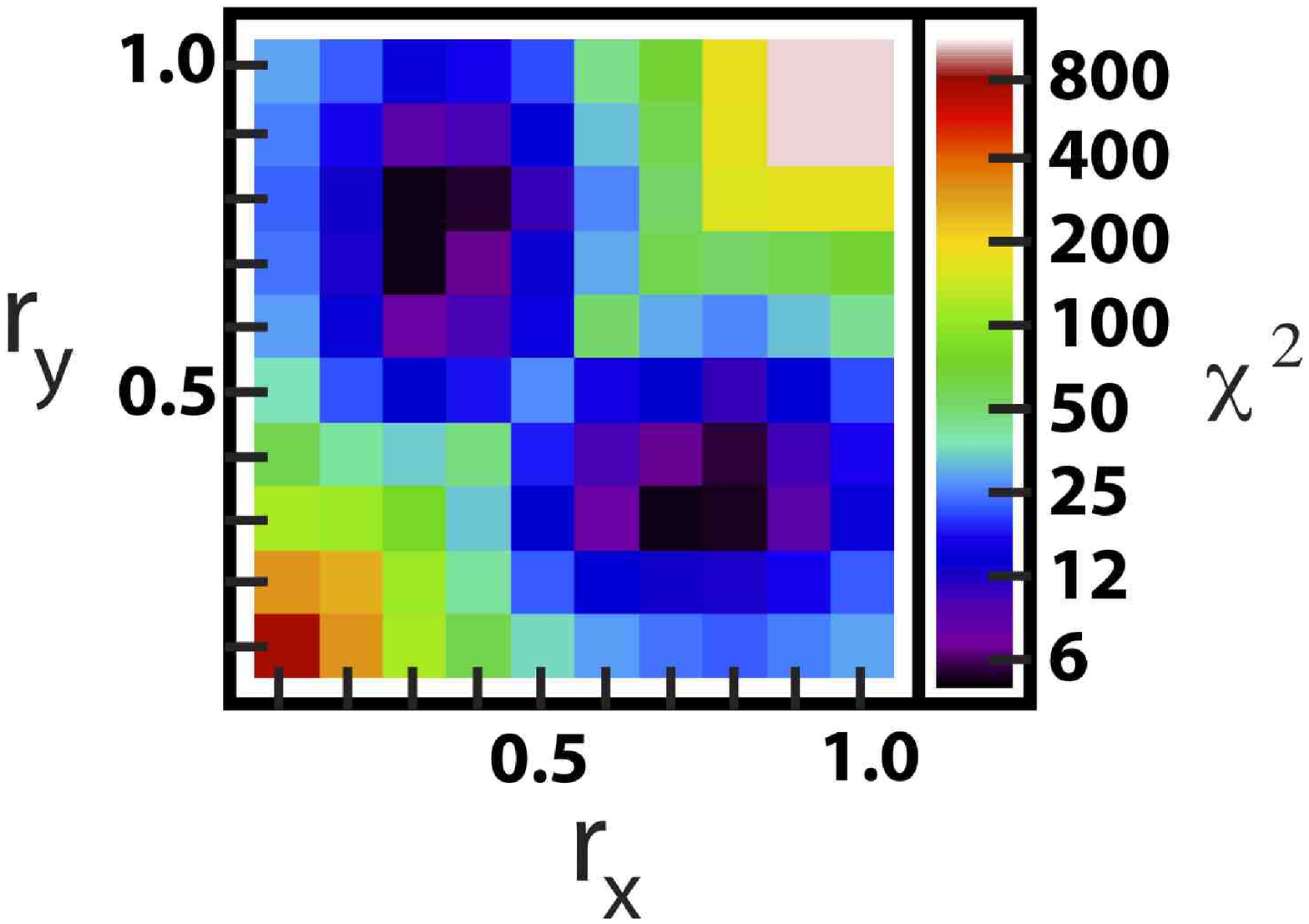}
\caption{Logarithmic plot of $\chi^2$ of the fit between model and observed distribution of axis ratios $b/a$ as a function of $r_x$ and $r_y$ scalelengths.
Since $r_z = 1.0$ is fixed, $r_x$ and $r_y$ effectively probe the range of intermediate/major and minor/major axis ratios in the range 0.1 - 1.0.  The plot is symmetric about
the line $r_x = r_y$ since these parameters are notationally interchangeable.  The minimum of $\chi^2$ is well-defined at $(r_x, r_y) = (0.7, 0.3)$ with $\chi^2 = 6.2$.
By comparison, the $\chi^2$ of the `saddle point' $r_x=r_y = 0.5$ is $\chi^2 = 29.0$.}
\label{chisq.fig}
\end{figure}

In Figure \ref{chisq.fig} we show a surface plot of $\chi^2$ as a function of $r_x$ and $r_y$.
The best agreement between model and observations clearly occurs for a well-defined region around ($r_x, r_y = 0.7, 0.3$).
The expected distribution of $b/a$ for ($r_x, r_y = 0.7, 0.3$)
is shown in Figure \ref{boahist.fig}.  While this is clearly a better description of the observations than the 
axisymmetric disk model (particularly in the expected number of systems with $b/a > 0.8$) it is still imperfect ($\chi^2 = 6.2$; $\nu=7$), 
predicting no galaxies with $b/a<0.3$ and a large excess with $b/a \sim 0.35-0.45$.
These remaining imperfections likely reflect the intrinsic range of morphologies within the galaxy sample- rather than every galaxy having 
an identical shape there is undoubtedly some range about these values.  
Permitting a more realistic distribution of axis ratios (i.e., picking $r_x$ and $r_y$ at random from gaussian distributions with mean 0.7 and 0.3,
and $1\sigma$ width 0.1 and 0.2 respectively), it is possible to reproduce the observed distribution of $b/a$ extremely well (solid red line in Figure \ref{boahist.fig}; $\chi^2 = 1.2$
with $\nu = 5$).

At present, it is meaningless  to distinguish between minor/major axis
ratios of 0.2 versus 0.3, or to state with certainty that a gaussian distribution of intrinsic axis ratios is appropriate.  Our fundamental conclusion, however, is that the majority
of $z=1.5-3.6$ star forming galaxies are best represented by triaxial systems rather than geometrically thick disks (as previously discussed by Ravindranath et al. 2006
and Elmegreen et al. 2005)
and it is worth asking what this means in a physical sense.

Given the overall similarity between the rest-UV and rest-optical morphology (\S \ref{uvopt.sec}) it may simply be that light from
clumpy (and asymmetrically distributed) star forming regions within galaxies (e.g., Bournaud et al. 2009)
dominates the emergent flux at both UV and optical wavelengths.
Alternatively, since we derived $b/a$ for the brightest subcomponent of each galaxy we may be
measuring the intrinsic shape distribution of individual giant star forming clumps.
However, the peaked distribution of $b/a$ persists if we restrict our attention to the most regular single-component systems (i.e., Type I galaxies) suggesting
that we are observing galaxy scale structures with characteristic radii $\sim 1-3$ kpc.
Similarly, the distribution of $b/a$ persists for galaxies with stellar masses greater than $10^{10} M_{\odot}$ in which  stellar continuum emission should be 
well detected in the WFC3 imaging data, suggesting that the stellar mass distribution itself is strongly asymmetric.

Combining our morphological results with observations (e.g., Law et al. 2007a, 2009; F{\"o}rster Schreiber et al. 2009)
that typical $z\sim 2-3$ star forming galaxies have large gas fractions, high velocity dispersions $> 50$ km s$^{-1}$,
and velocity fields that are in many cases inconsistent with rotationally-supported disk models
(especially at lower stellar masses; see discussion by Law et al. 2009), we suggest that the distribution of stars and gas in
these rapidly star-forming galaxies may be inherently triaxial rather than residing largely in a geometrically thick disk.
Such a distribution of gas would be gravitationally unstable, suggesting that the life cycle of $z\sim 2-3$ star forming galaxies may be continually passing in
and out of dynamical equilibrium (e.g., Ceverino et al. 2010).  In such a scenario, gas disks may be only short-lived and continuously forming from recently accreted gas
(whether acquired from mergers or hot/cold mode accretion; e.g., Dekel et al. 2009a, Kere{\v s} et al. 2009), rapidly becoming disrupted,
and reforming again until
the triaxial stellar component (perhaps a precursor of modern-day bulges) acquires sufficient mass to stabilize the growth of a long-lived and extended gas disk (e.g., Martig et al. 2010).
We discuss additional observational support for such a scenario based on low-ionization gas phase kinematics in a companion paper (Law et al. 2012, in preparation).

We note that both our results and conclusions are qualitatively consistent with those of Ravindranath et al. (2006),\footnote{More recently, see also Yuma et al. (2011).} 
who used {\it HST}/ACS imaging in the GOODS fields to demonstrate that the rest-UV
morphologies of star-forming galaxies at $z\sim 3-4$ also have 
a peaked distribution of ellipticities.  While we found the distribution for rest-optical morphologies to be peaked about $(b/a)_{\rm peak} \sim 0.6$ however, Ravindranath et al. (2006)
found $(b/a)_{\rm peak} \sim 0.5$/0.3 for galaxies at $z=3/4$ respectively as seen in the rest-UV.  This difference may be explained in part by the difference in rest-frame wavelength probed by
the two studies; it is perhaps unsurprising that the ellipticity of star forming galaxies in the young universe changes slightly from rest-frame $2000$ \AA (tracing the regions of most recent star formation)
to rest-frame $5000$ \AA (tracing the older stellar population).
In contrast, van der Wel et al. (2011) observed a relatively flat distribution of $b/a$ (above $b/a \sim 0.5$) for a sample of 14
massive ($M_{\ast} > 8 \times 10^{10} M_{\odot}$)  compact quiescent galaxies at $z \sim 2$.  While this may represent a fundamental structural difference between the star-forming and quiescent galaxy samples, we caution that the
quiescent galaxies are ten times more massive than the typical star forming galaxy in our survey, and note
that if increasing stellar mass stabilizes the formation of disks then star forming galaxies of similarly high mass may prove to have similarly disk-like ellipticities.

\subsection{Rest-Optical vs Rest-UV Morphologies}
\label{uvopt.sec}


One of our fields (Q1700+64) was imaged previously using {\it HST}/ACS with the F814W filter (GO-10581, PI: Shapley).  This filter ($\lambda_{\rm eff} = 8332$ \AA) traces rest-UV wavelengths
ranging from 2000-3000 \AA, depending on the redshift of the target galaxy.
The detailed morphologies resulting from this rest-UV imaging program have already been discussed
elsewhere (Peter et al. 2007).  Here we compare the rest-optical and rest-UV morphologies of galaxies overlapping with our WFC3/IR imaging.
For consistency we re-reduce the raw observational data from GO-10581, drizzling them to a 0.08 arcsec pixel scale and smoothing them to a FWHM of 0.18 arcsec
in order to match the observational characteristics of our WFC3/IR imaging data.
We calculate that the F814W image reaches a limiting depth of 28.7 AB for a $5\sigma$ detection within a 0.2 arcsec radius aperture, or $\sim$ 1 mag deeper than our WFC3/IR imaging data.

We show the morphologies of the 18 star forming galaxies that overlap between the two samples in Figure \ref{uvmorph.fig}.
Qualitatively, we note that the morphologies of most galaxies are similar in both rest-UV and rest-optical bandpasses; morphological irregularities or multiple components
visible in one bandpass are similarly visible in the other, resulting in a small morphological k-correction (see discussion by Conselice et al. 2011b).
The smallest variation is exhibited by galaxies of low stellar mass
(for which the light from young stars might reasonably
be expected to dominate both the rest-UV and rest-optical light of the galaxy), while high-mass galaxies exhibit greater differences consistent with the establishment of an evolved
stellar population.  In particular, the galaxies that were observed to be extremely low surface-brightness, red (${\cal R}-K_s \sim 3$ AB),
`wispy' systems in the rest-UV  tend to be high-mass systems that
are much brighter and well-nucleated in the rest-optical (e.g., Q1700-MD103, Q1700-BX767).
This result is similar to that found by Toft et al. (2005)
for a population of red star forming galaxies.

\begin{figure*}
\plotone{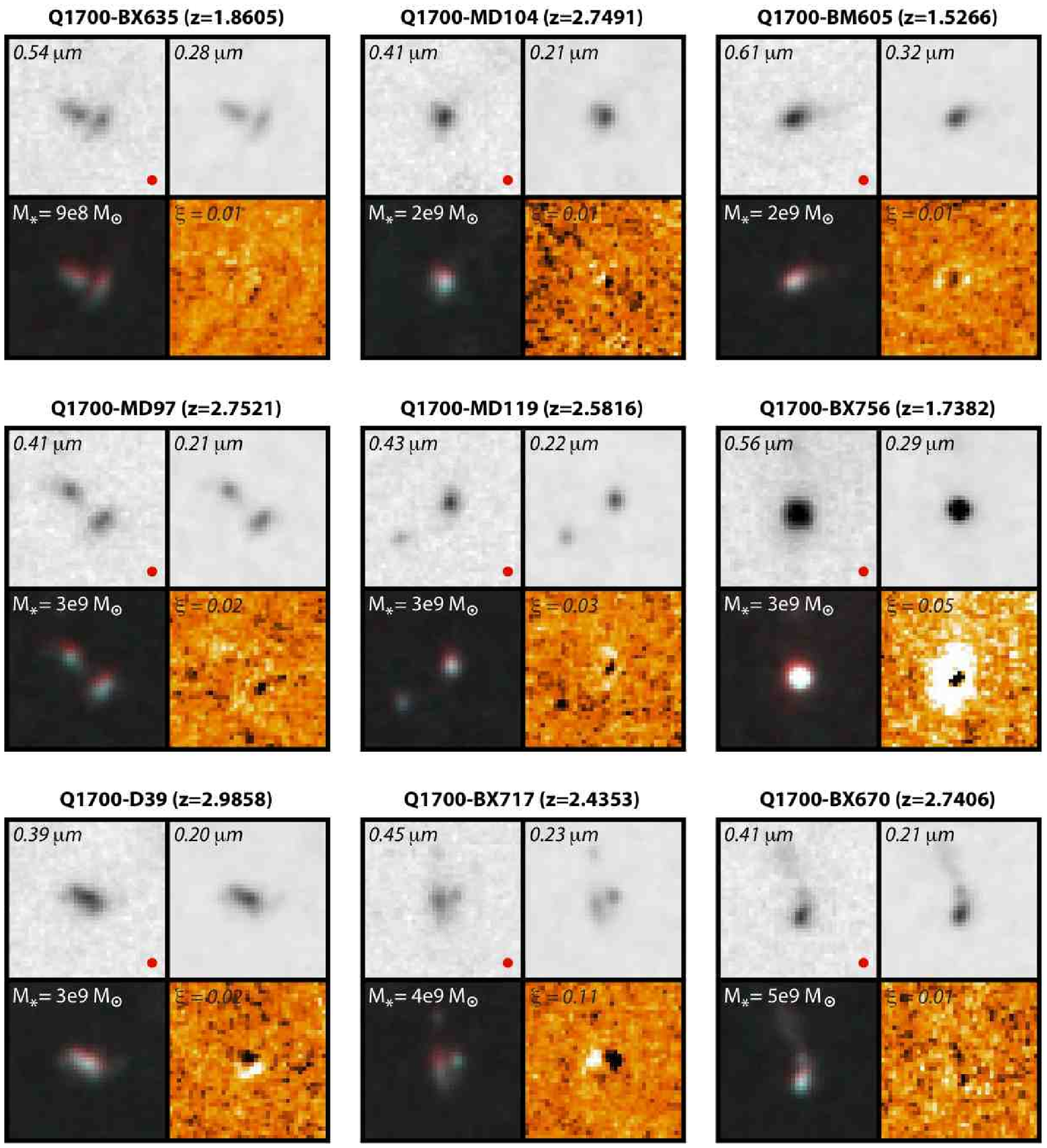}
\caption{Rest-frame UV and optical morphologies are shown (greyscale images) for 18 galaxies imaged with both {\it HST}/WFC3-IR and {\it HST}/ACS in the Q1700+64 field, sorted in order of increasing
stellar mass.  Both images use an arcsinh stretch with the blackpoint set to 21.8 AB arcsec$^{-2}$.
For each galaxy
we indicate the systemic redshift, stellar mass,  and the effective rest-frame wavelength probed by the F160W and F814W filters.  The lower left-hand panel for each galaxy represents a RGB color map
of the system (R=F160W, G=B=F814W), the lower-right hand panels are maps of the color dispersion $\xi$.}
\label{uvmorph.fig}
\end{figure*}

\addtocounter{figure}{-1}
\begin{figure*}
\plotone{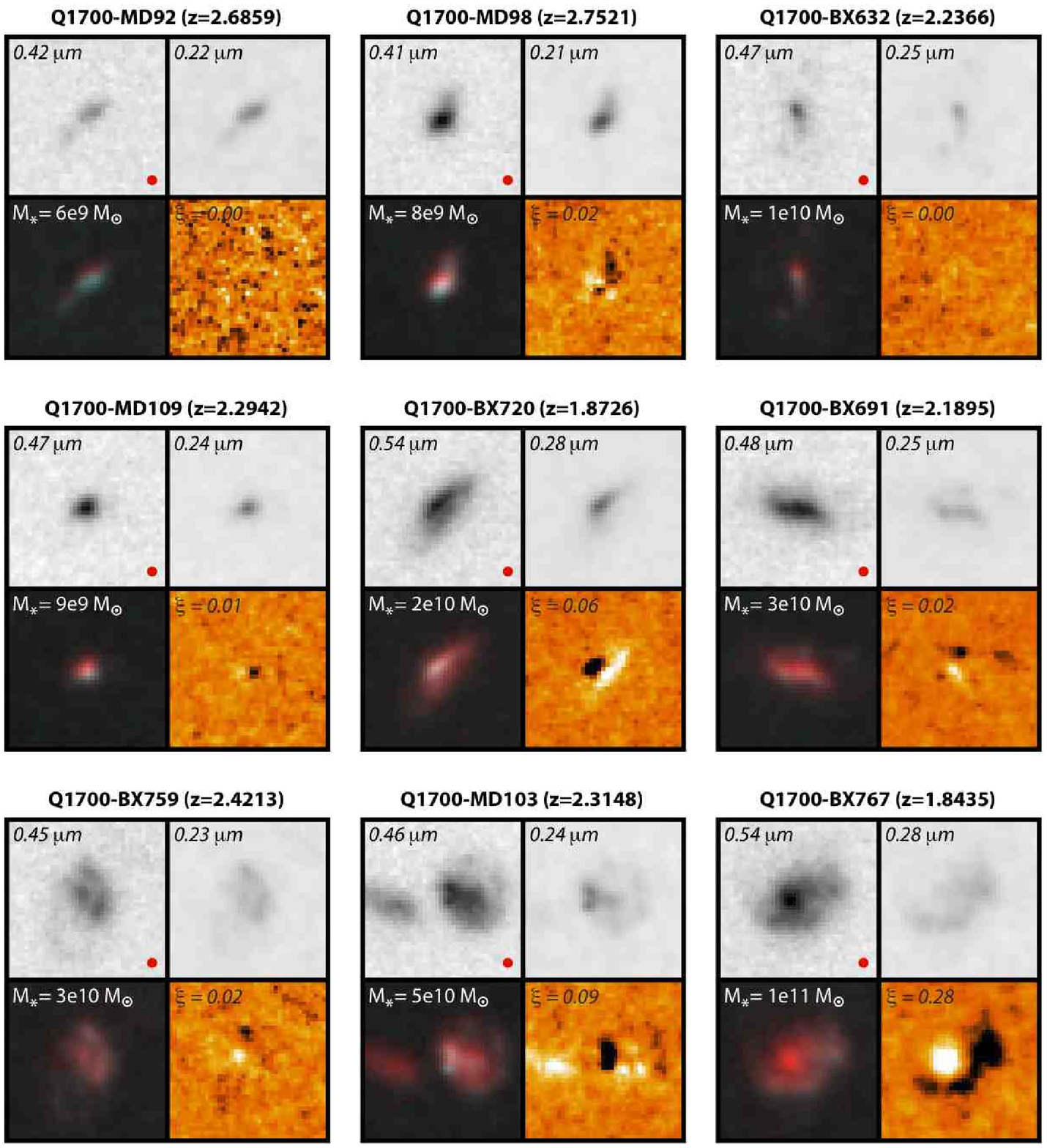}
\caption{{\it Continued}}
\end{figure*}

We quantify this morphological difference by calculating the internal color dispersion $\xi$ (Papovich et al. 2005) 
after carefully aligning the ACS and WFC3 images using the measured centroids of 10 stars.
\begin{equation}
\xi(I_1,I_2) = \frac{\sum(I_2 - \alpha I_1 - \beta)^2-\sum(B_2 - \alpha B_1)^2}{\sum(I_2-\beta)^2 - \sum(B_2 - \alpha B_1)^2}
\end{equation}
where $I_1$ and $I_2$  are the pixel fluxes in the F814W and F160W bandpasses, $\alpha= \sum(I_1 \, I_2)/\sum(I_1^2)$ 
is a scaling factor describing the overall color of the galaxy, $\beta$ adjusts
for the variable background level, and $B_1$ and $B_2$ represent blank background sky regions in each image.
$\alpha$ is set by minimizing the sum $\sum (I_2 - \alpha I_1)^2$, i.e. $\alpha = \sum(I_1  I_2)/\sum(I_1^2)$.
The sum is performed over all pixels in the F160W segmentation map.
The background sums were done by adopting the mean from the calculation performed on the segmentation map
grafted onto 1000 different regions of blank sky.

Values for $\xi$ calculated for each galaxy are quoted in Figure \ref{uvmorph.fig}, and confirm our visual impression that the UV and optical morphologies
differ more greatly for high-mass (alternatively, red) galaxies.  At the low-mass end ($M_{\ast} < 10^{10} M_{\odot}$) $\langle \xi \rangle = 0.02$, while for galaxies with $M_{\ast} > 10^{10} M_{\odot}$
we find $\langle \xi \rangle = 0.09$, peaking at $\xi = 0.28$ for the highest-mass galaxy Q1700-BX767 which displays a red core with a surrounding blue ring.
Similar trends were noted by Labb{\'e} et al. (2003), who found significant rest-UV to rest-optical morphological differences for a sample of six $K$-bright $z \sim 1.4-3$ disk galaxies,
and Papovich et al. (2005), who noted that their galaxies with the highest values of $\xi$ were those with the reddest colors.  We caution that there are relatively
few galaxies in our sample however, and recent work by Bond et al. (2011) looking at the rest-optical vs rest-UV morphologies of 117 ($1.4 < z < 2.9$) star forming galaxies in the GOODS-S
field found a similar mean $\xi = 0.02$ but no evidence for a correlation with galaxy color.
In the near future we anticipate that the relation between rest-UV and rest-optical morphology will be greatly refined by the 
large-area and multi-band CANDELS survey (Grogin et al. 2011; Koekemoer et al. 2011).

\begin{figure}
\plotone{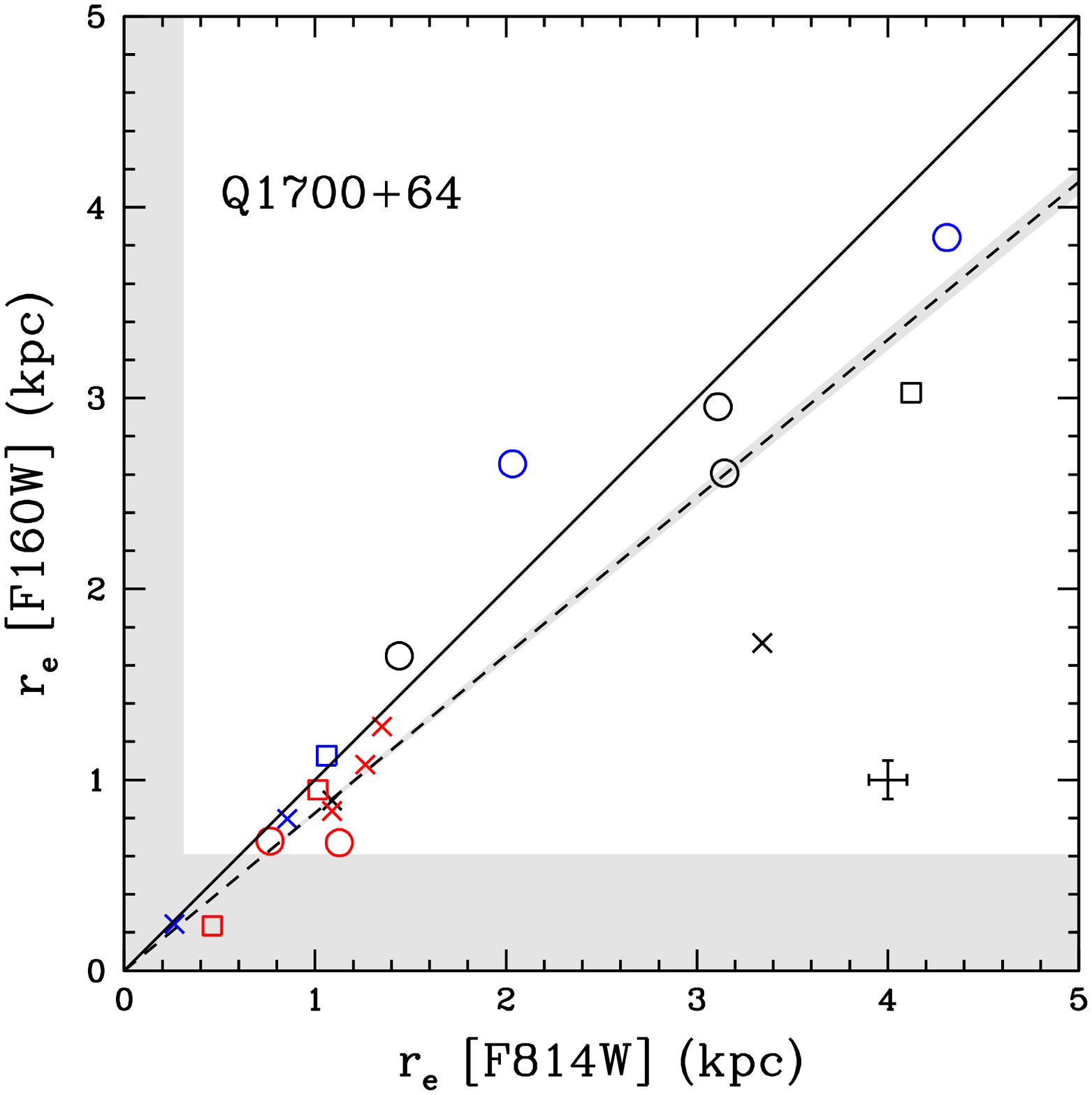}
\caption{Effective radius for galaxies in the Q1700+64  field observed with F160W (rest-frame $4000-5000$\AA)
 versus F814W (rest-frame $2000-3000$\AA).
Blue, black, and red points represent galaxies in the 
 $z=1.5-2.0$, $z=2.0-2.5$, and
$z=2.5-3.6$ redshift intervals respectively, while crosses, open boxes, and open circles represent galaxies 
of visual type I, II, and III.
The grey shaded regions indicate the $3\sigma$ limit on unresolved point sources in each bandpass.  The solid lines indicate 1-1 relations, while the dashed line
indicates rest-UV radii 21\% larger than the rest-optical.  A typical uncertainty is indicated by the symbol in the lower right corner.}
\label{uvoptrad.fig}
\end{figure}

It is also possible to compare the effective radii $r_e$ derived in each of the two bandpasses.  
Similarly to Dutton et al. (2010) and Barden et al. (2005), we find  (Figure \ref{uvoptrad.fig})
that the rest-UV sizes of these galaxies are $21\pm2$\% larger on average than their optical sizes, although there is increasing scatter in the relation at large radii (i.e., large mass) in part
because these galaxies are red (${\cal R}-K_s \sim 3$ AB) and poorly defined in the F814W data.
This relation is largely unchanged if the Sersic index $n$ of the radial profile is kept fixed between the F160W and F814W data.

\section{THE STELLAR MASS RADIUS RELATION}
\label{massrad.sec}

\subsection{Observed Relation}

\begin{figure*}
\plotone{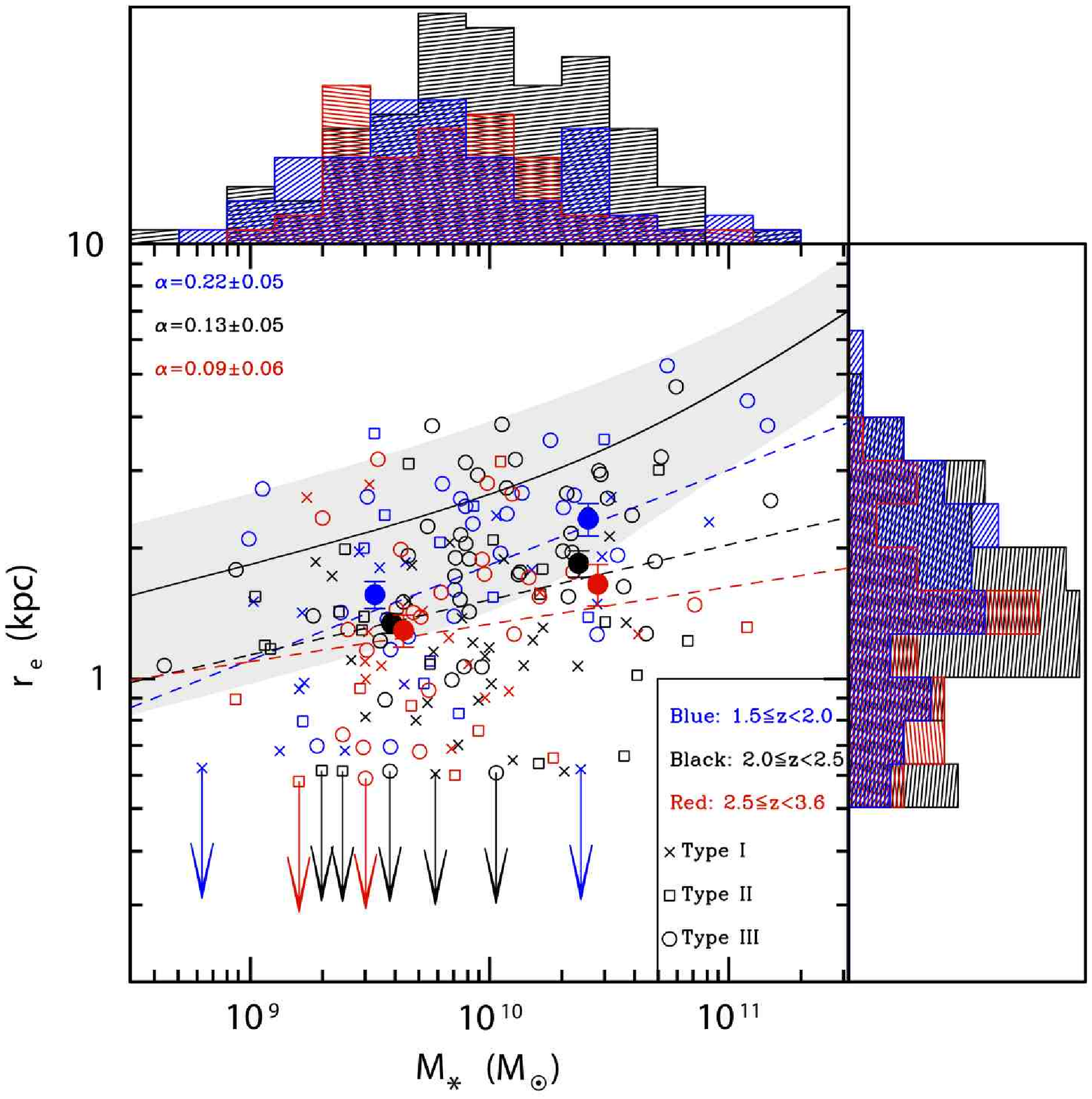}
\caption{Effective circularized radius $r_{\rm e}$ as a function of stellar mass $M_{\ast}$.
Symbols are as in Figure \ref{uvoptrad.fig},
upper limits for unresolved sources are denoted with arrows.  The filled circles and error bars represent the mean value and associated uncertainty for galaxies in each
redshift bin with stellar masses $M_{\ast} < 10^{10} M_{\odot}$ and $M_{\ast} \geq 10^{10} M_{\odot}$.  The solid black line and shaded grey region 
indicate the mean low-redshift relation and its $1\sigma$ scatter for late-type galaxies from
Shen et al. (2003), while the blue/black/red dashed lines indicate the best-fitting power-law relation of the form $r_{\rm e} \sim M_{\ast}^{\alpha}$ for the $z=1.5-2.0/2.0-2.5/2.5-3.6$
samples respectively.}
\label{mrfig.fig}
\end{figure*}

In Figure  \ref{mrfig.fig} we plot the effective circularized radius $r_e$ as a function of stellar mass for all galaxies 
with  $H_{\rm AB} \leq 24.0$ and  Sersic index $n<2.5$, constituting a sample
of 59/93/50 galaxies in the $z=1.5-2.0/2.0-2.5/2.5-3.6$ redshift ranges respectively.
Of these 202 galaxies, 9 ($\sim 4$\%) have effective radii consistent with an unresolved point source, and may represent either the compact end of the galaxy distribution
or faint AGN (albeit with no obvious signature in the UV spectra or broadband SED out to $\sim 7000$\AA\ rest frame).

\begin{deluxetable*}{rlccc}
\tablecolumns{5}
\tablewidth{0pc}
\tabletypesize{\scriptsize}
\tablecaption{Mean Circularized Effective Radii and Stellar Masses}
\tablehead{
\colhead{} & \colhead{z} &  \colhead{$M_{\ast} = 10^{9-11} M_{\odot}$} & \colhead{$M_{\ast} = 10^{9-10} M_{\odot}$} & \colhead{$M_{\ast} = 10^{10-11} M_{\odot}$}}
\startdata
$\langle r_{\rm e} \rangle$ &1.5-3.6 & $1.58\pm0.05$ kpc & $1.39\pm0.06$ kpc & $1.91\pm0.10$ kpc  \\
$\langle r_{\rm e} \rangle$ &1.5-2.0 & $1.82\pm0.11$ kpc  & $1.56\pm0.11$ kpc   & $2.33\pm0.20$ kpc  \\
$\langle r_{\rm e} \rangle$ & 2.0-2.5 & $1.60\pm0.07$ kpc  &  $1.34\pm0.07$ kpc  & $1.84\pm0.13$ kpc \\
$\langle r_{\rm e} \rangle$ & 2.5-3.6 & $1.35\pm0.09$ kpc  &  $1.29\pm0.11$ kpc  &  $1.65\pm0.18$ kpc  \\
\hline
$\langle r_{\rm e}/r_{\rm SDSS} \rangle$  & 1.5-3.6 & $0.60\pm0.02$ & $0.60\pm0.02$ & $0.59\pm0.03$  \\
$\langle r_{\rm e}/r_{\rm SDSS} \rangle$ & 1.5-2.0 & $0.70\pm0.04$  & $0.68\pm0.05$   & $0.73\pm0.06$  \\
$\langle r_{\rm e}/r_{\rm SDSS} \rangle$  & 2.0-2.5 & $0.59\pm0.03$  &  $0.60\pm0.03$   &  $0.57\pm0.04$  \\
$\langle r_{\rm e}/r_{\rm SDSS} \rangle$  & 2.5-3.6 & $0.45\pm0.02$  &  $0.46\pm0.03$  & $0.51\pm0.07$  \\
\hline
$\langle {\rm log}(M_{\ast}/M_{\odot}) \rangle$ &1.5-3.6 & 9.88 & 9.58 & 10.38  \\
$\langle  {\rm log}(M_{\ast}/M_{\odot})\rangle$ &1.5-2.0 & 9.81  & 9.52 &  10.41\\
$\langle  {\rm log}(M_{\ast}/M_{\odot})\rangle$ & 2.0-2.5 & 9.96  & 9.61 &  10.37\\ 
$\langle {\rm log}(M_{\ast}/M_{\odot}) \rangle$ & 2.5-3.6 &  9.85  & 9.62 &  10.45\\

\enddata
\label{mrtable.tab}
\end{deluxetable*}

Figure \ref{mrfig.fig} indicates that galaxies occupy a large range of effective radii at all redshifts $z=1.5-3.6$ and stellar masses
$M_{\ast} = 10^9 - 10^{11} M_{\odot}$ with the $1\sigma$ standard deviation of the distribution $\sim 0.2$ dex comparable to the scatter in the local 
star forming galaxy relation (e.g., Shen et al. 2003).
Despite the large width of the distribution in $r_e$, however, 
there is a mean mass-radius relation in place at early as $z\sim3$ that evolves with decreasing redshift.
Binning our sample by redshift and stellar mass we calculate\footnote{Values represent the $2.5\sigma$-clipped mean.} that
$\langle r_{\rm e} \rangle = 1.29 \pm 0.11$  ($1.65 \pm 0.18$) kpc for galaxies
in the mass range $M_{\ast} \leq10^{10}$ ($> 10^{10}) M_{\odot}$ respectively at redshift $z = 2.5 - 3.6$, increasing with cosmic time to
$\langle r_{\rm e} \rangle = 1.34 \pm 0.07$  ($1.84 \pm 0.13$) kpc by $z = 2.0 - 2.5$, and to 
$\langle r_{\rm e} \rangle = 1.56 \pm 0.11$ ($2.33 \pm 0.20$) kpc by $z = 1.5 - 2.0$ (see summary in Table \ref{mrtable.tab}).
These results are consistent with the early values calculated for a subset of our sample by Nagy et al. (2011)
to within the estimated uncertainty; 
the strongest evolution in effective radius with redshift occurs for higher mass galaxies $M_{\ast} > 10^{10} M_{\odot}$.
Parameterizing the stellar mass-radius relation as $r_{\rm e} \sim M_{\ast}^{\alpha}$ we
find that the best-fit value of the powerlaw index  $\alpha=0.22 \pm0.05$, $0.13\pm0.05$, and $0.09\pm0.06$ for the redshift $z=1.5-2.0$, $2.0-2.5$, and $2.5-3.6$ intervals
respectively.

\begin{figure}
\plotone{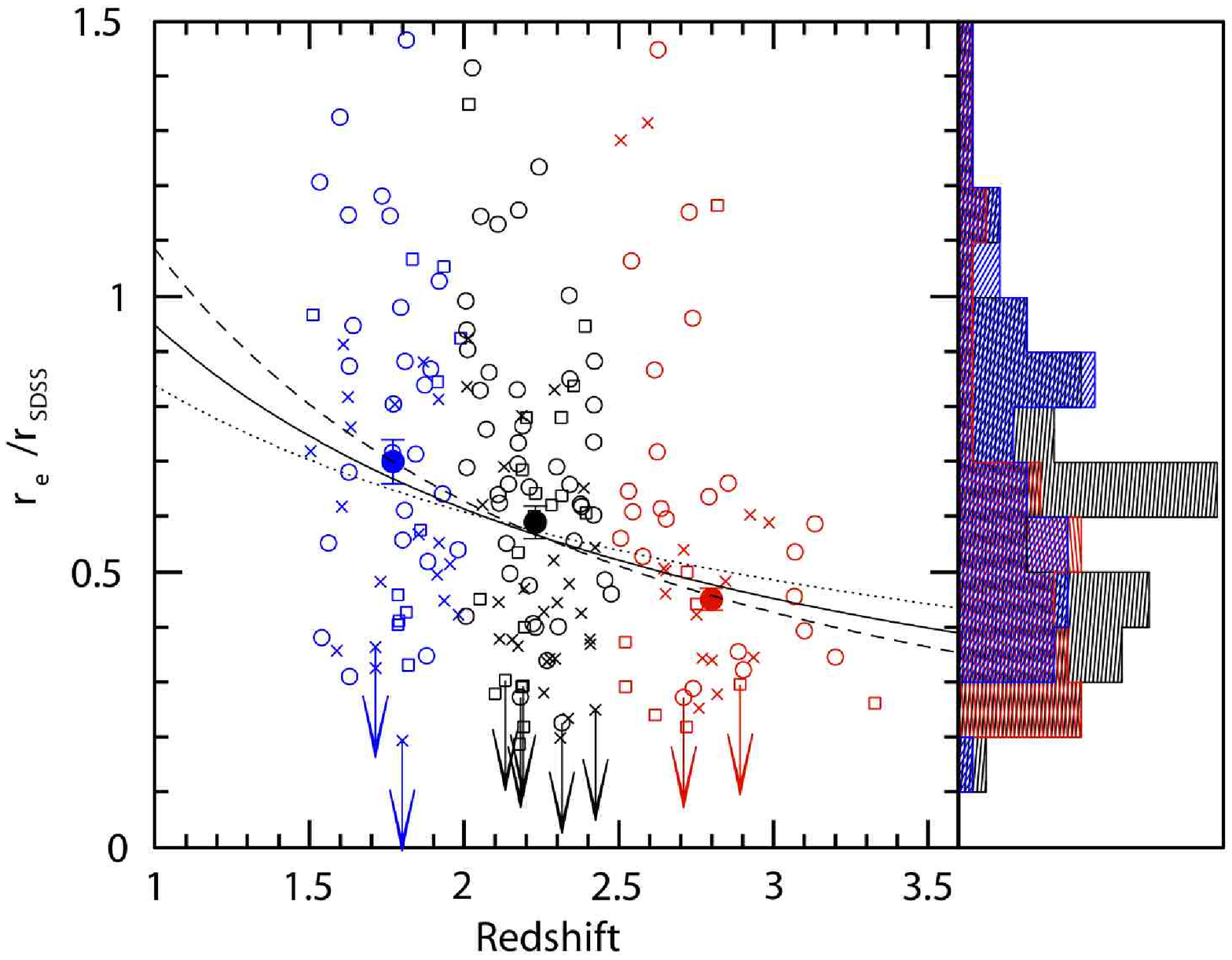}
\caption{Effective circularized radius $r_{\rm e}$ as a fraction of the local relation ($r_{\rm SDSS}$) at a given stellar mass as a function of redshift.  Symbols are as in Figure \ref{mrfig.fig}; filled circles
represent the mean and associated uncertainty of galaxies in each of the three redshift ranges.  Parametrizing the evolution of galaxy size with redshift
as $r_{\rm e} \sim (1+z)^{\gamma}$, the solid line indicates the best-fit value of $\gamma=-1.07 \pm 0.28$, while the the dashed/dotted lines represent the $1\sigma$ uncertainties on the index
of $\gamma=-1.35$ and $\gamma=-0.79$ respectively.}
\label{shenfig.fig}
\end{figure}

As indicated by the top histogram in Figure \ref{mrfig.fig} the
three redshift samples each probe galaxies with a slightly different range of stellar masses, and it is therefore useful to calculate a normalized quantity
$r_{\rm e}/r_{\rm SDSS}$ for each galaxy, where
$r_{\rm SDSS}$ as a function of $M_{\ast}$ (solid black line in Figure \ref{mrfig.fig})
is the mean effective circularized radius for late-type (i.e., $n<2.5$) low-redshift galaxies in the SDSS (Shen et al. 2003).
As indicated by Figure \ref{shenfig.fig}, typical star forming galaxies at fixed stellar mass were significantly smaller at $z>1.5$ than in the nearby universe,
with $\langle r_{\rm e}/r_{\rm SDSS} \rangle = 0.70\pm0.04$, $0.59\pm0.03$, and $0.45\pm0.02$ for the $z=1.5-2.0$, $2.0-2.5$, and $2.5-3.6$ samples respectively.\footnote{Since our
galaxies were selected from unresolved ground-based imaging data we do not expect intrinsic size to have an effect on our selection function.}
If galaxies at fixed stellar mass in the range $M_{\ast} = 10^9 - 10^{11} M_{\odot}$ can be assumed to grow with redshift as
$r_{\rm e} \sim (1+z)^{\gamma}$, a linear least squares fit to the data indicates that
$\gamma = -1.07 \pm 0.28$ between $z=3.6$ and $z=1.5$ (solid line in Figure \ref{shenfig.fig}).  
This is consistent with similar determinations $\gamma = -1.3$ and $\gamma = -1.11$ found for massive star forming galaxies by van Dokkum et al. (2010) and Mosleh et al. (2011) respectively.
Extrapolation of this power law suggests that actively star forming galaxies
in the young universe may evolve onto the local late-type mass-radius relation by $z\sim1$ (although see \S \ref{mrcaveats.sec}), consistent with recent evidence that the mass-radius relation
for star-forming galaxies evolves only weakly in the redshift interval $z = 0-1$ (Barden et al. 2005).

Individual star forming galaxies, however, grow in both stellar mass and radius simultaneously and eventually evolve into typical $\sim L^{\ast}$ galaxies
by the present day as indicated by clustering analyses (e.g., Conroy et al. 2008).  Given the shallow observed mass-radius relation for star forming galaxies at $z=1.5-3.6$, it is clearly not possible for individual galaxies
to evolve along this relation to match the local sample.  Rather, galaxies need to add mass at large radii via steeper growth of the form $r\sim M$ or $r\sim M^2$ as illustrated in Figure
\ref{mrcompare.fig} (see also Figure 8 of van Dokkum et al. 2010).  Such growth may be consistent with expectations for major and minor mergers respectively (e.g., Bezanson et al. 2009,
Naab et al. 2009 for early-type galaxies).

\subsection{Comparison with Previous Results}

\begin{figure*}
\epsscale{0.8}
\plotone{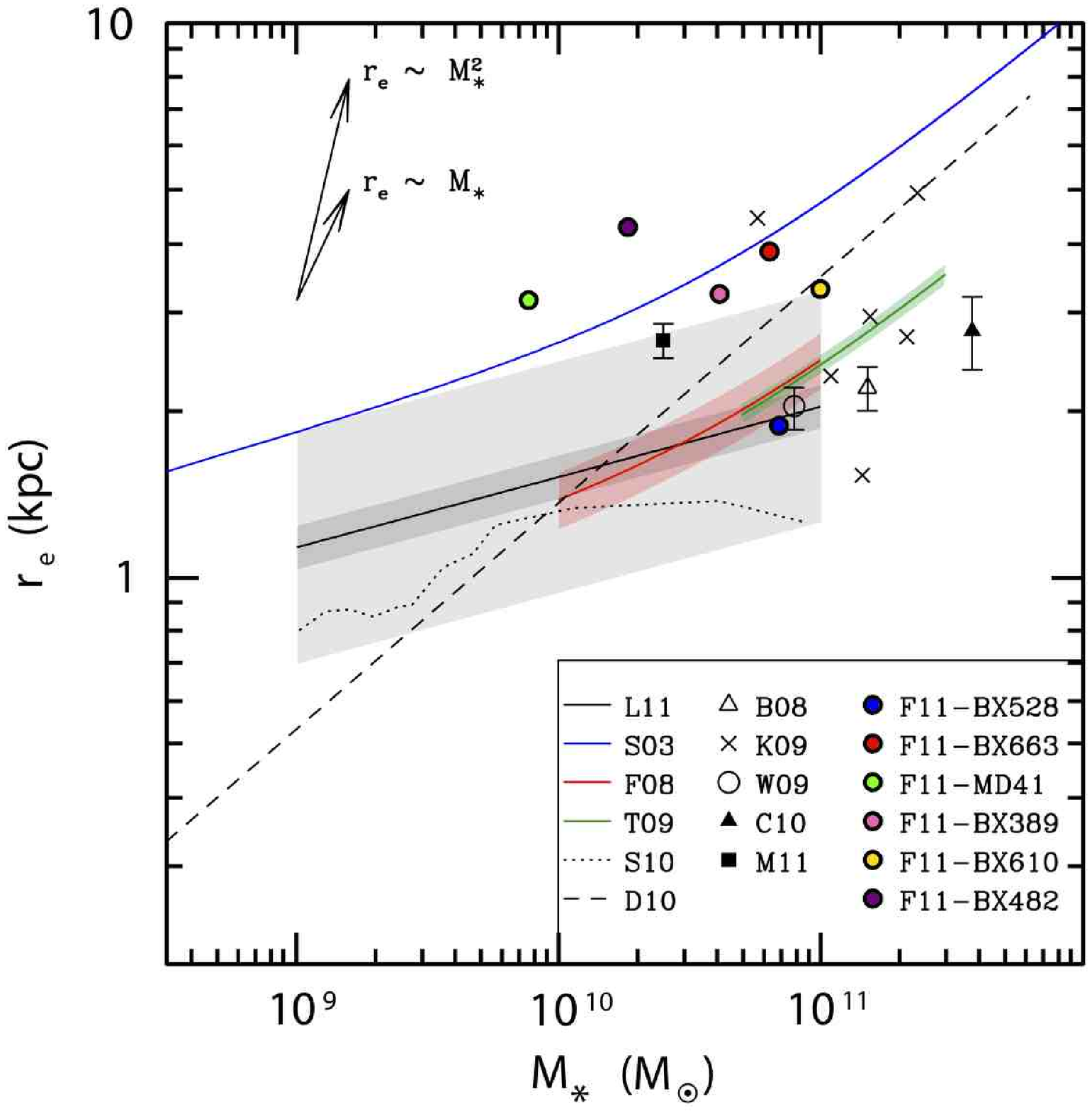}
\caption{Stellar mass vs. circularized effective radius compared to observational and theoretical results in the literature.  The solid line represents our power law fit to the $z=2.0-2.5$ relation
${\rm log} (r_{\rm e}/{\rm kpc}) = 0.13 {\rm log} (M_{\ast}/M_{\odot})  -1.09$, with 
the $1\sigma$ uncertainty in the mean and $1\sigma$ width of the distribution indicated by dark/light grey shaded regions respectively.  
The local late-type galaxy relation from Shen et al. (2003; S03) is indicated by a solid blue line.
Observational data correspond to Buitrago et al. (2008; B08),
 Franx et al. (2008; F08), Kriek et al. (2009; K09), Toft et al. (2009; T09), Williams et al. (2009; W09), Carrasco et al. (2010; C10), Mosleh et al. (2011; M11), and 
F{\"o}rster Schreiber et al. (2011; F11).  The selection criteria of each of these studies are discussed in the text,
the shaded regions for F08 and T09 correspond to the $1\sigma$ uncertainty in the mean.
We also plot the ``WF2Dec'' simulation of
Sales et al. (2010; dotted line), and the simulation of Dutton et al. (2011; dashed line).  The arrows represent growth of the form $r_{\rm e} \sim M_{\ast}$ and $r_{\rm e} \sim M_{\ast}^2$ 
for reference.}
\label{mrcompare.fig}
\end{figure*}

In Figure \ref{mrcompare.fig} we plot the best-fit power law model 
of the stellar mass --- radius relation
for our $z=2.0-2.5$ galaxy sample against a variety of previous observational samples available in the literature.\footnote{Where necessary, results have been 
converted to a Chabrier IMF and circularized effective half-light radii.}
Our results are generally consistent at the $1-2\sigma$ level with previous studies that, due to observational limitations, have typically been conducted for galaxies
with high stellar masses $M_{\ast} > 10^{10} M_{\odot}$ (e.g., Franx et al. 2008; Toft et al. 2009; Williams et al. 2010; Targett et al. 2011) and extend these previous results
down to $M_{\ast} \sim 10^{9} M_{\odot}$.

The most direct comparison can be made to Mosleh et al. (2011), who
used deep ground-based $K$-band imaging across the GOODS-N field to measure the characteristic sizes of 41 
massive ($M_{\ast} = 10^{10} - 10^{11} M_{\odot}$) BM/BX star-forming galaxies
in the redshift range $z=1.4-2.7$ and 4 LBGs in the redshift range $z=2.7-3.5$ for which spectroscopic redshifts have been made publicly available by Reddy et al. (2006).
Although these target galaxies were selected and spectroscopically confirmed in a manner identical to our own sample, we find
significant disagreement with respect to the mean $r_e$ as a function of stellar mass.
As given in their Table 3, the Mosleh et al. (2011) BM/BX galaxy sample has a median mass of ${\rm log} (M_{\ast}/M_{\odot}) = 10.4$
and median radius  of $r_{\rm e} = 2.68\pm0.19$ kpc, and the LBG sample a median mass  of  ${\rm log} (M_{\ast}/M_{\odot}) = 10.3$
and median radius of $r_{\rm e} = 2.22\pm0.61$ kpc.  Within the same ranges of redshift and stellar mass, our BM/BX and LBG samples have 
$2.5\sigma$-clipped mean radii of $r_{\rm e,c} = 1.91\pm0.10$ kpc and $r_{\rm e} = 1.62\pm0.28$ kpc respectively.
Although our WFC3 imaging data are significantly deeper ($\sim 2.5$ mag) and better resolved ($0.18$ arcsec vs. $\sim 0.5$ arcsec) than the ground-based $K$-band imaging,
our experience with the robustness of $r_{\rm e}$ (see \S \ref{robust.sec}) and tests degrading our images to the quality of the ground-based data
do not suggest an obvious instrumental reason for the large $\sim 4\sigma$ difference in the mean values.\footnote{Likewise, 
using the median instead of the sigma-clipped mean makes a negligible difference to our calculations.}

A particularly valuable comparison can also be made to 
F{\"o}rster Schreiber et al. (2011), who used {\it HST}/NICMOS F160W (PSF FWHM $\sim 0.14$ arcsec) to study the
rest-optical morphologies of 6 massive star-forming galaxies at $z=2.0-2.5$ selected from
the SINS H$\alpha$ integral-field kinematics survey (F{\"o}rster Schreiber et al. 2009).  Each of these six galaxies are plotted as colored circles in Figure \ref{mrcompare.fig};
we convert their measurements to circularized effective radii by multiplying by $\sqrt{b/a}$ as tabulated in their Table 4.\footnote{We plot the $r_{\rm e}$ of the brightest component 
from their two-component fit to the galaxy Q1623-BX528 for consistency with our procedure described in \S \ref{galfit.sec}.}
Two of these six galaxies were also observed as part of our HST/WFC3 imaging program: Q1623-BX528 and Q2343-BX389.
While our measured effective radii for Q1623-BX528 differ by $\sim 30$\% due to a different number of morphological components used to fit the complicated light distribution
(F{\"o}rster Schreiber et al. 2011 used two components, while we used three), our radii for
the single-component Q2343-BX389 agree to within 1\%, suggesting that there is negligible systematic difference between the radii calculated by the two surveys.
Except for the multicomponent Q1623-BX528 (which F{\"o}rster Schreiber et al. 2009 classify as a merger on the basis of kinematic data and multi-component rest-frame optical
continuum morphology) ,
all of the galaxies studied by F{\"o}rster Schreiber et al. (2011) have radii roughly twice the mean size of their parent color-selected, spectroscopically
confirmed galaxy population at a given stellar mass
and lie in the top 5\% of the $r_{\rm e}$ distribution for our observed sample of BM/BX galaxies at $z=2.0-2.5$.
This suggests that the subset of galaxies observed by F{\"o}rster Schreiber et al. (2011), 
the majority of which were selected to be the most disk-like within the SINS $z\sim2$ sample, falls among the high $r_{\rm e}$ extreme of the galaxy population
in the stellar mass range  $M_{\ast} \sim 10^{10} - 10^{11} M_{\odot}$ (see also discussion by Law et al. 2009; Dutton et al. 2010),
while following some of the general trends observed at this redshift between size, specific SFR, and stellar mass surface density (Franx et al. 2008).
We expand upon this discussion by relating the morphologies of these galaxies (plus 12 additional galaxies from the OSIRIS and/or SINS kinematic surveys that fell within our WFC3 imaging fields) 
to their ionized-gas kinematics in a forthcoming contribution (Law et al. 2012, in preparation).




\subsection{Comparison with Theoretical Simulations}

Although theoretical simulations of $z\sim2-3$ star forming galaxies are still in their infancy, the sizes predicted by such simulations are in rough agreement with our
observed values.  In Figure \ref{mrcompare.fig} (dotted and dashed lines) we illustrate the results of two such models from Sales et al. (2010) and Dutton et al. (2010) respectively.

Sales et al. (2010) use cosmological $N$-body/SPH simulations to model the growth of baryonic structures in galaxies for four different feedback prescriptions.
Of these four prescriptions, their ``WF2Dec'' model most closely matches both our observations and our physical understanding of these galaxies; in this model
relatively strong feedback from star forming regions
results in the efficient removal of gas from galaxies via an outflowing wind with velocity $\sim 600$ \kms.
Such peak outflow velocities are generally consistent with observations for our BM/BX/LBG galaxy sample (see, e.g., Steidel et al. 2010).
As discussed by Sales et al. (2010), as feedback strength increases it suppresses star formation so that galaxies of a given stellar mass 
tend to inhabit larger haloes and can thus have correspondingly larger characteristic sizes.
Assuming that their stellar half-mass radii roughly correspond to visible-band half-light radii, and converting to circularized values
by multiplying by  $\langle \sqrt{b/a} \rangle \approx 0.77$, we plot their predicted stellar mass --- radius relation in Figure \ref{mrcompare.fig} (dotted line).
The model is generally consistent with our observations, although it slightly underpredicts the typical galaxy size by $\sim 0.1$ dex.
In contrast, in models with no feedback the majority of stars form in dense systems at early times, resulting in mean circularized half-light radii $\sim 0.5$ kpc
at $M_{\ast} \sim 10^{10} M_{\odot}$ that disagree strongly with our observations.

Dutton et al. (2010) also study the evolution of scaling relationships with redshift using a series of semi-analytic models
that roughly reproduce the velocity-mass-radius relations at $z=0$.  In particular,
they focus on the evolution of the zero-point calibration of these relations, predicting that  the evolution from $z=2$ to $z=0$ shifts
the mass-radius relation upwards in radius by $\sim$ 0.3 dex.
As illustrated in Figure \ref{mrcompare.fig} (dashed line), the magnitude of this zeropoint shift is consistent with our observations
at $M_{\ast} \sim 10^{10} M_{\odot}$ (i.e., the mass at which the models also overlap the observed local relation).\footnote{The slope of the Dutton et al. (2010)
relation is too steep to match the observational data, but this is simply because their study was not intended to
address the mass dependence of the galaxy mass vs. halo mass fraction.}

\subsection{Caveats}
\label{mrcaveats.sec}

We close by discussing a few of the caveats and complications that can affect the mass-radius relation that we have derived.

First, the galaxies in the $z=2.5-3.6$ subsample have fainter $H_{160}$ magnitudes than galaxies in the lower redshift bins (Figure \ref{pophist.fig}), and Figure \ref{robfig.fig} demonstrated
(see discussion in \S \ref{magrobust.sec}) that the recovered value of $r_{\rm e}$ can vary as a function of total source magnitude.  However, 
this effect does not significantly influence our conclusions.  First, $r_{\rm e}$ is extremely stable for galaxies with isolated morphologies and small radii
characteristic of much of the observational sample.  While $r_{\rm e}$ is less robust to $H_{160}$ magnitude for larger and more irregular galaxies, the majority of the variation occurs
for magnitudes $H_{160}>24.0$ which we deliberately exclude from our analysis.  
The mean observed magnitudes of our $1.5\leq z < 2.0$, $2.0 \leq z < 2.5$, and $2.5 \leq z < 3.6$ samples are
$\langle H_{160} \rangle = 23.1, 23.2, 23.5$ respectively.  Across such a small range $\Delta H_{160} = 0.4$ mag the change in radius for all morphological types is $\lesssim 4$\%, comparable to the 
statistical uncertainty in the quoted  $\langle r_{\rm e}/r_{\rm SDSS} \rangle$.
Indeed, even were 
we to include faint galaxies with $H_{160}>24.0$ in our analysis we find that the mean values of $\langle r_{\rm e}/r_{\rm SDSS} \rangle$ change by $\lesssim 1\sigma$.

It is also possible that our results may be biased due to our assumption that the radius $r_{\rm e}$ of a multi-component system may be characterized by the radius
of the brightest individual component, while our stellar masses (derived from seeing-limited ground-based photometry and similarly confused {\it Spitzer}/IRAC photometry)
represent the integral over the light of all of the components.
If we repeat our previous analyses instead assuming that the stellar mass of these systems is proportional to the fraction of the $H_{160}$ flux in the primary component,
or simply omitting galaxies with multiple well-defined individual components from our analysis, we find that 
values for $\langle r_{\rm e}/r_{\rm SDSS} \rangle$ in each of the three redshift bins are consistent with their previously calculated values
to within $\sim 1\sigma$.
We are therefore confident that our results are not significantly affected by our assumption of how to define $r_{\rm e}$ for multi-component systems.

Some of the apparent evolution in characteristic radius at fixed stellar mass from $z\sim3$ to $z\leq 2$ may also be
due to the variable $K$-correction in our fixed observational bandpass.  With an effective wavelength of $\lambda_{\rm eff} = 15369$\AA, the F160W filter probes
rest frame 5548, 4758, and 4044 \AA\ emission at the mean redshift of three samples ($\langle z \rangle = 1.77, 2.23, 2.80$).
However, we note that the effective radii derived for our galaxies in the Q1700+64 field varied by only $\sim 20$\% from rest-frame 5000\AA\ to rest-frame 2500\AA;
linear interpolation suggests that the change from 5000\AA\ to 4000\AA\ would be much smaller, $\lesssim 8$\%.
Similarly, Dutton et al. (2010) make theoretical predictions for the difference in effective radius between a variety of optical/NIR bandpasses; interpolating their results
suggests that we might expect a systematic increase of $0.04\pm0.03$ dex in log($r_{\rm e}$) from the lowest to highest redshift sample (i.e., sizes measured at longer wavelengths
are smaller than those measured at shorter wavelengths, corresponding to inside-out disk growth) due to such bandshifting.
This is comparable to the formal uncertainty on our measured $\langle r_{\rm e}/r_{\rm SDSS} \rangle$ in each of the 3 redshift bins, and would represent only a minor correction.
Likewise, the results of Barden et al. (2005; see their Fig. 2) suggest that the correction factor would be $\lesssim 2$\%, which is much smaller than our $\sim 5-10$\% uncertainty on
$\langle r_{\rm e} \rangle$ in each of our redshift bins.

Finally, we caution that the precise values derived for the size evolution of galaxies compared to their low-redshift counterparts at similar stellar mass is complicated by uncertainties
in the local relation.  Although we adopted the Shen et al. (2003) estimate of the local mass-radius relation for late type galaxies, we note that numerous authors (e.g., Barden et al. 2005;
Trujillo et al. 2006; Guo et al. 2009) find that Shen et al. (2003) underestimate their effective radii.  
This discrepancy is due in part to systematic differences in analysis techniques (GALFIT modeling vs 1-dimensional radial profile fitting), definition of early vs late-type galaxies ($n < 3.5$ versus
$n<2.5$), and effective wavelength ($r$ vs $z$ band) of the observations.  Although the measured discrepancy among radii is less pronounced for low Sersic indices similar to those of
our galaxy sample (Guo et al. 2009), these varied effects may considerably complicate  interpretations of the evolution of the high-redshift mass radius relation to the present day.




\section{QUANTIFYING MERGERS IN THE STAR FORMING GALAXY SAMPLE}
\label{mergers.sec}

While the irregular and clumpy morphologies of galaxies at $z>1.5$ may be interpreted as arising from dynamical instabilities within gas-rich systems
(e.g., Bournaud et al. 2008; Dekel et al. 2009b; Genzel et al. 2011), they have also commonly been taken as
indicators of ongoing mergers by numerous authors 
(e.g., Conselice et al. 2011b; Lotz et al. 2008a; and references therein).
In this section, we discuss the properties of galaxies that can be identified as mergers via three common morphological criteria
(the quantitative statistics $G-M_{20}$ and $A$, and the observed fraction of close pairs)
 and assess how their relative abundance
evolves throughout the redshift range $z\sim 1.5-3$.
Additionally, we discuss the association of putative mergers with physical quantities such as stellar mass, SFR, and gas-phase kinematics,
finding (similar to Law et al. 2007b) that whether or not a galaxy {\it looks} like a merger makes little difference to many of
its physical properties.

Since our $H_{160}<24.0$ apparent magnitude cut (adopted to ensure robustness of the morphological statistics)
introduces a redshift-dependent bias in the absolute magnitudes of our galaxies, all numerical values for the merger fraction (and/or merger rate) 
are calculated for a
mass-limited subsample of galaxies with 
$M_{\ast} > 10^{10} M_{\odot}$ for which  $>90\%$ of galaxies at all redshifts also fulfill the $H_{160}<24.0$ criterion.


\subsection{Defining the Mergers}

\subsubsection{Quantitative Morphologies}

\begin{figure*}
\plotone{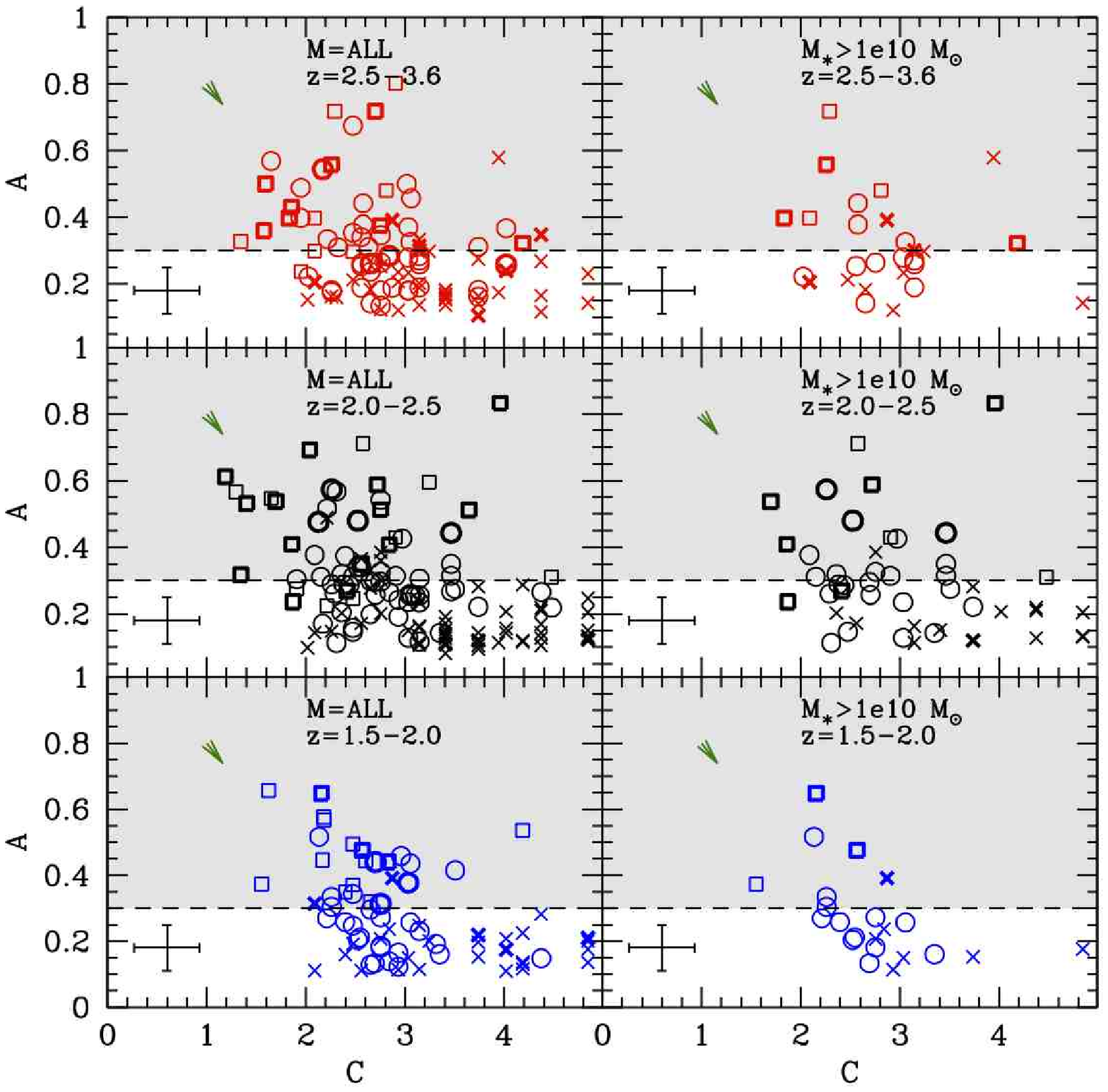}
\caption{Concentration ($C$) vs asymmetry ($A$) for all target galaxies with $H_{160}<24.0$ (left panels) and for the subset with
stellar masses $M_{\ast} > 10^{10} M_{\odot}$ (right panels).  Morphological statistics are given with respect to the Lotz et al. (2004, 2006) reference frame using the transformation
equations given in \S \ref{segmapsA.sec}.  
Point colors and types are as given in Figure \ref{mrfig.fig}, the error bar in the lower left corner of each panel indicates the typical uncertainty in individual points based on Monte Carlo simulations.
The green arrow indicates the approximate vector that the points would move along if converted
to the Conselice et al. (2000, 2008) reference frame.  
The shaded region above the dashed line indicates the merger regime.  Systems that are indicated to be mergers according to the $G-M_{20}$ criterion (Figure \ref{gm20.fig}) are 
highlighted in bold type.}
\label{caplot.fig}
\end{figure*}

One common way of identifying mergers is to use their morphological asymmetry $A$, as discussed extensively in the literature by (e.g.) Conselice et al. (2000, 2003, 2008, 2009),
Lotz et al. (2008b, 2010ab), Papovich et al. (2005), and Scarlata et al. (2007).
In Figure \ref{caplot.fig} we plot $C$ versus $A$ for our magnitude-limited sample of galaxies ($H_{160} < 24.0$, left panels) and for a mass-limited subsample
($M_{\ast} >10^{10} M_{\odot}$, right panels).
At low redshifts ongoing mergers have typically been identified by the criterion $A>0.35$ (e.g., Conselice et al. 2003), although
for less well resolved, lower surface-brightness galaxies similar to those of our sample Lotz et al. (2008b) find that $A>0.30$ is more appropriate.
Adopting the $A>0.30$ criterion, we find that the merger fraction (for $M_{\ast} > 10^{10} M_{\odot}$) is 
$0.32\pm0.06, 0.43\pm0.04, 0.41\pm0.07$ in the $1.5 \leq z < 2.0$, $2.0 \leq z < 2.5$, and $2.5 \leq z < 3.6$ samples respectively.\footnote{Uncertainties 
are estimated by a Monte Carlo technique randomizing the individual values of $A$ based on a Gaussian probability distribution about the measured values.  The
$1\sigma$ width of this distribution combines the uncertainty in the measured value of $A$ and the  scatter about the mean relation in our transformation to the
Lotz et al. (2006) reference frame.}

\begin{figure*}
\plotone{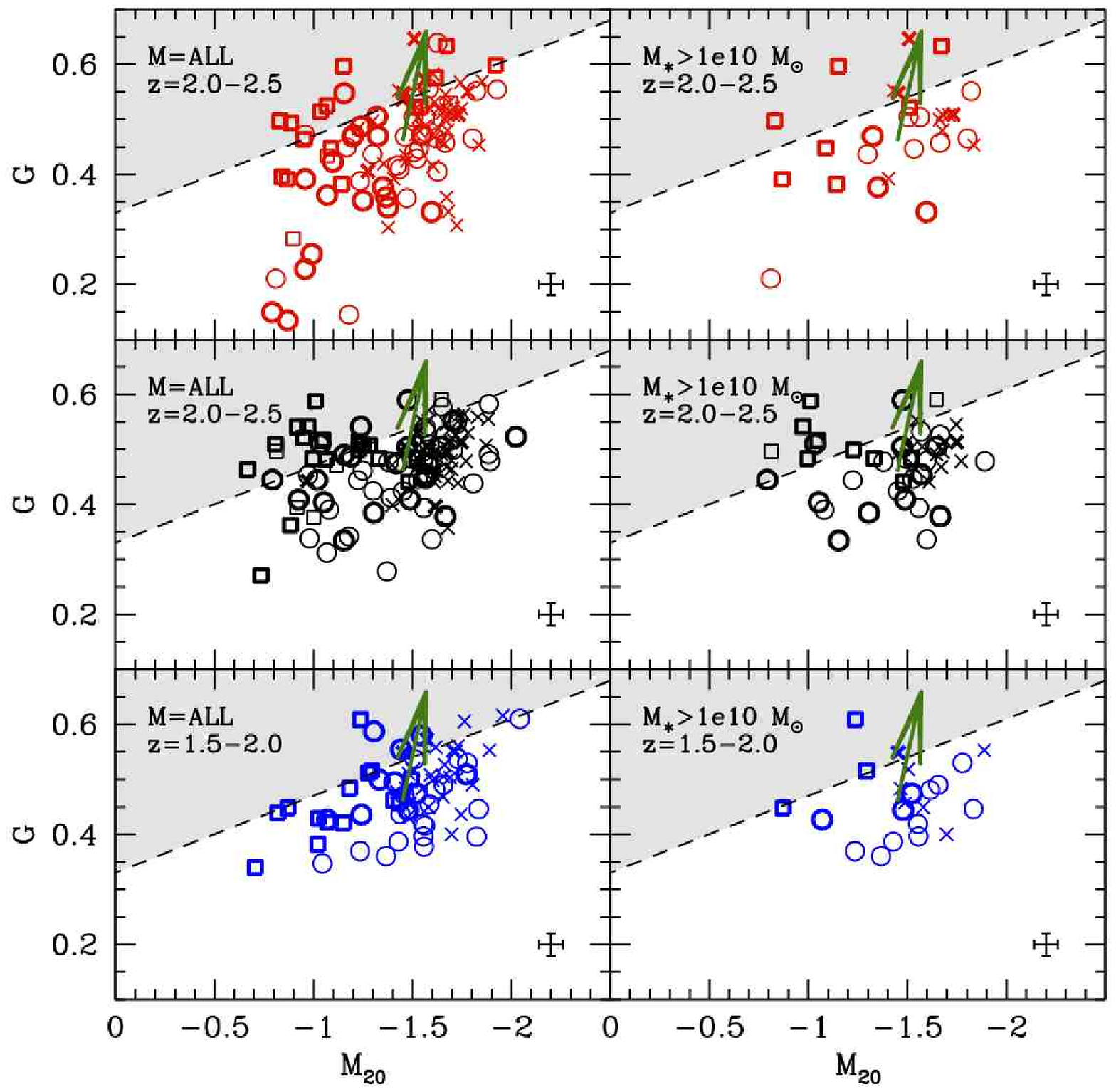}
\caption{Gini vs $M_{20}$ for all target galaxies with $H_{160}<24.0$ (left panels) and for the subset with
stellar masses $M_{\ast} > 10^{10} M_{\odot}$ (right panels).  
Morphological statistics are given with respect to the Lotz et al. (2004, 2006) reference frame using the transformation
equations given in \S \ref{segmapsA.sec}.  
Point colors and types are as given in Figure \ref{mrfig.fig}, the error bar in the lower right corner of each panel indicates the typical uncertainty in individual points based on Monte Carlo simulations.
The green arrow indicates the approximate vector that the points would move along if converted
to the Conselice et al. (2000, 2008) reference frame.  
The shaded region above the dashed line indicates the merger regime.  Systems that are indicated to be mergers according to the asymmetry criterion (Figure \ref{caplot.fig}) are 
highlighted in bold type.}
\label{gm20.fig}
\end{figure*}

Another common method of identifying mergers is by their location in $G-M_{20}$ space, as originally defined by Lotz et al. (2004, 2006).
In Figure \ref{gm20.fig} we plot $G$ versus $M_{20}$ for our magnitude-limited sample of galaxies ($H_{160} < 24.0$, left panels) and for a mass-limited subsample
($M_{\ast} >10^{10} M_{\odot}$, right panels).
The merger criterion defined by Lotz et al. (2008b) for high-redshift galaxies\footnote{There is no merger criterion tailored specifically to our galaxy sample and angular resolution of WFC3/IR; we
adopt the Lotz et al. (2008b) definition as an approximation given its popularity in the literature.}
\begin{equation}
G > -0.14 M_{20} + 0.33
\end{equation}
gives a merger fraction of $0.14\pm0.04, 0.23\pm0.03,0.24\pm0.05$ at
$1.5 \leq z < 2.0$, $2.0 \leq z < 2.5$, and $2.5 \leq z < 3.6$ respectively.

Figures \ref{caplot.fig} and \ref{gm20.fig} demonstrate the necessity for caution when estimating the merger fraction using different segmentation maps: If we had calculated the 
morphological statistics using a segmentation map modeled on the methods of Conselice et al. (2009), typical points in these figures would be offset in the direction indicated by the green arrows.
While the effect in the $C-A$ plane is fairly minimal, values of $G$ can change drastically, pushing a large number of points over the merger/non-merger dividing line
and resulting in a wildly different derived merger fraction if the
merger/non-merger division is not made appropriately.
As discussed by Lisker (2008), this offset may in large part  account  for the discrepancy in the number of mergers identified in similar observational samples at $z\sim1$ using the $G-M_{20}$ technique by
Lotz et al. (2008a; see their Figure 10) and Conselice et al. (2008; see their Figure 8).

\subsubsection{Nearby pairs}
\label{pairs.sec}

Another method of identifying mergers is to count the number of systems with close physical pairs.  In practice, we consider systems with multiple distinct clumps 
of comparable $H_{160}$ flux ($\sim$ 3:1 - 1:1)
in their light profiles, colors consistent with the rest-UV selection criteria, and well-defined separations in the range 5 $<r<$ 16 kpc
 (i.e., are classified as Type II galaxies) as physical pair candidates.\footnote{Of course, not all pairs at redshifts $z=1.5-3.6$ will be in our spectroscopic sample, 
 but we do not expect this to bias the derived pair {\it fraction} because the spectroscopic targets
 were chosen independently of whether or not they appeared to be in angular pairs.}
For galaxies with $H_{160} < 24.0$ and $M_{\ast} > 10^{10} M_{\odot}$ we find that the fraction of pairs is
 $0.14^{+ 0.10}_{- 0.05}$,  $0.23^{+0.07}_{-0.06}$, and $0.24^{+0.12}_{-0.08}$
at $1.5 \leq z < 2.0$, $2.0 \leq z < 2.5$, and $2.5 \leq z < 3.6$ respectively.\footnote{Uncertainties are estimated using Bayesian binomial confidence intervals (see discussion by Cameron 2011).}
Some fraction of these candidates will not be physical pairs however, but simply
projected angular pairs of galaxies with different redshifts and no physical association.

\begin{figure}
\plotone{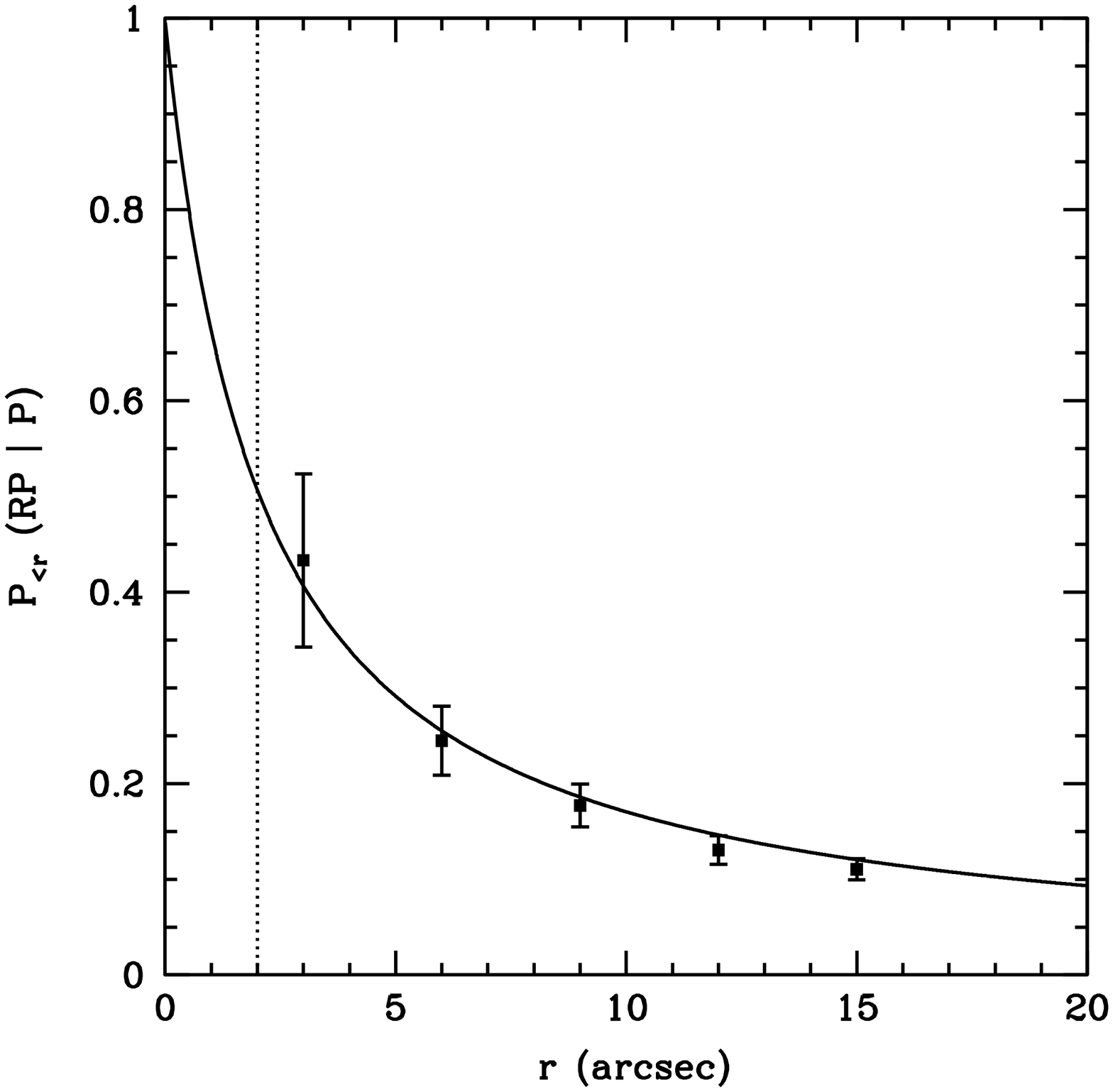}
\caption{Probability $P_{<r}$ (RP $|$ P) that a BM/BX/LBG galaxy with spectroscopic redshift $z_2$ observed within radius $r$ of another
BM/BX/LBG galaxy with spectroscopic redshift $z_1$ is a real physical pair with $|z_2 - z_1| < 0.01$.  Error bars represent Poissonian uncertainty based on the number
of galaxies observed in each bin out of a total spectroscopic sample of 2874 galaxies in the redshift range $1.5 < z < 3.5$.  The solid black line represents a numerical fit 
based on the observed
number of physical/apparent pairs as a function of radius.  The vertical dotted line indicates the maximum radius of pairs identified by the WFC3 morphological sample.}
\label{pairs_larger.fig}
\end{figure}

One effort to constrain the incidence rate of false pairs can be made by extrapolating the false pair fraction observed at larger distances for which spectroscopic redshifts
can be obtained for individual objects.  Considering the 2874 galaxies (across 19 different fields) in our catalog with spectroscopic redshifts in the range $1.5 < z < 3.5$,
we count the number of distinct angular pairs as a function of separation in comparison to the number of genuine physical pairs whose spectroscopic redshifts lie within
$\Delta z = 0.01$ of each other.  As illustrated by Figure \ref{pairs_larger.fig}, extrapolation of this relation to the radii probed by our WFC3 data suggests that $\sim50$\% of
our observed pairs 
should correspond to genuine physical pairs.

Alternatively, we can also estimate the false pair fraction based on the statistical distribution of objects 
in the WFC3 imaging fields.  Using our Source Extractor catalogs, we evaluate the number of unique pairs with primary magnitudes in the range $H_{160} = 22.0 -24.0$
and secondary magnitudes within 1 magnitude of the primary as a function of their separation radius.  Assuming that the majority of such pairs in the WFC3 fields
are false pairs, we estimate that $7\pm1$\% of galaxies have false pairs
within $r < 16$ kpc.
Subtracting this 0.07 false pair fraction from  the angular pair fraction calculated above, we obtain the true physical pair fractions
 $0.07^{+ 0.10}_{- 0.05}$,  $0.16^{+0.07}_{-0.06}$, and $0.17^{+0.12}_{-0.08}$
at $1.5 \leq z < 2.0$, $2.0 \leq z < 2.5$, and $2.5 \leq z < 3.6$ respectively for separations in the range 5 kpc $< r < $ 16 kpc.
We note that these values are consistent to with observational uncertainty with what would be derived had we simply assumed that 50\% of angular pairs were false pairs.

\subsection{Evolution with Redshift}

\begin{figure*}
\plotone{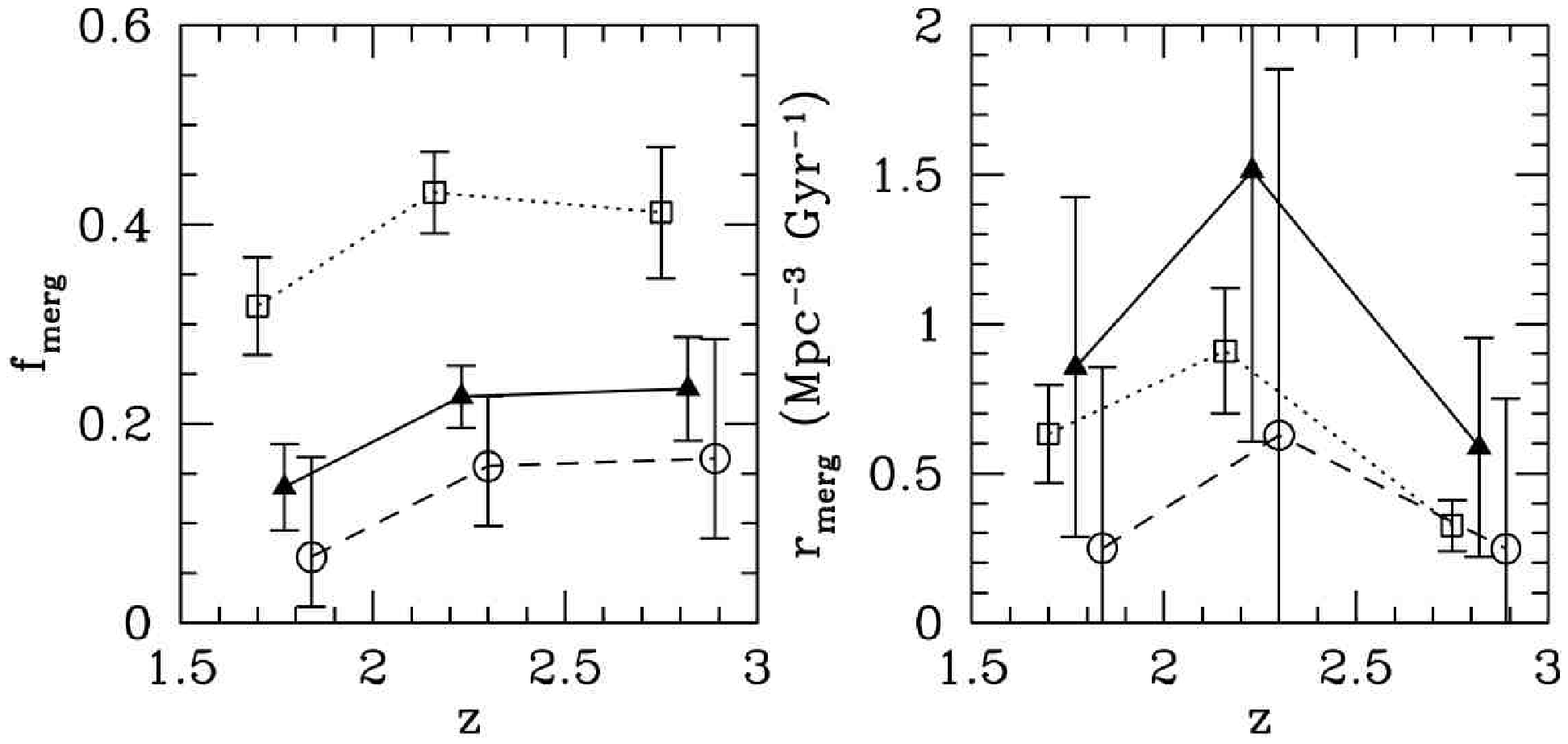}
\caption{Evolution of the merger fraction (left-hand panel) and merger rate (right-hand panel) with redshift for the star forming galaxy sample with $M_{\ast} > 10^{10} M_{\odot}$.
Filled triangles represent mergers identified according to the $G-M_{20}$ criterion, open boxes according to the $A> 0.30$ criterion, and open circles according to the morphological
pair within $5 < r < 16$ kpc criterion.}
\label{mergerrate.fig}
\end{figure*}

As detailed above, estimates of the merger fraction $f_{\rm merg}$ derived from all three methods are roughly constant across our three redshift ranges, albeit with mild evidence 
(at the $\sim 1-2\sigma$ level) for a decline in the merger fraction at $z<2$ (see Figure \ref{mergerrate.fig}, left-hand panel).
In order to construct the merger {\it rate} from the merger {\it fraction} it is necessary to combine the merger fractions with the estimated timescale $T$ for visibility and the comoving
space density $n(z)$ of the target sample (see, e.g., Lotz et al. 2008a):
\begin{equation}
N_{\rm merg} = n(z) \, f_{\rm merg} / T
\end{equation}

Estimating the comoving space densities by integrating the mass functions for the star forming galaxy sample given by Reddy \& Steidel (2009) above 
$M_{\ast} = 10^{10} M_{\odot}$, and 
adopting $T_{GM20} = 0.24 \pm 0.14$ Gyr, $T_A = 0.76\pm0.16$ Gyr, and $T_{\rm pair} = 0.20 \pm0.38$ Gyr (see discussion in \S \ref{mergerprop.sec}),
we obtain estimates of the merger rate as shown in Figure \ref{mergerrate.fig} (right-hand panel).
Clearly the actual merger rate of our galaxies is highly uncertain, and for the small number of galaxies observed in the present sample it is not possible
to comment meaningfully on the evolution of the merger fraction with redshift (although our results are consistent with those derived for similar populations of galaxies in other studies;
see, e.g., Conselice et al. 2011a and references therein).  Even for significantly larger galaxy samples (e.g., Faber et al. 2011) it may prove difficult to constrain the merger rate
given the large uncertainty in  observability timescales that require  numerical simulations to constrain.

\subsection{Physical Properties of the Mergers}
\label{mergerprop.sec}

It is not obvious whether it is meaningful from a physical sense to identify galaxies as mergers on the basis of their rest-frame optical morphology.
As argued by some authors  (e.g., Bournaud et al. 2008; Genzel et al. 2011) irregular morphologies may instead
arise from dynamical instabilities within gas-rich systems.  Additionally, as we demonstrated in Law et al. (2007b) and expand upon below, merger-like morphologies are poorly
correlated with other physical observables.

There are significant differences between the subsamples of $M_{\ast} > 10^{10} M_{\odot}$ galaxies from our survey  selected as mergers at $z=2.0-2.5$ according to different criteria.
$43\pm4$\% of such galaxies are identified as mergers on the basis of their morphological asymmetry, while $23\pm3$\% are identified using the $G-M_{20}$ selection criterion,
and $16^{+7}_{-6}$\% using pair statistics.
As illustrated in Figures \ref{caplot.fig} and \ref{gm20.fig}, 76\% of galaxies selected as mergers according to $G-M_{20}$ are also selected as mergers using $A>0.30$,
but only 39\% of galaxies selected as mergers using $A>0.30$ are also selected as mergers according to $G-M_{20}$.
Similarly, 59\% (45\%) of mergers identified by $G-M_{20}$ (asymmetry)  are also identified as mergers based on the presence of a nearby angular pair.
Clearly, while there is a significant overlap between the galaxy samples, there are also a significant number of galaxies uniquely selected by each technique.

This difference is unsurprising given that the various morphological selection criteria may isolate mergers with different mass ratios and in a different range of evolutionary phases.
Lotz et al. (2008b, 2010ab) performed a series of hydrodynamic simulations to explore the timescales and visibility of disk galaxy mergers
as a function of morphological selection criterion, mass ratio, and gas content.
Dividing their mergers into six stages (pre-merger, first-pass, maximal separation, final merger, post merger, remnant) these authors found
that the observability of mergers at rest-frame 4686 \AA\ can vary dramatically from stage to stage.  For the `G3gf1' model,\footnote{Stellar and gas masses
$\sim 2 \times 10^{10} M_{\odot}$.} Lotz et al. (2010b) find that $G-M_{20}$ and pair criteria tend to have observability timescales
$T_{G M20} = 0.24\pm0.14$ Gyr and  $T_{\rm pair} = 0.20 \pm0.38$ Gyr (predominantly identifying first-passage mergers)
while the $A>0.30$ criterion has a longer observability timescale $T_{A} = 0.76 \pm0.16$ (identifying both first-passage and final mergers; see also Conselice et al. 2006).
The greater fraction of galaxies that we identify as mergers based on  their asymmetry than by the other two methods (Figure \ref{mergerrate.fig}) may therefore simply reflect
this large difference in observability timescales.
Further, Lotz et al. (2010a) find that while $A$ is most sensitive to major mergers like those identified using our pair selection criteria ($\sim 3:1 - 1:1 H_{160}$ flux ratio),
 $G-M_{20}$ detects both major and minor mergers, potentially explaining why we identify more mergers using $G-M_{20}$ than with pair selection.

In Figure \ref{MphHist1.fig} we plot histograms of various physical properties for galaxies classified as mergers/non-mergers according to the $G-M_{20}$, $A$, and pair criteria
and use a KS test to evaluate the significance of the null hypothesis that both sets of galaxies (mergers and non-mergers) were drawn from
the same distribution.  We conclude that for almost all physical parameters (stellar mass, SFR, rest-frame $U-B$ color\footnote{Estimated from the best-fit SED.}, etc.)
there is no significant difference (confidence in the null hypothesis $>5$\%) between putative mergers and non-mergers.
Similarly, there is no obvious difference in the gas-phase kinematics between mergers and non-mergers, although our sample size of 35 galaxies with systemic H$\alpha$
redshifts and high-quality UV spectra is too small to conclusively rule out association.
The one notable exception is that galaxies
identified as mergers via the $G-M_{20}$ or pair classification schemes have significantly smaller radii and
correspondingly higher $\Sigma_{\rm SFR}$ than non-mergers.
This may suggest either that $\Sigma_{\rm SFR}$ peaks around the first-passage during a major merger event, or that the
 $G-M_{20}$ and pair classification schemes are simply effective at finding galaxies with small radii.

The lack of correlation observed between morphology and these physical observables may be unsurprising in light of both numerical uncertainties in
our morphologies (i.e., exactly where the dividing line between mergers and non-mergers lies) and
expectations (e.g., Lotz et al. 2010ab) that 
star formation may typically peak after the major morphological disturbances have subsided.
Regardless, it is unclear whether it is physically meaningful to classify $z\sim 2-3$ galaxies as mergers
on the basis of morphology alone given that there appears to be little to distinguish these
systems (whether observed in the rest-optical or the rest-UV; see discussion by Law et al. 2007b, see also Swinbank et al. 2010 for a similar
discussion of submillimeter galaxies) from their non-merging counterparts.
Rather, it may simply be that {\it most}  $z\sim 2-3$ star forming galaxies are dynamically unstable systems driven by the accretion of large
quantities of gas, whether this gas is acquired through mergers, cold-mode, or hot-mode accretion processes.

\begin{figure*}
\plotone{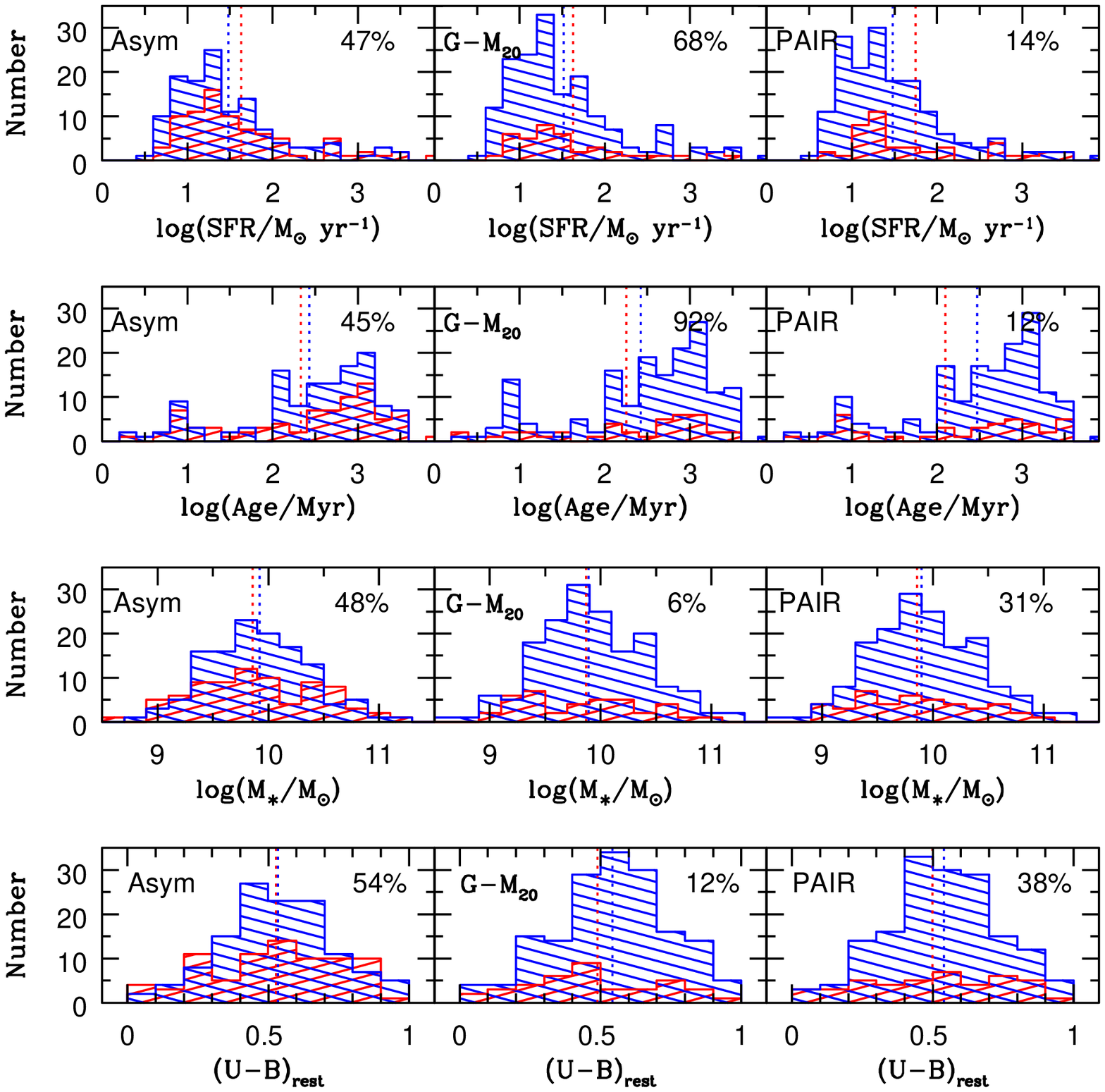}
\caption{Physical properties of mergers vs non-mergers.  Red/blue histograms represent mergers/non-mergers selected according to the asymmetry (left panel), $G-M_{20}$ (middle panel), 
and nearby pair (right panel) criteria.  Vertical dotted lines represent the mean value in each case, the percentage given in the upper right corner of each panel indicates the significance of the
null hypothesis that the merger/non-merger galaxies are drawn from the same parent distribution.}
\label{MphHist1.fig}
\end{figure*}

\addtocounter{figure}{-1}
\begin{figure*}
\plotone{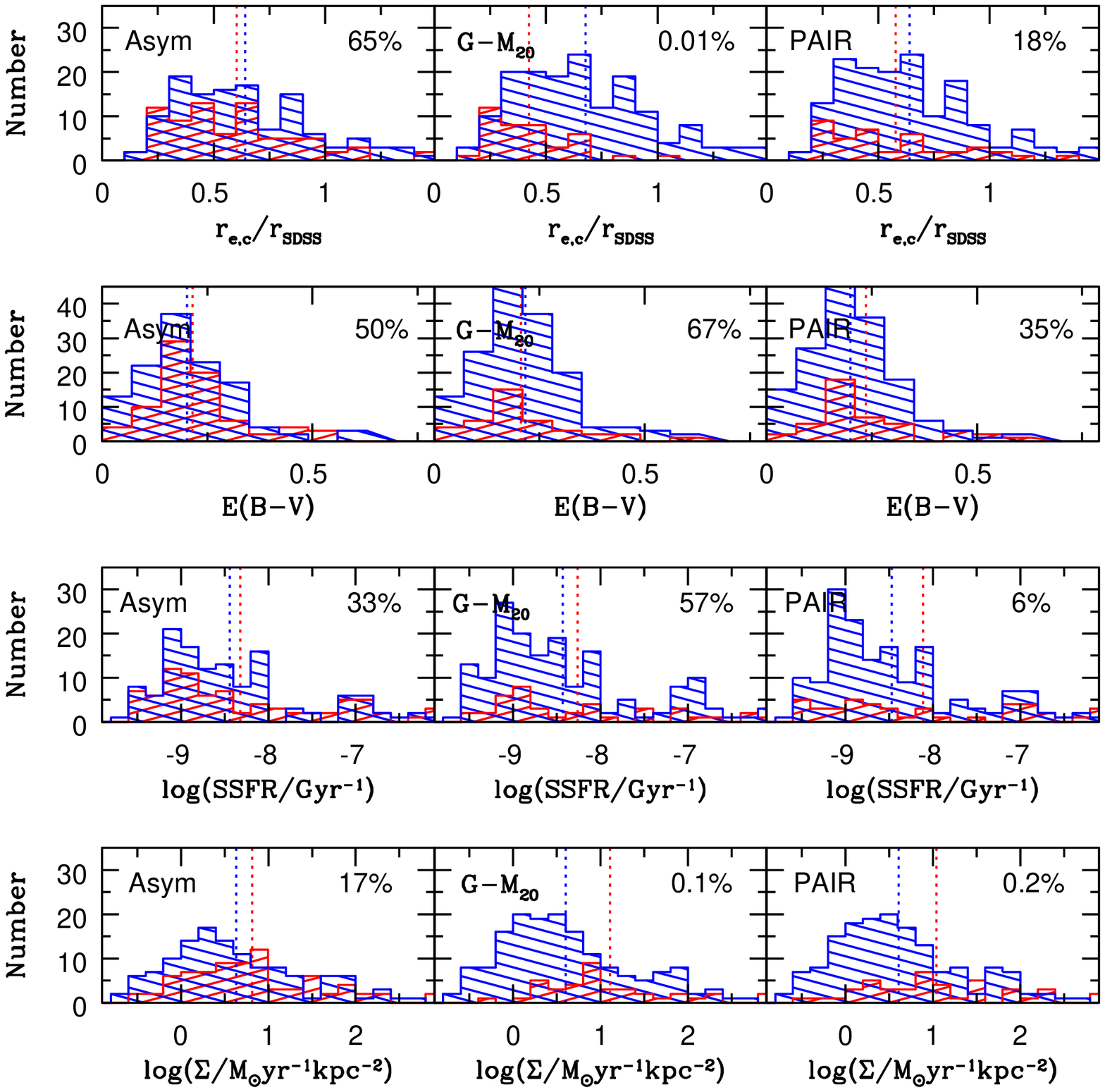}
\caption{{\it Continued}}
\end{figure*}


\section{SUMMARY}
\label{discussion.sec}

We have presented rest-optical morphologies for a sample of 306 spectroscopically confirmed  $z = 1.5-3.6$ star forming galaxies 
with stellar masses in the range $M_{\ast} = 10^9 - 10^{11} M_{\odot}$.
Since these galaxies were distributed among 10 different fields widely separated on the sky the effects of 
sample variance are expected to be greatly reduced
compared to surveys over contiguous regions of similar total area.
We summarize our principle scientific conclusions as follows:

\begin{enumerate}

\item Typical $z \sim 1.5-3.6$ star forming galaxies have circularized effective radii $r_e \approx 0.7-3$ kpc and a projected $n\sim1$ exponential surface
brightness profile that extends out to $>6 r_e$ in stacked galaxy images.
The observed sizes are consistent with 
previous observational estimates (e.g., Buitrago et al. 2008; Kriek et al. 2009) for high-mass galaxy populations and with
numerical simulations (e.g., Sales et al. 2010) that assume strong stellar feedback.

\item A stellar mass - radius relation for star forming galaxies 
is observed to exist as early as $z\sim3$; at fixed mass typical sizes evolve with redshift as $\sim (1+z)^{-1.07 \pm0.28}$ in the interval $z \sim 3$ to $z\sim 1.5$.  
These galaxies must grow at least as fast as $r \sim M_{\ast}$ in order  to evolve onto the local late-type galaxy relation
by the present day.

\item The distribution of axis ratios $b/a$ is strongly 
inconsistent with a population of axisymmetric thick exponential disks
and more consistent with a population of triaxial ellipsoids
with intrinsic minor/major and intermediate/major axis ratios $0.3\pm0.2$ and $0.7\pm0.1$ respectively.
The typical ellipticity is qualitatively similar to that previously found by Ravindranath et al. (2006), but there may be mild evidence
for evolution with wavelength.
The ellipsoidal nature of these galaxies 
indicates at minimum that the distribution of stellar mass within them is markedly asymmetric, and (in combination with their high
gas fractions and velocity dispersions) 
may further suggest that they are not in stable dynamical equilibrium with short-lived gas disks (e.g., Ceverino et al. 2010) continually forming and re-forming from recently accreted gas
until stabilized (e.g., Martig et al. 2010) by a sufficiently massive triaxial stellar component.

\item Consistent with previous studies (e.g., Dickinson 2000; Papovich et al. 2005),
rest-optical ($\lambda \sim 4000-5000$\AA) and rest-UV ($\lambda \sim 2000-3000$ \AA) morphology for $z\sim2$ star forming galaxies is generally similar
with typical color dispersion $\xi \sim 0.02$ (although rest-UV radii are larger by $21\pm2$\% on average), while
high mass ($M_{\ast} > 3 \times 10^{10} M_{\odot}$) galaxies tend to exhibit greater morphological differences with $\xi$ as large as $0.28$ (although c.f. Bond et al. 2011).
The most massive galaxies in our sample are typically bright and well nucleated at rest-optical wavelengths but faint and diffuse in the rest-UV.

\item Finally, we demonstrate that while the nonparametric morphological statistics $G$, $M_{20}$, $C$, $A$, and $\Psi$ calculated using different segmentation maps commonly
adopted in the literature are strongly correlated with each other, there can be systematic offsets that are important to account for when comparing values between samples
or estimating merger fractions.  Merger fractions estimated according to the $G-M_{20}$, $A$, or pair criteria are consistent with recent determinations in the literature
(e.g., Conselice et al. 2011a), with evidence
for at most mild evolution with redshift.  There is moderate overlap between galaxies selected as mergers with these three criteria, but in general mergers and non-mergers
have statistically indistinguishable distributions of measured and inferred properties (SFR, stellar mass, etc.), with the exception that mergers selected by the pair
and $G-M_{20}$ statistics have smaller effective radii and correspondingly larger $\Sigma_{\rm SFR}$.
We suggest that  most  $z\sim 2-3$ star forming galaxies may be dynamically unstable systems driven by the accretion of large
quantities of gas, whether this gas is acquired through mergers, cold-mode, or hot-mode accretion processes.

\end{enumerate}

In general, our observations are consistent with inside-out growth of star forming galaxies in the young universe.  We suggest that mass growth 
proceeds according to the following qualitative picture:
`Typical' $z\sim2$ star forming galaxies appear to be gas-rich,
compact, triaxial systems systems that are dominated by velocity dispersion between individual star forming regions rather than systemic rotation, and whose high $\Sigma_{\rm SFR}$
drives strong outflows into the surrounding IGM.  As these galaxies mature they gain stellar mass, stabilizing the formation of extended (albeit still thick) gaseous disks
in which rotational support plays an increasing role.  As the star formation migrates from central regions 
into these extended disks 
the $\Sigma_{\rm SFR}$ drops, and the disk component superimposes
a zero-velocity component atop the outflowing absorption line gas (Law et al. 2011, in preparation).


\acknowledgements

DRL, CCS, and SRN have been supported by grant GO-11694 from the Space Telescope Science Institute.
Support for DRL and NAR was also provided by NASA through Hubble Fellowship grant \# HF-51244.01
and HF-01223.01
awarded by the Space Telescope Science Institute, which is operated by the Association of Universities for Research in Astronomy, Inc., for NASA, under contract NAS 5-26555.
CCS has been supported by the US National Science Foundation through grants AST-0606912 and AST-0908805
AES acknowledges support from the David and Lucile Packard Foundation.
DRL appreciates productive conversations with  A. Dutton and E. Bell, and thanks the referee (J. Lotz) for insightful comments that improved the final version of this manuscript.
Finally, we extend thanks to those of Hawaiian ancestry on whose sacred mountain we are privileged to be guests.


\appendix
\section{A.  ROBUSTNESS OF THE MORPHOLOGICAL PARAMETERS}
\label{robust.sec}

\subsection{A.1 Choice of Segmentation Map}
\label{segmapsA.sec}

Three methods have been generally adopted in the literature for defining robust segmentation maps:

\begin{enumerate}
\item Law et al. (2007b) and Peter et al. (2007) used a scaled surface brightness method to select galaxy pixels whose flux is at least
$n\sigma$, where $\sigma$ is the standard deviation of the sky pixels, and $n$ scales with source redshift as $n = 3 \left(\frac{1+z}{1+z_{\rm max}}\right)^{-3}$,
where $z_{\rm max} =3.0$.\footnote{Although the maximum redshift of our sample is $z=3.6$, we adopt $z_{\rm max} = 3.0$ for consistency with Law et al. 2007b.  This corresponds
to a selection threshold of $2\sigma$ at $z=3.6$, and $12\sigma$ at $z=1.5$.}  This method 
is independent of galaxy morphology and compensates
for cosmological surface brightness dimming (which scales as $(1+z)^{-3}$ for a fixed observational bandpass), therefore giving consistent results across a given redshift interval.
However, it is explicitly tied to the noise characteristics of the observational data, and can yield morphological parameters
that vary systematically with total flux for galaxies with identical morphological profiles but different total luminosities (see, e.g., Figure 9 of Law et al. 2007b, Lotz et al. 2008).

\item Many authors (e.g., Conselice et al. 2000, 2008; Lotz et al. 2004, 2006) start with simple Source Extractor segmentation maps to either pre-select galaxy 
pixels or mask foreground/background objects, and apply either a circular or elliptical Petrosian (1976) selection technique
to select pixels independent of total galaxy flux or background noise characteristics.
Conselice et al. (2000, 2008) include in their segmentation map all pixels within 1.5 Petrosian radii ($r_{\rm P}$, i.e., the radius at which the surface brightness is some fraction $\eta$
of the enclosed surface brightness), while Lotz et al. (2004, 2006) include only pixels with flux greater than the surface brightness at the Petrosian radius (but following the potentially
irregular isophotal contours).
While robust to total source magnitude, cosmological dimming, and observational noise characteristics, these method can sometimes yield suboptimal results
when applied to the often-irregular morphologies of $z \sim 2$ galaxies (e.g., Figure \ref{postage.fig}) because of their ill-defined Petrosian radii.

\item Abraham et al. (2007) generalized the Petrosian pixel selection method to work equally well for galaxies of arbitrary shapes whose flux components
are not necessarily contiguous.  As outlined by Abraham et al. (2007), all pixels in the preliminary segmentation map calculated using Source Extractor are sorted in decreasing
order of flux into the array $f_i$, which is then used to construct the cumulative flux array $F_i = \sum_{j=1}^{i} f_j$.  The quasi-Petrosian isophote is
set by determining the pixel index $i$ at which $f_i = \eta (F_i/i)$ where $F_i/i$ is the cumulative mean surface brightness.  This quasi-Petrosian segmentation map
preserves the advantages
of Petrosian-based methods (i.e., robustness to source magnitude, cosmological dimming, and observational noise) while being applicable to arbitrary morphology.

\end{enumerate}

We adopt the quasi-Petrosian method of Abraham et al. (2007) as our baseline segmentation method.
As described by these authors, the failure mode of this approach is graceful in that, if
the isophotal Petrosian threshold  $\eta$ is below the surface-brightness threshold of the initial 
Source Extractor segmentation map, it simply defaults to the initial map (\S \ref{initialmap.sec}).  It is not desirable for
this to occur frequently however, since it eliminates the advantages of the Petrosian pixel selection.  The isophotal threshold $\eta$ must therefore be set sufficiently high that it is more
restrictive than a simple $1.5\sigma$  surface brightness cut, but sufficiently low that it rejects as little information (i.e., pixels) as possible from the final segmentation maps.

\begin{figure}
\plotone{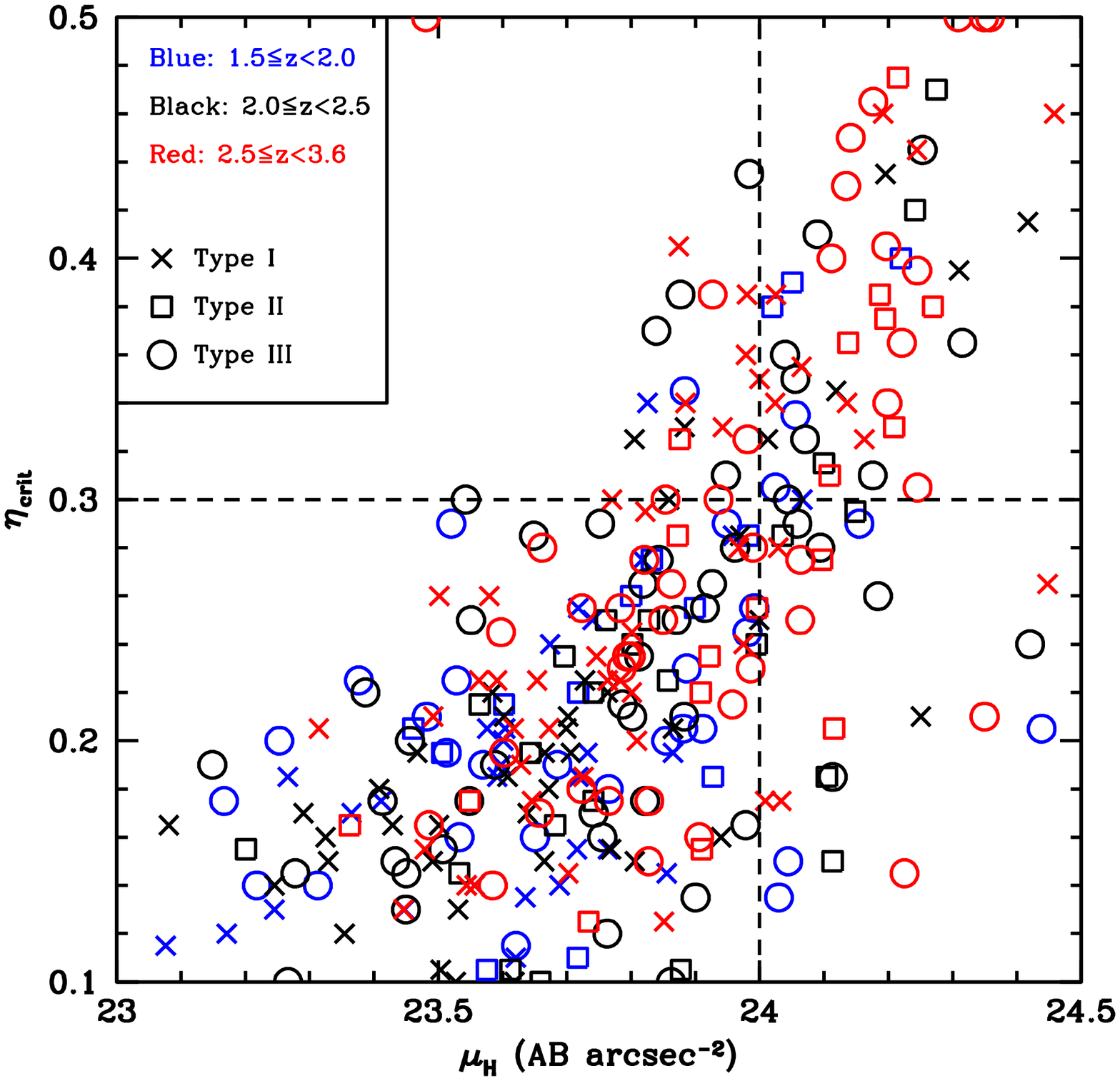}
\caption{Critical value $\eta_{\rm crit}$ for the pseudo-Petrosian pixel selection method to produce a pixel map that is more restrictive than the initial surface-brightness
bounded segmentation map for the 306 star forming galaxies in our survey as a function of average surface brightness $\mu_H$.
Higher values of $\eta_{\rm crit}$ correspond to galaxies with less dynamic range in surface brightness above the noise floor of the images.
Symbols correspond to different visual morphological types (\S \ref{vclass.sec}) and redshifts as given in the legend.}
\label{eta.fig}
\end{figure}

In Figure \ref{eta.fig}, we plot the critical surface brightness threshold $\eta_{\rm crit}$ at which the pseudo-Petrosian algorithm produces a segmentation map that is more restrictive
than the initial Source Extractor $1.5\sigma$ surface-brightness segmentation map for each of our 306 galaxies.
Intuitively, there is a strong correlation with mean apparent surface brightness $\mu_H$ (defined as the $H_{160}$ magnitude divided by the Source Extractor segmentation map area).
We note that for the lowest mean surface brightness objects (disproportionately galaxies of Type III and/or at redshifts $z>2.5$) $\eta_{\rm crit}$ is relatively high; that is, the galaxy surface brightness decreases only slightly to $\sim 40$\% of its mean value
before reaching the  $1.5\sigma$ sky background.  In contrast, for higher surface brightness objects the dynamic range of the galaxy is greater and can decrease to $\sim 20$\% or less of its mean
value before reaching the sky background.  Given these results, a traditional choice of 
$\eta = 0.2$ would result in an unsatisfactorily high $\sim 60$\% of our galaxies defaulting to a simple $1.5\sigma$ isophotal pixel selection.
We therefore take $\eta = 0.3$ instead, for which only $\sim 23$\% of galaxies default to the surface-brightness limited segmentation map.
This fraction decreases to $\sim 15$\% when we reject from consideration galaxies with $H_{160} > 24.0$, for which we find that quantitative morphological statistics are not robust regardless
of segmentation map (see \S \ref{magrobust.sec}).

\begin{figure}
\epsscale{1.0}
\plotone{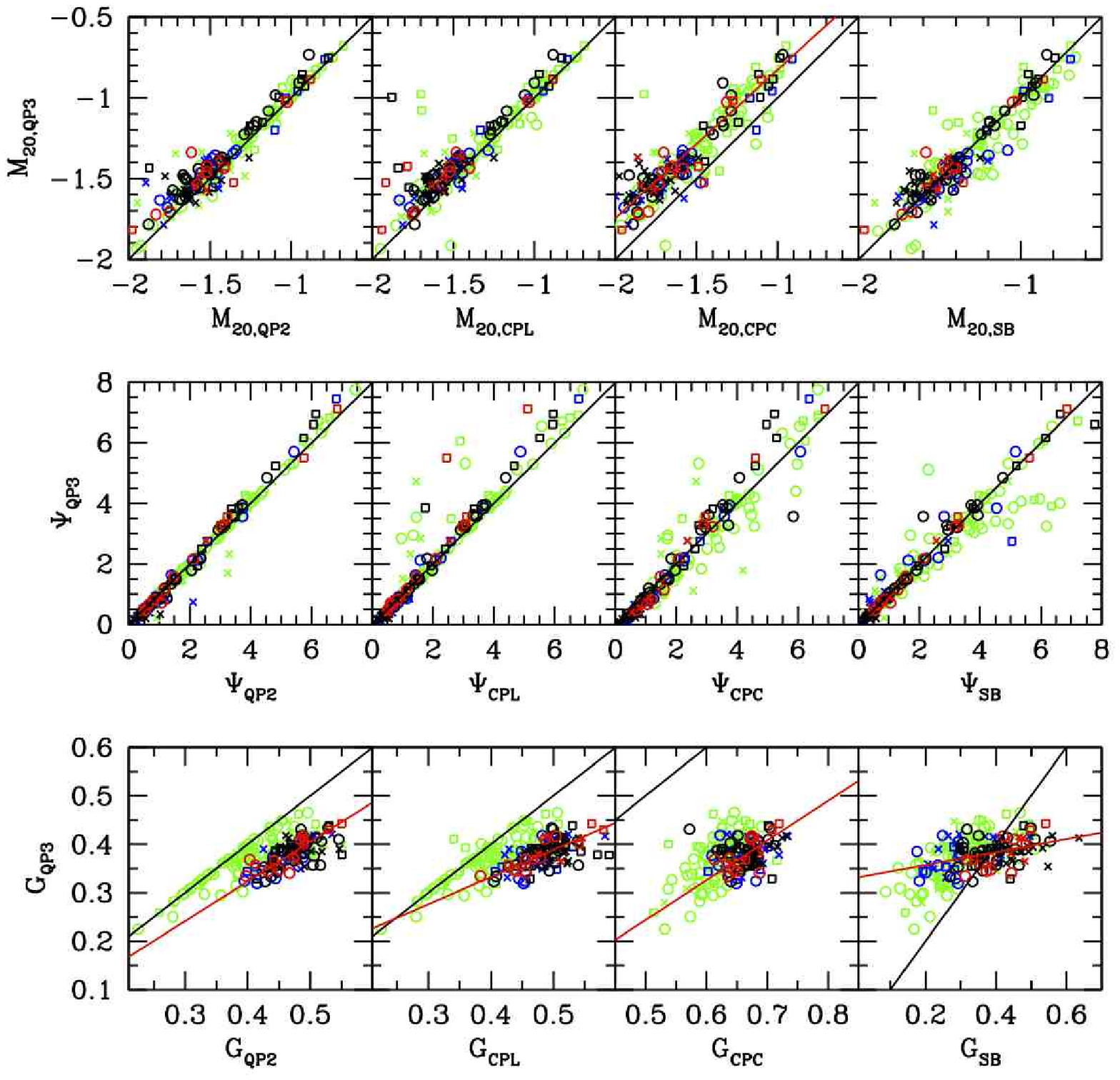}
\caption{Figure comparing $G$, $\Psi$, and $M_{20}$ morphologies computed using the `QP3', `QP2', `CPL', `CPC', and `SB' segmentation map techniques.
Only galaxies with $H_{160} < 24.0$ are shown since the morphological statistics are less reliable at fainter magnitudes.
Symbol color and types are as given in Figure \ref{eta.fig}, except that green colored points represent galaxies with $\eta_{\rm crit} > 0.2$ for which the 
$\eta=0.2$ isophote is ill-defined.
Black lines indicate 1-1 correspondence, the red lines represent the linear least-squares fit to relations that depart significantly from unity.}
\label{robfig_seg.fig}
\end{figure}

\begin{figure}
\plotone{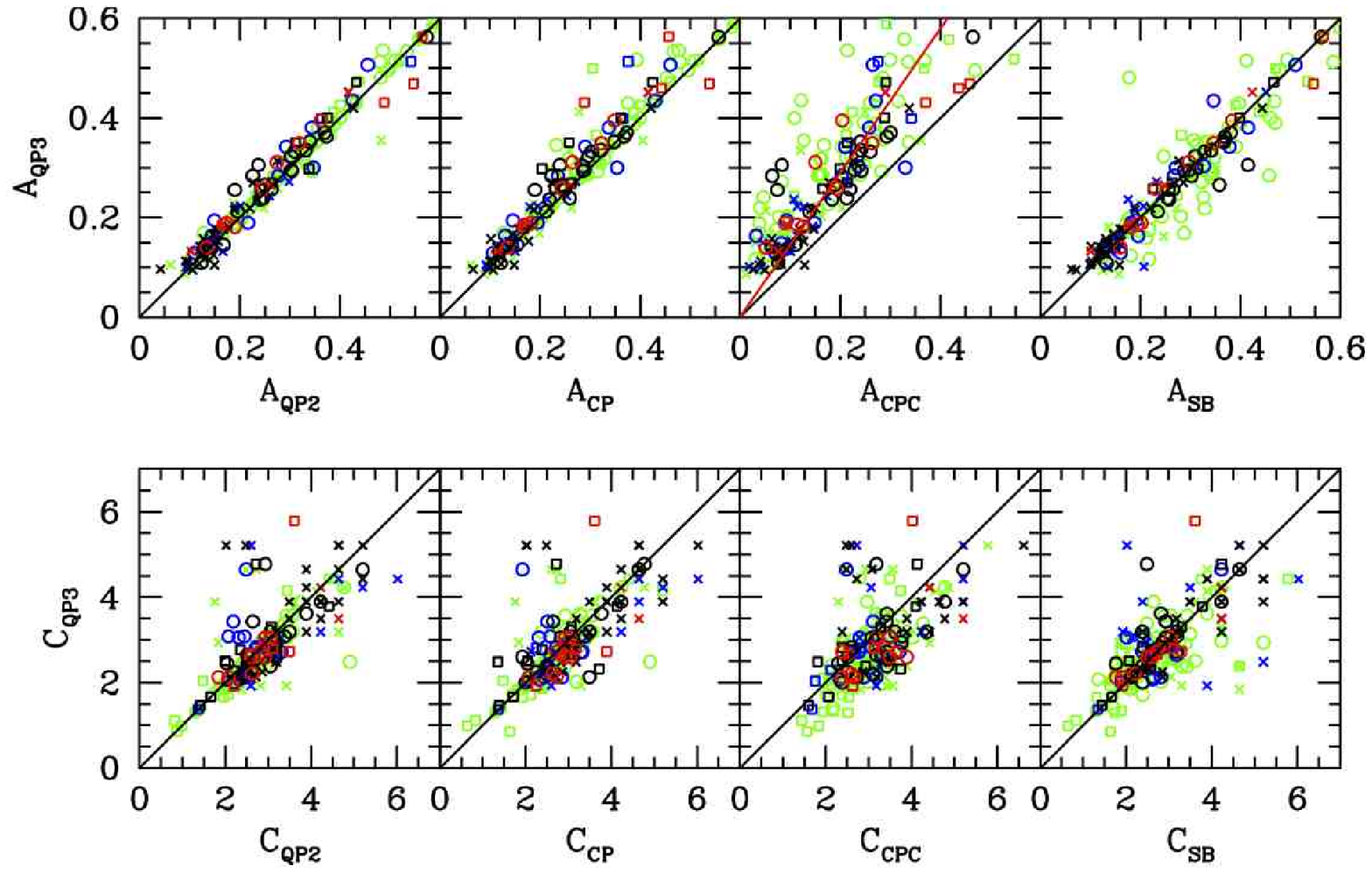}
\caption{As Figure \ref{robfig_seg.fig}, but for the $C$ and $A$ statistics.}
\label{robfig_seg2.fig}
\end{figure}

In Figures \ref{robfig_seg.fig} and \ref{robfig_seg2.fig} we compare morphological statistics calculated using the following five segmentation maps:
\begin{description}

\item[1 (`QP3'): ] Quasi-Petrosian segmentation map with threshold $\eta=0.3$, this is the default segmentation map.
We denote statistics calculated using this map with subscripts of the form (e.g.) $G_{\rm QP3}$.

\item[2 (`QP2'): ] Quasi-Petrosian segmentation map with threshold $\eta=0.2$.
We denote statistics calculated using this map with subscripts of the form (e.g.) $G_{\rm QP2}$.

\item[3 (`CPL'): ] Elliptical Petrosian segmentation map with threshold $\eta=0.2$ that includes  all pixels with flux greater than the surface brightness at the Petrosian radius (Lotz et al. 2004, 2006)
but following the potentially irregular isophotal contours.
We denote statistics calculated using this map with subscripts of the form (e.g.) $G_{\rm CPL}$.

\item[4 (`CPC'): ] Circular Petrosian segmentation map with threshold $\eta=0.2$ that includes all pixels within 1.5 Petrosian radii irrespective of flux (Conselice et al. 2000, 2008).
We denote statistics calculated using this map with subscripts of the form (e.g.) $G_{\rm CPC}$.

\item[5 (`SB'): ] Scaled isophotal (surface brightness) segmentation map of the form adopted by Law et al. (2007b) and Peter et al. (2007).
We denote statistics calculated using this map with subscripts of the form (e.g.) $G_{\rm SB}$.
\end{description}
There is generally good correlation among the nonparametric statistics derived using each of
these segmentation maps, especially when restricting our attention to the higher surface brightness systems for which the $\eta = 0.2$ threshold is well-defined (blue, black,
and red points).  In order to aid comparison between morphological properties derived by different groups in the literature, we present below a series of transformations that 
relate values calculated using different segmentation maps.
In determining these relations we consider only those galaxies for which the $\eta = 0.2$ threshhold is well-defined and the morphologies robust to statistical uncertainties
(i.e., we require $H_{160}<24.0$ and $\eta_{\rm crit} \leq 0.2$), and perform a linear least squares fit with uniform uncertainties in both quantities.

The greatest variation occurs in the Gini parameter $G$, which is extremely sensitive to the pixels included in the segmentation map (see also a previous analysis by
Lisker 2008).
As illustrated by Figure \ref{robfig_seg.fig} (bottom left-hand panel), simply adopting a Petrosian threshold of $\eta=0.2$ significantly increases $G$ over the $\eta=0.3$ case by including
more low-flux pixels in the segmentation map (note that this effect is not as noticeable for the green-colored points, for which the dynamic range of the galaxy surface brightness did not
permit the Petrosian algorithm to reach the 20\% flux threshold, and defaulted instead to the initial Source Extractor $1.5\sigma$ isophotal segmentation map).
Similarly, the CPL map also results in systematically higher values of $G$ than calculated by our default $\eta=0.3$ quasi-Petrosian algorithm.  This effect is even more noticeable in
the CPC segmentation map; this map increases the mean value of $G$ significantly by including many more low-surface brightness pixels than the other segmentation maps, and also
compresses the dynamic range of $G$ among the galaxy sample.  In contrast, the scaled surface-brightness selection technique (SB) stretches the dynamic range of $G$, but corresponds
poorly to estimates obtained using other segmentation maps.
We find that the key transformations between these segmentation maps are given on average by:
\begin{equation}
G_{\rm CPL} = 1.78 G_{\rm QP3} - 0.19
\end{equation}
\begin{equation}
G_{\rm CPL} = 2.02 G_{\rm CPC} -0.87
\end{equation}

The second order moment of the light distribution ($M_{20}$) is tightly correlated amongst all 5 segmentation maps, although values calculated using the CPC segmentation map
lie at systematically lower values due to the inclusion of additional low surface-brightness pixels.  The key transformations are described by:
\begin{equation}
M_{\rm 20, CPL}  = 1.04 M_{\rm 20, QP3} -0.03
\end{equation}
\begin{equation}
M_{\rm 20, CPL} = 0.96 M_{\rm 20, CPC}  + 0.06
\end{equation}

The concentration parameter $C$ exhibits minimal systematic offsets between segmentation maps, but considerably more scatter than the other morphological statistics.
As discussed in \S \ref{psrobust.sec} this is primarily due to the poor sampling of the inner 20\% of the light profile in the $z\sim2-3$ galaxies, 
which are small and poorly resolved in comparison to nearby galaxy samples (see discussion by Bershady et al. 2000).
The key transformations are described by:
\begin{equation}
C_{\rm CPL} = 0.84 C_{\rm QP3} + 0.47
\end{equation}
\begin{equation}
C_{\rm CPL} = 1.11 C_{\rm CPC} - 0.51
\end{equation}

The asymmetry parameter $A$ is also well correlated between different segmentation maps, with relatively little scatter and no significant systemic shifts
amongst 4 of the 5 segmentation maps.
There is more scatter and a systematic offset however when comparing estimates to the CPC segmentation map,
with $A_{\rm CPC}$ systematically lower compared to the other 4 segmentation maps.
We find that the mean relations are governed by the equations
\begin{equation}
A_{\rm CPL} = 0.80 A_{\rm QP3} + 0.03
\end{equation}
\begin{equation}
A_{\rm CPL} = 1.15 A_{\rm CPC} + 0.03
\end{equation}

The tightest correlation is found for $\Psi$, for which all segmentation maps produce nearly identical values with only minimal scatter about the 1-1 relation.


\subsection{A.2 Source Magnitude}
\label{magrobust.sec}

\begin{figure}
\epsscale{0.9}
\plotone{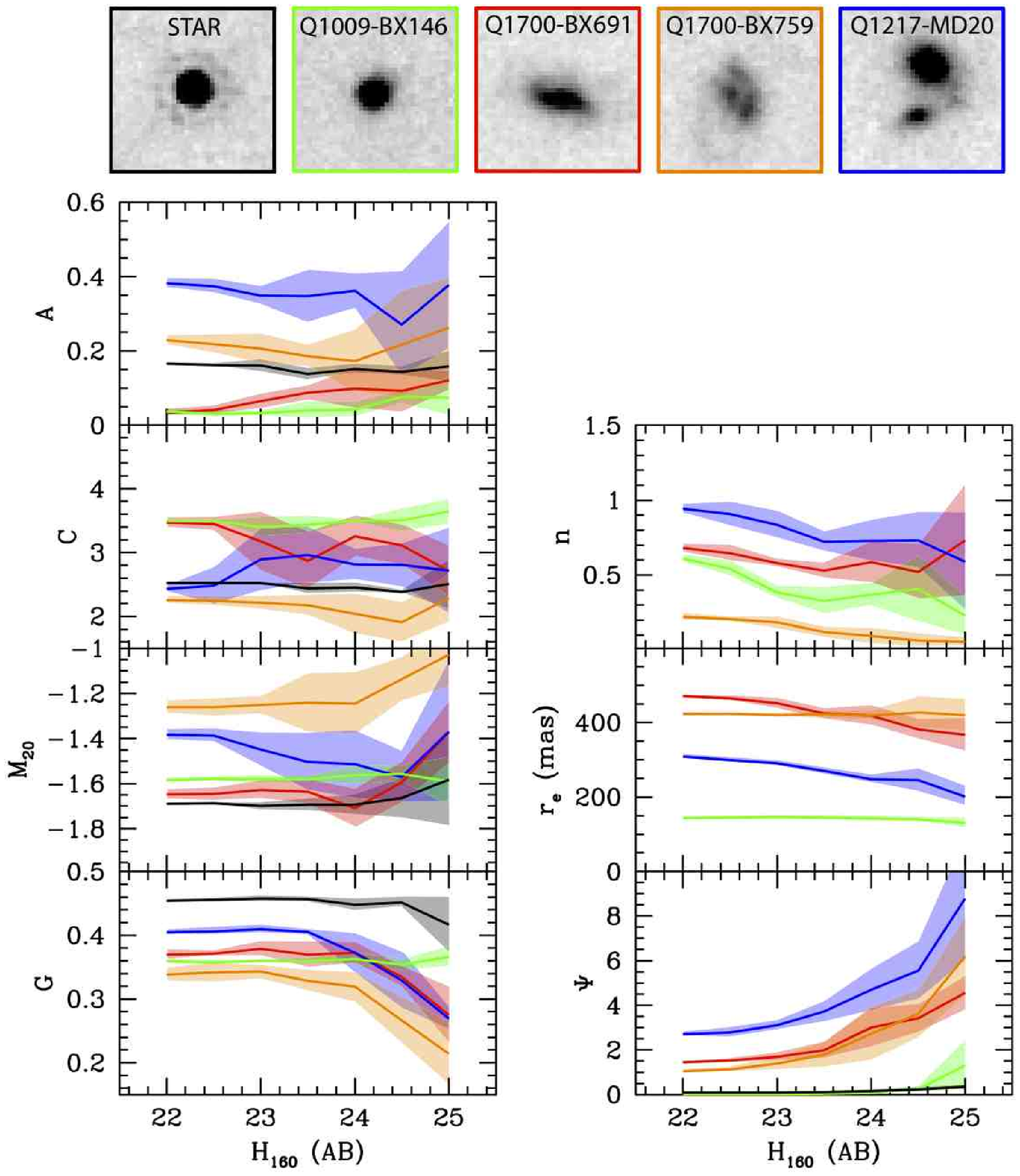}
\caption{Robustness of the morphological parameters $G$, $M_{20}$, $C$, $A$, $\Psi$ and the GALFIT indices $r_{\rm e}$ and $n$ 
to total source magnitude $H_{160}$
for five sources selected to span the typical range of morphologies. 
Solid lines indicate the mean value derived from 10 Monte Carlo realizations of the noise; shaded regions indicate the $1\sigma$ deviation about the mean.
Postage stamps (3x3 arcsec) showing each of the five test sources are shown at the top, with a colored
border corresponding to their respective lines in the lower panels.  Note that the stellar source is not shown in the $r_{\rm e}$ and $n$ panels since it is not well reproduced by
a Sersic model.}
\label{robfig.fig}
\end{figure}

We perform Monte Carlo tests to quantify the mean and standard deviation of the morphological statistics
for sources of fixed structure with different total magnitudes.  We choose five sources (four galaxies and reference star; see top row of Figure \ref{robfig.fig}) 
that are roughly representative of the range of morphologies found within
our galaxy sample, and construct morphological models of them using GALFIT.  We scale the total flux of these models to
$H_{160} = 22.0$ AB, and insert ten copies of each into randomly selected blank-field regions of the WFC3 images  in order to obtain multiple realizations of the background
noise.  For each copy, we compute the segmentation maps and morphological parameters as described above in \S \ref{stats.sec}.  This exercise
is repeated every 0.5 mag in the range $H_{160} = 22 - 25$ AB.

Figure \ref{robfig.fig} suggests that while  the morphological parameters are relatively robust for $H_{160} \leq 24$ AB, at $H_{160} > 24$ AB 
most start to break down either in the sense that their
mean values deviate significantly from the mean values derived at brighter magnitudes, or the variance among different realizations of the background noise becomes large
relative to the mean.  This behavior for each statistic can be summarized as follows:

\begin{itemize}

\item $M_{20}$, $C$, and $A$ exhibit $\sim$ 1/2/15\% uncertainty (averaged over the 4 star forming galaxy models)
at the bright end of the sample ($H_{160} \sim 22$ AB), increasing to 13/15/45\% at the faint
end ($H_{160} \sim 25$ AB).  Mean uncertainty at the average magnitude of the $H_{160} \leq 24$ AB sample is 4/11/22\% respectively; in the case of $C$ this is sufficient to confuse the
relative ordering between galaxies of different morphologies.
There are no systematic variations with magnitude.

\item $G$ has a bright-end uncertainty $\sim 2$\%, a faint-end uncertainty $\sim 13$\%, and
a mean uncertainty at the average magnitude of the $H_{160} \leq 24$ AB sample of 3\%.
While the mean value of $G$ remains relatively constant down to
$H_{160} \sim 24$ AB, it decreases systematically at lower magnitudes.  This systematic decline is because the $\eta=30$\% surface brightness threshold for the quasi-Petrosian pixel
decreases below the $1.5\sigma$ Source Extractor threshold, resulting in effective loss of the lowest-flux pixels from the segmentation map.

\item $\Psi$ has a bright-end uncertainty $\sim 7$\%, a faint-end uncertainty $\sim 41$\%, and a mean uncertainty at the average magnitude of the $H_{160} > 24$ AB sample of 21\%.
$\Psi$ systematically increases for $H_{160} > 23$ AB, most noticeably
for $H_{160} > 24$ AB.  

\item The Sersic index $n$ and effective circularized radius $r_{\rm e}$ have mean bright-end uncertainties of
$\sim 5/1$\%, faint-end uncertainties of $\sim 54/11$\%,
and mean uncertainties at the average magnitude of the $H_{160} \leq 24$ AB sample of 15/2\% respectively.
Both $r_{\rm e}$ and $n$ decline systematically with magnitude as it becomes progressively more difficult to distinguish faint outer regions
of the galaxies from the background sky (see also Gray et al. 2009).  These effects are particularly pronounced for $H_{160} > 24$ AB.  The significance of the decline in $r_{\rm e}$ varies as a function
of morphology; regular symmetric objects show negligible variation across the full range $H_{160} = 22-25$ AB, while more irregular multi-component galaxies may decline by as much as 30\%.

\end{itemize}

\begin{figure}
\plotone{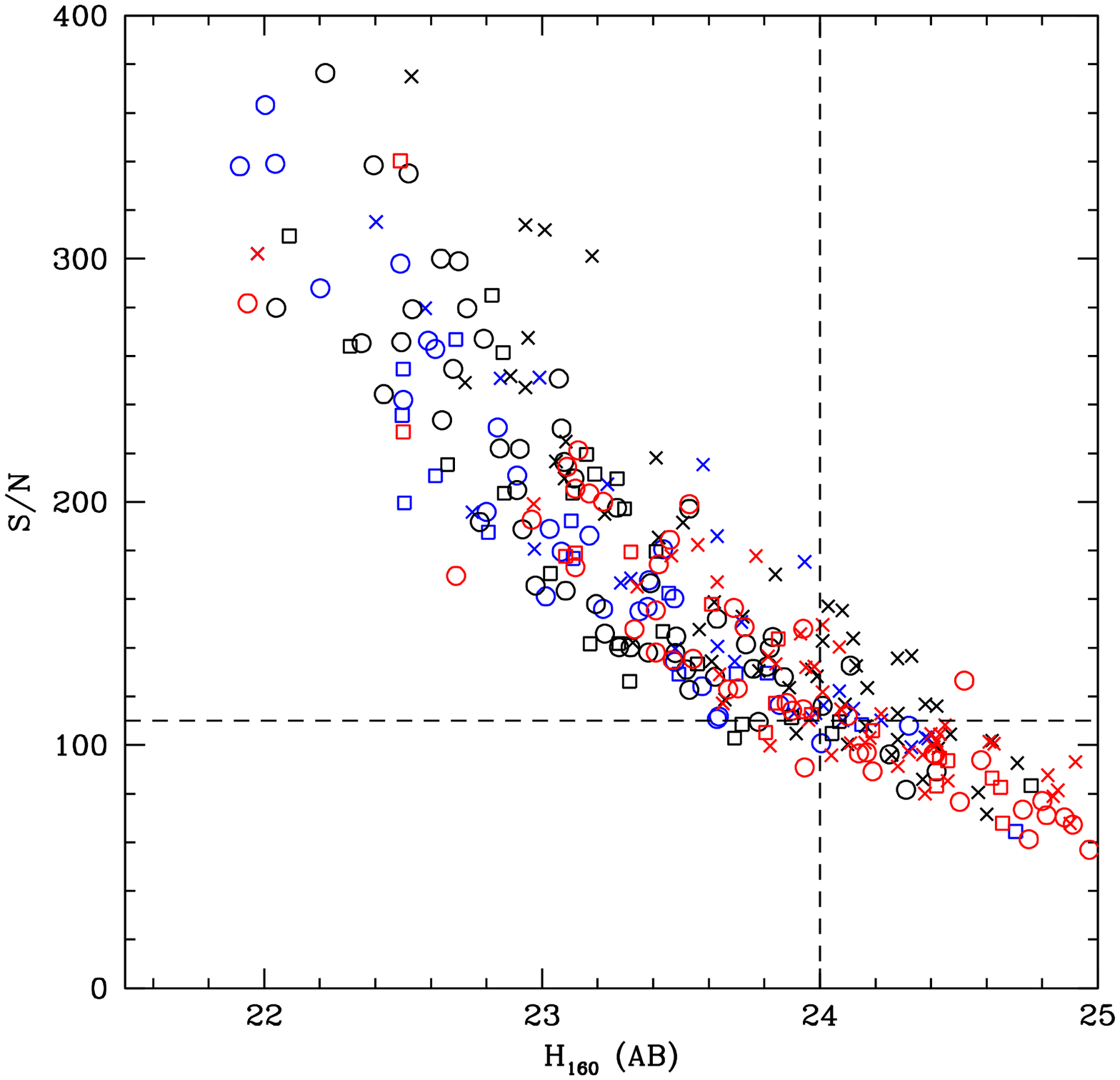}
\caption{Signal-to-noise ratio $S/N$ of the galaxies as a function of $H_{160}$ magnitude.  Dashed lines denote 
the close correspondence between cuts of the form $H_{160} < 24$ AB and $S/N > 110$.  Symbol color and types are as given in Figure \ref{eta.fig}.}
\label{snr.fig}
\end{figure}

In the interests of measuring physically meaningful morphological statistics, we therefore impose an apparent magnitude cut on our sample of galaxies
at $H_{160}  \leq 24$ AB, corresponding closely  to a signal-to-noise ratio cut $S/N>110$ (see Figure \ref{snr.fig}).
This is consistent with the analyses of Conselice et al. (2000) and Lisker et al. (2008), who found (respectively) that $A$ became dominated by the background noise 
and that $G$ becomes less robust below $S/N \sim 100$.

\subsection{A.3 Pixel Scale}
\label{psrobust.sec}

\begin{figure}
\plotone{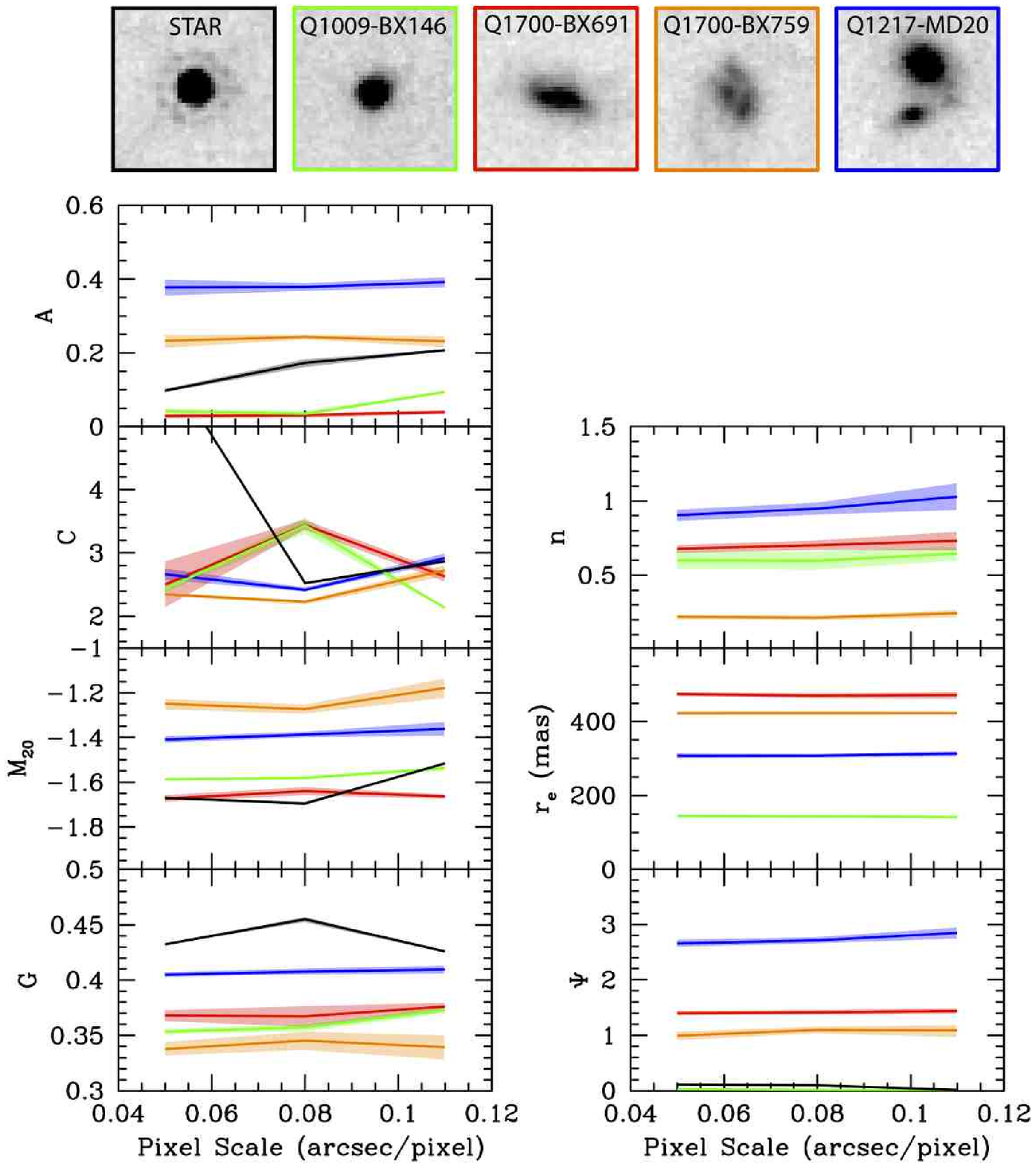}
\caption{As Figure \ref{robfig.fig}, but showing robustness to choice of angular pixel scale.}
\label{robfigPS.fig}
\end{figure}

It is also worth investigating the dependence of the calculated morphological statistics on the pixel scale that we selected to drizzle our WFC3 data onto.
We therefore inserted the GALFIT galaxy models (normalized to $H_{\rm AB} \sim 22$) from \S \ref{magrobust.sec} into random blank regions of the WFC3 fields, and rebinned
the data using linear interpolation (using the IDL routine {\it CONGRID}, with conservation of total flux) to a variety of pixel scales that we could realistically have chosen.
As illustrated by Figure \ref{robfigPS.fig}, the recovered morphologies are extremely robust to variations $\sim$ a factor of two in pixel scale, with the exception of the 
stellar source model (for $A$, $C$, $M_{20}$, and $G$) and the concentration parameter $C$ (for the stellar source, Q1009-BX146, and Q1700-BX691).
That is, all of the morphological parameters can vary unsatisfactorily with choice of sampling scale for unresolved objects, or (in the case of $C$) for objects
in which the innermost region containing 20\% of the light is poorly sampled.  Indeed, we note that while $C$ is not robust for Q1009-BX146 and Q1700-BX691, it is more so for
Q1700-BX759 and Q1217-MD20 because these two galaxies are significantly more spatially extended and the central region containing 20\% of the total light correspondingly better sampled.
This resolution dependence of $C$ for the poorly sampled inner radius $r_{20}$ is well-known in the literature
(see, e.g., Figure 9 of Bershady et al. 2000).

Given the general robustness of the morphological parameters to the choice of angular sampling scale, we do not make any corrections to the measured morphologies
due to the small ($< 10$\%) change in angular size subtended by a physical kpc across the redshift interval $1.5 < z < 3.6$, but we choose not to use the concentration parameter $C$
in our analyses.

\subsection{A.4 Point Spread Function}

We also explore the robustness of the morphological statistics to the width of the observational PSF, which affects the degree to which spatial structures are resolved.  In order to reliably
trace structures resolved on scales smaller than the WFC3/IR PSF we repeat the analysis from \S \ref{magrobust.sec}, but using GALFIT models of five sources in the Q1700+64 field
observed with HST/ACS F814W as part of program GO-10581 (PI: Shapley; see description in Peter et al. 2007).  These models are convolved with 2d Gaussian profiles
(using the IDL routine {\it FILTER\_IMAGE}) to mimic
observations with PSF FWHM ranging from 0.1 (i.e., native resolution for the F814W imaging data) - 0.3 arcsec.

\begin{figure}
\plotone{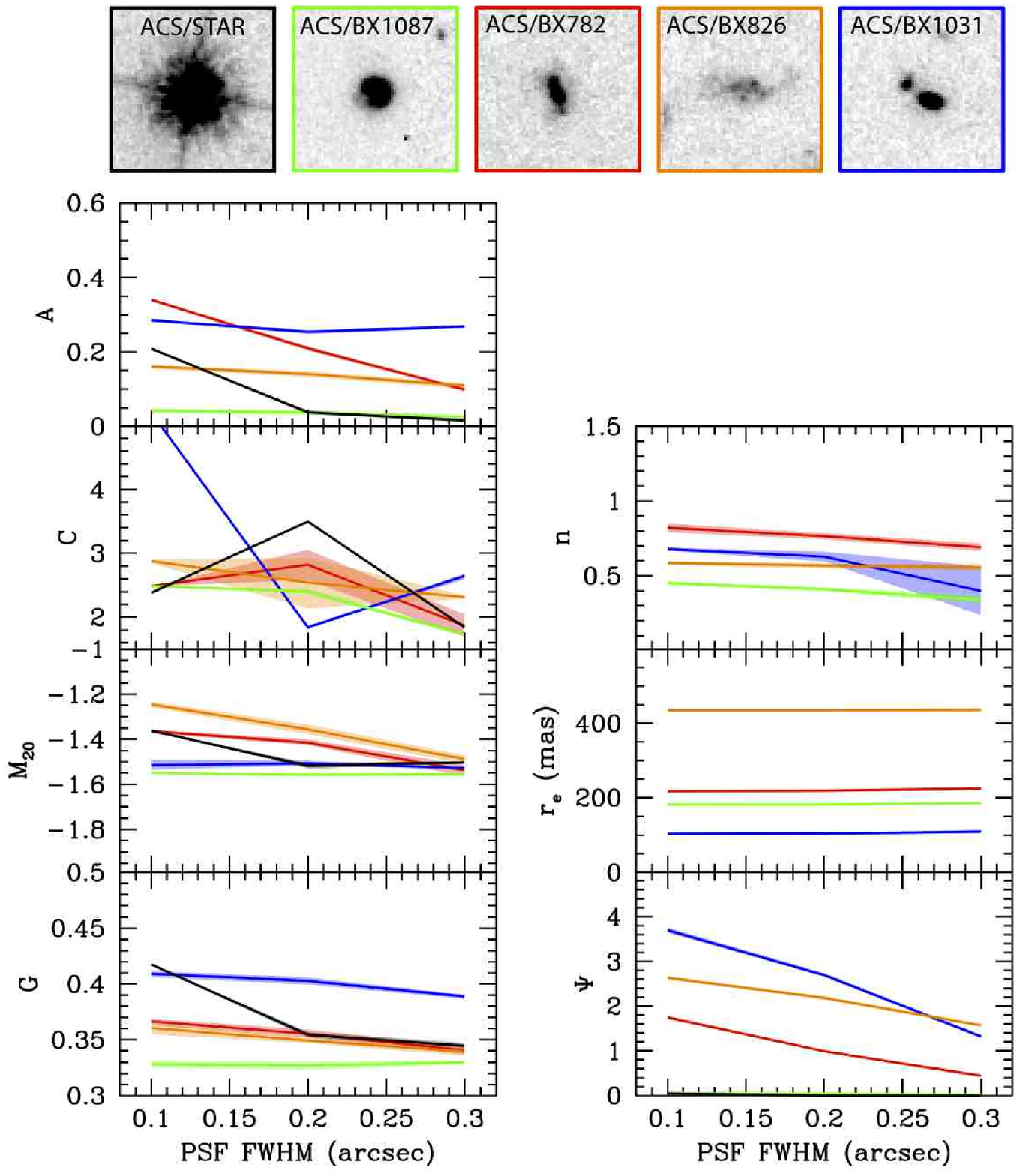}
\caption{As Figure \ref{robfig.fig}, but showing robustness to PSF FWHM.  In order to realistically discuss PSFs smaller than that of the WFC3/IR data we base our models on 
four $z\sim2$ star forming galaxies and one reference star observed with HST/ACS F814W as part of program GO-10581 (PI: Shapley).}
\label{robfigPSF.fig}
\end{figure}

As illustrated in Figure \ref{robfigPSF.fig}, there is little change in the statistical uncertainty of the morphological measurements with PSF FWHM, but most exhibit systematic
variations (as noted previously by, e.g., Lotz et al. 2004, 2008b).  As discussed in \S \ref{psrobust.sec}, $C$ is poorly behaved since the inner 20\% of the light profile is poorly sampled.
As PSF FWHM increases, the dynamic range of $A$, $M_{20}$, $G$, and $\Psi$ decreases, approaching the limit that when the PSF is large compared to the size of the galaxies
all objects will be unresolved and have indistinguishable morphologies.  The compression of the dynamical range is less pronounced for objects such as ACS/BX1031 which have two well-separated
components that require a more substantial change in the PSF to lose information about the double morphological structure.  Similarly, the parametric statistics $r_{\rm e}$ and $n$ are
generally quite stable to variations in the PSF since GALFIT incorporates the observational PSF in its fitting algorithm, although $n$ declines by a few percent from FWHM $\sim 0.1$ arcsec to
$\sim 0.3$ arcsec.  This effect was previously noted by Buitrago et al. (2008) who found that Sersic indices measured in the infrared
with NICMOS were $13\pm12$\% smaller than measured in the optical with ACS.

Since all of our galaxies have been observed with uniform coverage and a PSF that varies by less than $\sim$ 4\%,
the trends illustrated in Figure \ref{robfigPSF.fig}
will be unimportant for internal comparisons between the morphologies of galaxies in our sample.  These trends will be important, however, 
to keep in mind when comparing any of our galaxies to 
low-redshift samples, or to similar $z\sim2$ galaxies observed with {\it HST} in bandpasses tracing the rest-frame UV.

\clearpage
\section{B. THE PROJECTED AXIS RATIO OF A TRIAXIAL ELLIPSOID VIEWED IN AN ARBITRARY ORIENTATION}
\label{batriax.sec}

Let the ellipsoid be a surface in 3-space characterized by the scale lengths $r_x$, $r_y$, and $r_z$.
Adopting spherical polar coordinates, the surface of the ellipsoid is defined by
\begin{equation}
\vec{P}(\alpha,\beta) = (r_x {\rm cos}\alpha \, {\rm cos} \beta, r_y {\rm sin}\alpha \, {\rm cos}\beta, r_z {\rm sin} \beta)
\end{equation}
where $\alpha$ is the azimuthal angle $0 \leq \alpha < 2\pi$ and $\beta$ the polar angle $-\pi/2 \leq \beta \leq \pi/2$.

Rather than rotating the ellipsoid, consider the identical problem in which the viewer is located at a large distance along the direction described by $\theta$ and $\phi$, the azimuthal
and polar angles respectively.  The unit vector in the direction of the viewer $\hat{f}$ may be written as
\begin{equation}
\hat{f} = ({\rm cos}\theta \, {\rm cos} \phi, {\rm sin}\theta \, {\rm cos} \phi, {\rm sin}\phi)
\end{equation}

At each point on the surface of the ellipsoid there is a corresponding tangent plane; by definition, the `edge' of the figure as seen by the viewer is located where the unit vector towards
the viewer is parallel to the tangent plane.  If $\vec{n}(\alpha,\beta)$ is the normal to the tangent plane, then 
the projected ellipse observed by the viewer is described by the set of $\alpha,\beta$ such that
\begin{equation}
\vec{n} \cdot \hat{f} = 0
\end{equation}

Since the tangent plane to the ellipsoid is described by the partial derivatives $\frac{\partial \vec{P}}{\partial \alpha}$ and $\frac{\partial \vec{P}}{\partial \beta}$, the normal to the tangent
plane may be constructed by the cross product of these vectors:
\begin{equation}
\vec{n} = \frac{\partial \vec{P}}{\partial \alpha} \times \frac{\partial \vec{P}}{\partial \beta}\\
= (r_y r_z {\rm cos}\alpha \, {\rm cos}^2\beta, r_x r_z {\rm sin}\alpha \, {\rm cos}^2 \beta, r_x  r_y {\rm sin}\beta \, {\rm cos}\beta)
\end{equation}

Setting $\vec{n} \cdot \hat{f} = 0$ we obtain the relation 
\begin{equation}
{\rm tan}\beta= - \left( \frac{r_y r_z {\rm cos}\alpha \, {\rm cos} \theta \, {\rm cos}\phi + r_x r_z {\rm sin}\alpha \, {\rm sin}\theta \, {\rm cos}\phi}{r_x r_y {\rm sin}\phi} \right )
\label{beta.eqn}
\end{equation}
For all $\phi \neq 0$ (i.e., viewing the ellipsoid perfectly edge-on), $\alpha$ along the projected ellipse attains all values
in the range $0 - 2\pi$.  By setting $\alpha=0^{\circ}$, $0.1^{\circ}$, ..., $359.9^{\circ}$ and calculating the corresponding $\beta$ from Equation \ref{beta.eqn} it is possible to obtain
a set of ($\alpha,\beta$) pairs fully describing the projected ellipse.

The projected radius $q$ of the ellipse, as seen by the viewer, at each ($\alpha,\beta$) is given by the magnitude of the cross product of the vector $\vec{P}$ to the point on the surface with
the unit vector on the line of sight to the viewer $\hat{f}$:
\begin{equation}
\vec{q} = \vec{P}(\alpha,\beta) \times \hat{f}
\end{equation}
Some computation gives the vector components of $\vec{q} = (q_x, q_y, q_z)$ as
\begin{eqnarray}
q_x = r_y {\rm sin}\alpha \, {\rm cos}\beta \, {\rm sin}\phi - r_z {\rm sin}\beta \, {\rm sin}\theta \, {\rm cos}\phi  \\
q_y = r_z {\rm sin}\beta \, {\rm cos}\theta \, {\rm cos}\phi - r_x {\rm cos}\alpha \, {\rm cos}\beta \, {\rm sin}\phi  \\
q_z = r_x {\rm cos}\alpha \, {\rm cos}\beta \, {\rm sin}\theta \, {\rm cos}\phi - r_y {\rm sin}\alpha \, {\rm cos}\beta \, {\rm cos}\theta \, {\rm cos}\phi
\end{eqnarray}

The magnitude $q$ is then simply
\begin{equation}
q = \sqrt{q_x q_x +q_y q_y + q_z q_z}
\end{equation}
and the axis ratio of the projected ellipse may be trivially calculated as the ratio of the minimum and maximum value of $q$ for all ($\alpha,\beta$) coordinate pairs:
\begin{equation}
b/a = \frac{{\rm min} (q)}{{\rm max} (q)}
\label{baellip.eqn}
\end{equation}

It can be verified that in the limit where two axes have equal length Equation \ref{baellip.eqn} gives identical results to the thick-disk case derived by Hubble (1926) and
described by Equation \ref{hubble.eqn}.


\begin{thebibliography}


\bibitem[Abraham et al.(1994)]{1994ApJ...432...75A} Abraham, R.~G., Valdes, 
F., Yee, H.~K.~C., \& van den Bergh, S.\ 1994, \apj, 432, 75 



\bibitem[Abraham et al.(1996)]{1996ApJS..107....1A} Abraham, R.~G., van den 
Bergh, S., Glazebrook, K., Ellis, R.~S., Santiago, B.~X., Surma, P., 
\& Griffiths, R.~E.\ 1996, \apjs, 107, 1 

\bibitem[Abraham et al.(2003)]{2003ApJ...588..218A} Abraham, R.~G., van den 
Bergh, S., \& Nair, P.\ 2003, \apj, 588, 218

\bibitem[Abraham et al.(2007)]{2007ApJ...669..184A} Abraham, R.~G., et al.\ 
2007, \apj, 669, 184 

\bibitem[Adelberger et al.(2004)]{2004ApJ...607..226A} Adelberger, K.~L., 
Steidel, C.~C., Shapley, A.~E., Hunt, M.~P., Erb, D.~K., Reddy, N.~A., 
\& Pettini, M.\ 2004, \apj, 607, 226 

\bibitem[Barden et al.(2005)]{2005ApJ...635..959B} Barden, M., et al.\ 
2005, \apj, 635, 959 

\bibitem[Bell et al.(2003)]{2003MNRAS.343..367B} Bell, E.~F., Baugh, C.~M., 
Cole, S., Frenk, C.~S., \& Lacey, C.~G.\ 2003, \mnras, 343, 367 



\bibitem[Bershady et al.(2000)]{2000AJ....119.2645B} Bershady, M.~A., 
Jangren, A., \& Conselice, C.~J.\ 2000, \aj, 119, 2645 




\bibitem[Bertin 
\& Arnouts(1996)]{1996A&AS..117..393B} Bertin, E., \& Arnouts, S.\ 1996, \aaps, 117, 393 


\bibitem[Bezanson et al.(2009)]{2009ApJ...697.1290B} Bezanson, R., van 
Dokkum, P.~G., Tal, T., Marchesini, D., Kriek, M., Franx, M., 
\& Coppi, P.\ 2009, \apj, 697, 1290 




\bibitem[Bond et al.(2011)]{2011ApJ...729...48B} Bond, N.~A., Gawiser, E., 
\& Koekemoer, A.~M.\ 2011, \apj, 729, 48 




\bibitem[Bournaud et al.(2007)]{2007ApJ...670..237B} Bournaud, F., 
Elmegreen, B.~G., \& Elmegreen, D.~M.\ 2007, \apj, 670, 237 

\bibitem[Bournaud et 
al.(2008)]{2008A&A...486..741B} Bournaud, F., et al.\ 2008, \aap, 486, 741 




\bibitem[Bournaud 
\& Elmegreen(2009)]{2009ApJ...694L.158B} Bournaud, F., \& Elmegreen, B.~G.\ 2009, \apjl, 694, L158 





\bibitem[Bouwens et al.(2004)]{2004ApJ...611L...1B} Bouwens, R.~J., 
Illingworth, G.~D., Blakeslee, J.~P., Broadhurst, T.~J., 
\& Franx, M.\ 2004, \apjl, 611, L1 

\bibitem[Bouwens et al.(2010)]{2010ApJ...709L.133B} Bouwens, R.~J., et al.\ 
2010, \apjl, 709, L133 

\bibitem[Buitrago et al.(2008)]{2008ApJ...687L..61B} Buitrago, F., 
Trujillo, I., Conselice, C.~J., Bouwens, R.~J., Dickinson, M., 
\& Yan, H.\ 2008, \apjl, 687, L61 




\bibitem[Calzetti et al.(2000)]{2000ApJ...533..682C} Calzetti, D., Armus, 
L., Bohlin, R.~C., Kinney, A.~L., Koornneef, J., 
\& Storchi-Bergmann, T.\ 2000, \apj, 533, 682 



\bibitem[Cameron et al.(2010)]{2010arXiv1007.2422C} Cameron, E., Carollo, 
C.~M., Oesch, P.~A., Bouwens, R.~J., Illingworth, G.~D., Trenti, M., Labbe, 
I., \& Magee, D.\ 2010, arXiv:1007.2422 


\bibitem[Cameron(2011)]{2011PASA...28..128C} Cameron, E.\ 2011, 
Publications of the Astronomical Society of Australia, 28, 
128 




\bibitem[Carrasco et al.(2010)]{2010MNRAS.405.2253C} Carrasco, E.~R., 
Conselice, C.~J., \& Trujillo, I.\ 2010, \mnras, 405, 2253 

\bibitem[Casertano et al.(2000)]{2000AJ....120.2747C} Casertano, S., et 
al.\ 2000, \aj, 120, 2747 




\bibitem[Cassata et al.(2010)]{2010ApJ...714L..79C} Cassata, P., et al.\ 
2010, \apjl, 714, L79 

\bibitem[Ceverino et al.(2010)]{2010MNRAS.404.2151C} Ceverino, D., Dekel, 
A., \& Bournaud, F.\ 2010, \mnras, 404, 2151 


\bibitem[Chabrier(2003)]{2003PASP..115..763C} Chabrier, G.\ 2003, \pasp, 
115, 763 



\bibitem[Colley et al.(1996)]{1996ApJ...473L..63C} Colley, W.~N., Rhoads, 
J.~E., Ostriker, J.~P., \& Spergel, D.~N.\ 1996, \apjl, 473, L63 


\bibitem[Conroy et al.(2008)]{2008ApJ...679.1192C} Conroy, C., Shapley, 
A.~E., Tinker, J.~L., Santos, M.~R., \& Lemson, G.\ 2008, \apj, 679, 1192 




\bibitem[Conselice et al.(2000)]{2000ApJ...529..886C} Conselice, C.~J., 
Bershady, M.~A., \& Jangren, A.\ 2000, \apj, 529, 886 

\bibitem[Conselice(2003)]{2003ApJS..147....1C} Conselice, C.~J.\ 2003, 
\apjs, 147, 1 

\bibitem[Conselice et al.(2003)]{2003AJ....126.1183C} Conselice, C.~J., 
Bershady, M.~A., Dickinson, M., \& Papovich, C.\ 2003, \aj, 126, 1183 

\bibitem[Conselice et al.(2004)]{2004ApJ...600L.139C} Conselice, C.~J., et 
al.\ 2004, \apjl, 600, L139 

\bibitem[Conselice et al.(2005)]{2005ApJ...628..160C} Conselice, C.~J., 
Bundy, K., Ellis, R.~S., Brichmann, J., Vogt, N.~P., 
\& Phillips, A.~C.\ 2005, \apj, 628, 160 

\bibitem[Conselice(2006)]{2006MNRAS.373.1389C} Conselice, C.~J.\ 2006, 
\mnras, 373, 1389 




\bibitem[Conselice et al.(2008)]{2008MNRAS.386..909C} Conselice, C.~J., 
Rajgor, S., \& Myers, R.\ 2008, \mnras, 386, 909 



\bibitem[Conselice et al.(2009)]{2009MNRAS.394.1956C} Conselice, C.~J., 
Yang, C., \& Bluck, A.~F.~L.\ 2009, \mnras, 394, 1956 

\bibitem[Conselice et al.(2011)]{2011MNRAS.413...80C} Conselice, C.~J., et 
al.\ 2011a, \mnras, 413, 80 

\bibitem[Conselice et al.(2011)]{2011arXiv1105.2522C} Conselice, C.~J., 
Bluck, A.~F.~L., Ravindranath, S., Mortlock, A., Koekemoer, A., Buitrago, 
F., Gr{\"u}tzbauch, R., \& Penny, S.\ 2011b, arXiv:1105.2522 




\bibitem[Cowie 
\& Barger(2008)]{2008ApJ...686...72C} Cowie, L.~L., \& Barger, A.~J.\ 2008, \apj, 686, 72 

\bibitem[Dalcanton 
\& Bernstein(2002)]{2002AJ....124.1328D} Dalcanton, J.~J., \& Bernstein, R.~A.\ 2002, \aj, 124, 1328 

\bibitem[Dekel 
\& Birnboim(2006)]{2006MNRAS.368....2D} Dekel, A., \& Birnboim, Y.\ 2006, \mnras, 368, 2 

\bibitem[Dekel et al.(2009)]{2009Natur.457..451D} Dekel, A., Birnboim, Y., 
Engel, G., et al.\ 2009a, \nat, 457, 451 

\bibitem[Dekel et al.(2009)]{2009ApJ...703..785D} Dekel, A., Sari, R., 
\& Ceverino, D.\ 2009b, \apj, 703, 785 







\bibitem[Dickinson(2000)]{2000RSPTA.358.2001D} Dickinson, M.\ 2000, 
Philos. Trans. R. Soc. London A, 358, 2001 




\bibitem[Dickinson et al.(2003)]{2003ApJ...587...25D} Dickinson, M., 
Papovich, C., Ferguson, H.~C., \& Budav{\'a}ri, T.\ 2003, \apj, 587, 25 

\bibitem[Dutton et al.(2011)]{2011MNRAS.410.1660D} Dutton, A.~A., et al.\ 
2011, \mnras, 410, 1660 

\bibitem [Elmegreen et al.(2005)] {2005ApJ...631...85E} Elmegreen, D.~M., Elmegreen, B.~G., Rubin, 
D.~S., \& Schaffer, M.~A. 2005, \apj, 631, 85





\bibitem[Erb et al.(2006a)]{2006ApJ...644..813E} Erb, D.~K., Shapley, A.~E., 
Pettini, M., Steidel, C.~C., Reddy, N.~A., \& Adelberger, K.~L.\ 2006a, 
\apj, 644, 813

\bibitem[Erb et al.(2006b)]{2006ApJ...647..128E} Erb, D.~K., Steidel, C.~C., 
Shapley, A.~E., Pettini, M., Reddy, N.~A., \& Adelberger, K.~L.\ 2006b, 
\apj, 647, 128

\bibitem[Erb et al.(2006c)]{2006ApJ...646..107E} Erb, D.~K., Steidel, C.~C., 
Shapley, A.~E., Pettini, M., Reddy, N.~A., \& Adelberger, K.~L.\ 2006c, 
\apj, 646, 107

\bibitem[Erb et al.(2010)]{2010ApJ...719.1168E} Erb, D.~K., Pettini, M., 
Shapley, A.~E., Steidel, C.~C., Law, D.~R., 
\& Reddy, N.~A.\ 2010, \apj, 719, 1168 




\bibitem[F{\"o}rster Schreiber et al.(2009)]{2009ApJ...706.1364F} 
F{\"o}rster Schreiber, N.~M., et al.\ 2009, \apj, 706, 1364 

\bibitem[F{\"o}rster Schreiber et al.(2011)]{2011ApJ...731...65F} 
F{\"o}rster Schreiber, N.~M., Shapley, A.~E., Erb, D.~K., Genzel, R., 
Steidel, C.~C., Bouch{\'e}, N., Cresci, G., 
\& Davies, R.\ 2011, \apj, 731, 65 




\bibitem[Franx et al.(2008)]{2008ApJ...688..770F} Franx, M., van Dokkum, 
P.~G., Schreiber, N.~M.~F., Wuyts, S., Labb{\'e}, I., 
\& Toft, S.\ 2008, \apj, 688, 770 

\bibitem[Genzel et al.(2008)]{2008ApJ...687...59G} Genzel, R., et al.\ 
2008, \apj, 687, 59 

\bibitem[Genzel et al.(2011)]{2011ApJ...733..101G} Genzel, R., et al.\ 
2011, \apj, 733, 101 



\bibitem[Giavalisco et al.(1996)]{1996ApJ...470..189G} Giavalisco, M., 
Steidel, C.~C., \& Macchetto, F.~D.\ 1996, \apj, 470, 189 



\bibitem[Gini (1912)]{1329} Gini, C. 1912, reprinted in Memorie di Metodologia Statistica, ed. E. Pizetti \& T. Salvemini (1955; Rome: Libreria Eredi Virgilio Veschi).

\bibitem[Glasser (1962)]{1331} Glasser, G. J. 1962, J. Am. Stat. Assoc., 57, 648.

\bibitem[Gray et al.(2009)]{2009MNRAS.393.1275G} Gray, M.~E., et al.\ 2009, 
\mnras, 393, 1275 

\bibitem[Grogin et al.(2011)]{2011arXiv1105.3753G} Grogin, N.~A., et al.\ 
2011, arXiv:1105.3753 



\bibitem[Guo et al.(2009)]{2009MNRAS.398.1129G} Guo, Y., et al.\ 2009, 
\mnras, 398, 1129 


\bibitem[Guthrie(1992)]{1992A&AS...93..255G} Guthrie, B.~N.~G.\ 1992, \aaps, 93, 255 



\bibitem[Hubble(1926)]{1926ApJ....64..321H} Hubble, E.~P.\ 1926, \apj, 64, 
321 

\bibitem[Jones et al.(2010)]{2010MNRAS.404.1247J} Jones, T.~A., Swinbank, 
A.~M., Ellis, R.~S., Richard, J., \& Stark, D.~P.\ 2010, \mnras, 404, 1247 

\bibitem[Kennicutt(1998)]{1998ARA&A..36..189K} Kennicutt, R.~C., Jr.\ 1998, \araa, 36, 189 

\bibitem[Kent(1985)]{1985ApJS...59..115K} Kent, S.~M.\ 1985, \apjs, 59, 115 



\bibitem[Kere{\v s} et al.(2005)]{2005MNRAS.363....2K} Kere{\v s}, D., 
Katz, N., Weinberg, D.~H., \& Dav{\'e}, R.\ 2005, \mnras, 363, 2 

\bibitem[Kere{\v s} et al.(2009)]{2009MNRAS.395..160K} Kere{\v s}, D., 
Katz, N., Fardal, M., Dav{\'e}, R., 
\& Weinberg, D.~H.\ 2009, \mnras, 395, 160 



\bibitem[Koekemoer et al.(2002)]{2002hstc.conf..337K} Koekemoer, A.~M., 
Fruchter, A.~S., Hook, R.~N., 
\& Hack, W.\ 2002, The 2002 HST Calibration Workshop : Hubble after the Installation of the ACS and the NICMOS Cooling System, 337 


\bibitem[Koekemoer et al.(2011)]{2011arXiv1105.3754K} Koekemoer, A.~M., 
Faber, S.~M., Ferguson, H.~C., et al.\ 2011, arXiv:1105.3754 


\bibitem[Komatsu et al.(2011)]{2011ApJS..192...18K} Komatsu, E., et al.\ 
2011, \apjs, 192, 18 


\bibitem[Kormendy et al.(2009)]{2009ApJS..182..216K} Kormendy, J., Fisher, 
D.~B., Cornell, M.~E., \& Bender, R.\ 2009, \apjs, 182, 216 





\bibitem[Kriek et al.(2009)]{2009ApJ...705L..71K} Kriek, M., van Dokkum, 
P.~G., Franx, M., Illingworth, G.~D., 
\& Magee, D.~K.\ 2009, \apjl, 705, L71 

\bibitem[Kuchinski et al.(2000)]{2000ApJS..131..441K} Kuchinski, L.~E., et 
al.\ 2000, \apjs, 131, 441 


\bibitem[Labb{\'e} et al.(2003)]{2003ApJ...591L..95L} Labb{\'e}, I., 
Rudnick, G., Franx, M., et al.\ 2003, \apjl, 591, L95 



\bibitem[Law et al.(2007)]{2007ApJ...656....1L} Law, D.~R., Steidel, C.~C., 
Erb, D.~K., Pettini, M., Reddy, N.~A., Shapley, A.~E., Adelberger, K.~L., 
\& Simenc, D.~J.\ 2007b, \apj, 656, 1 

\bibitem[Law et al.(2007)]{2007ApJ...669..929L} Law, D.~R., Steidel, C.~C., 
Erb, D.~K., Larkin, J.~E., Pettini, M., Shapley, A.~E., 
\& Wright, S.~A.\ 2007a, \apj, 669, 929 

\bibitem[Law et al.(2009)]{2009ApJ...697.2057L} Law, D.~R., Steidel, C.~C., 
Erb, D.~K., Larkin, J.~E., Pettini, M., Shapley, A.~E., 
\& Wright, S.~A.\ 2009, \apj, 697, 2057 

\bibitem[Lisker(2008)]{2008ApJS..179..319L} Lisker, T.\ 2008, \apjs, 179, 
319 


\bibitem[Lotz et al.(2004)]{2004AJ....128..163L} Lotz, J.~M., Primack, J., 
\& Madau, P.\ 2004, \aj, 128, 163 

\bibitem[Lotz et al.(2006)]{2006ApJ...636..592L} Lotz, J.~M., Madau, P., 
Giavalisco, M., Primack, J., \& Ferguson, H.~C.\ 2006, \apj, 636, 592 

\bibitem[Lotz et al.(2008a)]{2008ApJ...672..177L} Lotz, J.~M., et al.\ 2008a, 
\apj, 672, 177 

\bibitem[Lotz et al.(2008b)]{2008MNRAS.391.1137L} Lotz, J.~M., Jonsson, P., 
Cox, T.~J., \& Primack, J.~R.\ 2008b, \mnras, 391, 1137 

\bibitem[Lotz et al.(2010a)]{2010MNRAS.404..575L} Lotz, J.~M., Jonsson, P., 
Cox, T.~J., \& Primack, J.~R.\ 2010a, \mnras, 404, 575 

\bibitem[Lotz et al.(2010b)]{2010MNRAS.404..590L} Lotz, J.~M., Jonsson, P., 
Cox, T.~J., \& Primack, J.~R.\ 2010b, \mnras, 404, 590 




\bibitem[Lowenthal et al.(1997)]{1997ApJ...481..673L} Lowenthal, J.~D., et 
al.\ 1997, \apj, 481, 673 




\bibitem[Maiolino et 
al.(2008)]{2008A&A...488..463M} Maiolino, R., et al.\ 2008, \aap, 488, 463 

\bibitem[Martig 
\& Bournaud(2010)]{2010ApJ...714L.275M} Martig, M., \& Bournaud, F.\ 2010, \apjl, 714, L275 


\bibitem[McLure et al.(2011)]{2011arXiv1102.4881M} McLure, R.~J., et al.\ 
2011, arXiv:1102.4881 




\bibitem[Melbourne et al.(2008)]{2008AJ....135.1207M} Melbourne, J., et 
al.\ 2008a, \aj, 135, 1207 

\bibitem[Melbourne et al.(2008)]{2008AJ....136.1110M} Melbourne, J., et 
al.\ 2008b, \aj, 136, 1110 



\bibitem[Melbourne et al.(2011)]{2011AJ....141..141M} Melbourne, J., et 
al.\ 2011, \aj, 141, 141 

\bibitem[Meurer et al.(1999)]{1999ApJ...521...64M} Meurer, G.~R., Heckman, 
T.~M., \& Calzetti, D.\ 1999, \apj, 521, 64 

\bibitem[Miyaji et 
al.(2000)]{2000A&A...353...25M} Miyaji, T., Hasinger, G., \& Schmidt, M.\ 2000, \aap, 353, 25 




\bibitem[Mosleh et al.(2011)]{2011ApJ...727....5M} Mosleh, M., Williams, 
R.~J., Franx, M., \& Kriek, M.\ 2011, \apj, 727, 5 

\bibitem[Naab et al.(2009)]{2009ApJ...699L.178N} Naab, T., Johansson, 
P.~H., \& Ostriker, J.~P.\ 2009, \apjl, 699, L178 




\bibitem[Nagy et al.(2011)]{2011arXiv1105.3954N} Nagy, S.~R., Law, D.~R., 
Shapley, A.~E., \& Steidel, C.~C.\ 2011, arXiv:1105.3954 

\bibitem[Oesch et al.(2010)]{2010ApJ...714L..47O} Oesch, P.~A., et al.\ 
2010, \apjl, 714, L47 



\bibitem[Padilla 
\& Strauss(2008)]{2008MNRAS.388.1321P} Padilla, N.~D., \& Strauss, M.~A.\ 2008, \mnras, 388, 1321 


\bibitem[Papovich et al.(2005)]{2005ApJ...631..101P} Papovich, C., 
Dickinson, M., Giavalisco, M., Conselice, C.~J., 
\& Ferguson, H.~C.\ 2005, \apj, 631, 101 

\bibitem[Peng et al.(2002)]{2002AJ....124..266P} Peng, C.~Y., Ho, L.~C., 
Impey, C.~D., \& Rix, H.-W.\ 2002, \aj, 124, 266 



\bibitem[Peng et al.(2010)]{2010AJ....139.2097P} Peng, C.~Y., Ho, L.~C., 
Impey, C.~D., \& Rix, H.-W.\ 2010, \aj, 139, 2097 





\bibitem[Peter et al.(2007)]{2007ApJ...668...23P} Peter, A.~H.~G., Shapley, 
A.~E., Law, D.~R., Steidel, C.~C., Erb, D.~K., Reddy, N.~A., 
\& Pettini, M.\ 2007, \apj, 668, 23 

\bibitem[Petrosian(1976)]{1976ApJ...209L...1P} Petrosian, V.\ 1976, \apjl, 
209, L1

\bibitem[Ravindranath et al.(2006)]{2006ApJ...652..963R} Ravindranath, S., 
et al.\ 2006, \apj, 652, 963 

\bibitem[Reddy et al.(2006)]{2006ApJ...653.1004R} Reddy, N.~A., Steidel, 
C.~C., Erb, D.~K., Shapley, A.~E., \& Pettini, M.\ 2006, \apj, 653, 1004 




\bibitem[Reddy et al.(2008)]{2008ApJS..175...48R} Reddy, N.~A., Steidel, 
C.~C., Pettini, M., Adelberger, K.~L., Shapley, A.~E., Erb, D.~K., 
\& Dickinson, M.\ 2008, \apjs, 175, 48 



\bibitem[Reddy 
\& Steidel(2009)]{2009ApJ...692..778R} Reddy, N.~A., \& Steidel, C.~C.\ 2009, \apj, 692, 778 


\bibitem[Reddy et al.(2010)]{2010ApJ...712.1070R} Reddy, N.~A., Erb, D.~K., 
Pettini, M., Steidel, C.~C., \& Shapley, A.~E.\ 2010, \apj, 712, 1070 

\bibitem[Rix et al.(2007)]{2007ApJ...670...15R} Rix, S.~A., Pettini, M., 
Steidel, C.~C., Reddy, N.~A., Adelberger, K.~L., Erb, D.~K., 
\& Shapley, A.~E.\ 2007, \apj, 670, 15 




\bibitem[Ryden(2006)]{2006ApJ...641..773R} Ryden, B.~S.\ 2006, \apj, 641, 
773 


\bibitem[Sales et al.(2010)]{2010MNRAS.409.1541S} Sales, L.~V., Navarro, 
J.~F., Schaye, J., Dalla Vecchia, C., Springel, V., 
\& Booth, C.~M.\ 2010, \mnras, 409, 1541 


\bibitem[Scarlata et al.(2007)]{2007ApJS..172..406S} Scarlata, C., et al.\ 
2007, \apjs, 172, 406 

\bibitem[Schade et al.(1995)]{1995ApJ...451L...1S} Schade, D., Lilly, 
S.~J., Crampton, D., Hammer, F., Le Fevre, O., 
\& Tresse, L.\ 1995, \apjl, 451, L1 


\bibitem[S{\'e}rsic(1963)]{1963BAAA....6...41S} S{\'e}rsic, J.~L.\ 1963, 
Boletin de la Asociacion Argentina de Astronomia La Plata Argentina, 6, 41 

\bibitem[Shapley et al.(2001)]{2001ApJ...562...95S} Shapley, A.~E., 
Steidel, C.~C., Adelberger, K.~L., Dickinson, M., Giavalisco, M., 
\& Pettini, M.\ 2001, \apj, 562, 95 



\bibitem[Shapley et al.(2005)]{2005ApJ...626..698S} Shapley, A.~E., 
Steidel, C.~C., Erb, D.~K., Reddy, N.~A., Adelberger, K.~L., Pettini, M., 
Barmby, P., \& Huang, J.\ 2005, \apj, 626, 698 

\bibitem[Shen et al.(2003)]{2003MNRAS.343..978S} Shen, S., Mo, H.~J., 
White, S.~D.~M., Blanton, M.~R., Kauffmann, G., Voges, W., Brinkmann, J., 
\& Csabai, I.\ 2003, \mnras, 343, 978 




\bibitem[Skrutskie et al.(2006)]{2006AJ....131.1163S} Skrutskie, M.~F., et 
al.\ 2006, \aj, 131, 1163 


\bibitem[Stark et al.(2008)]{2008Natur.455..775S} Stark, D.~P., Swinbank, 
A.~M., Ellis, R.~S., Dye, S., Smail, I.~R., 
\& Richard, J.\ 2008, \nat, 455, 775 



\bibitem[Stark et al.(2009)]{2009ApJ...697.1493S} Stark, D.~P., Ellis, 
R.~S., Bunker, A., Bundy, K., Targett, T., Benson, A., 
\& Lacy, M.\ 2009, \apj, 697, 1493 


\bibitem[Steidel et al.(2003)]{2003ApJ...592..728S} Steidel, C.~C., 
Adelberger, K.~L., Shapley, A.~E., Pettini, M., Dickinson, M., 
\& Giavalisco, M.\ 2003, \apj, 592, 728 

\bibitem[Steidel et al.(2004)]{2004ApJ...604..534S} Steidel, C.~C., 
Shapley, A.~E., Pettini, M., Adelberger, K.~L., Erb, D.~K., Reddy, N.~A., 
\& Hunt, M.~P.\ 2004, \apj, 604, 534 

\bibitem[Steidel et al.(2010)]{2010ApJ...717..289S} Steidel, C.~C., Erb, 
D.~K., Shapley, A.~E., Pettini, M., Reddy, N., Bogosavljevi{\'c}, M., 
Rudie, G.~C., \& Rakic, O.\ 2010, \apj, 717, 289 

\bibitem[Swinbank et al.(2010)]{2010MNRAS.405..234S} Swinbank, A.~M., et 
al.\ 2010, \mnras, 405, 234 




\bibitem[Szomoru et al.(2010)]{2010ApJ...714L.244S} Szomoru, D., et al.\ 
2010, \apjl, 714, L244 

\bibitem[Targett et al.(2011)]{2011MNRAS.412..295T} Targett, T.~A., Dunlop, 
J.~S., McLure, R.~J., et al.\ 2011, \mnras, 412, 295 




\bibitem[Toft et al.(2005)]{2005ApJ...624L...9T} Toft, S., van Dokkum, P., 
Franx, M., Thompson, R.~I., Illingworth, G.~D., Bouwens, R.~J., 
\& Kriek, M.\ 2005, \apjl, 624, L9 




\bibitem[Toft et al.(2007)]{2007ApJ...671..285T} Toft, S., et al.\ 2007, 
\apj, 671, 285 




\bibitem[Toft et al.(2009)]{2009ApJ...705..255T} Toft, S., Franx, M., van 
Dokkum, P., F{\"o}rster Schreiber, N.~M., Labbe, I., Wuyts, S., 
\& Marchesini, D.\ 2009, \apj, 705, 255 

\bibitem[Trujillo et al.(2006)]{2006ApJ...650...18T} Trujillo, I., et al.\ 
2006, \apj, 650, 18 




\bibitem[Trujillo et al.(2007)]{2007MNRAS.382..109T} Trujillo, I., 
Conselice, C.~J., Bundy, K., Cooper, M.~C., Eisenhardt, P., 
\& Ellis, R.~S.\ 2007, \mnras, 382, 109 

\bibitem[Tully 
\& Fisher(1977)]{1977A&A....54..661T} Tully, R.~B., \& Fisher, J.~R.\ 1977, \aap, 54, 661 



\bibitem[van den Bergh(1988)]{1988PASP..100..344V} van den Bergh, S.\ 1988, 
\pasp, 100, 344 

\bibitem[van der Wel et al.(2011)]{2011ApJ...730...38V} van der Wel, A., et 
al.\ 2011, \apj, 730, 38 


\bibitem[van Dokkum et al.(2009)]{2009PASP..121....2V} van Dokkum, P.~G., 
et al.\ 2009, \pasp, 121, 2 



\bibitem[van Dokkum et al.(2010)]{2010ApJ...709.1018V} van Dokkum, P.~G., 
et al.\ 2010, \apj, 709, 1018 





\bibitem[Williams et al.(2010)]{2010ApJ...713..738W} Williams, R.~J., 
Quadri, R.~F., Franx, M., van Dokkum, P., Toft, S., Kriek, M., 
\& Labb{\'e}, I.\ 2010, \apj, 713, 738 

\bibitem[Windhorst et al.(2011)]{2011ApJS..193...27W} Windhorst, R.~A., et 
al.\ 2011, \apjs, 193, 27 






\bibitem[Wright et al.(2009)]{2009ApJ...699..421W} Wright, S.~A., Larkin, 
J.~E., Law, D.~R., Steidel, C.~C., Shapley, A.~E., 
\& Erb, D.~K.\ 2009, \apj, 699, 421 

\bibitem[Yuma et al.(2011)]{2011ApJ...736...92Y} Yuma, S., Ohta, K., Yabe, 
K., Kajisawa, M., \& Ichikawa, T.\ 2011, \apj, 736, 92 




\end{thebibliography}
\end{document}